\documentclass{JFM-FLM_Au}

\usepackage{microtype}
\usepackage{array}
\usepackage{float}
\usepackage{placeins}
\usepackage{jfm-manuscript}
\graphicspath{{figures/}}
\hypersetup{hidelinks}
\renewcommand{\doi}[1]{\href{https://doi.org/#1}{doi:\,\nolinkurl{#1}}}

\lefttitle{}
\righttitle{Journal of Fluid Mechanics}
\title{Extracting target-related flow structures using a time-lagged cross-correlation ranking criterion}
\author[Bao and Jin]{Xuran Bao\aff{1,2} and Zhongkun Jin\aff{1,2}}
\affiliation{\aff{1}Institute of Acoustics, Chinese Academy of Sciences, Beijing 100190, PR China
\aff{2}University of the Chinese Academy of Sciences, Beijing 100190, PR China}
\corresau{Zhongkun Jin, \href{mailto:jinzhongkun@mail.ioa.ac.cn}{\textcolor{blue}{jinzhongkun@mail.ioa.ac.cn}}}

\begin{document}
\maketitle

\begin{abstract}
Structures with large global contributions to a flow field are not necessarily those that are most closely associated with a specified engineering observable. We propose a time-lagged cross-correlation (TCC) ranking criterion to extract target-related dynamical structures from a fixed modal expansion while preserving existing spatial modes and temporal coefficients. The criterion ranks mode groups by the mean squared norm of cross-responses between their flow contributions and the observable over a lag window. It is tested separately using dynamic mode decomposition (DMD) and proper orthogonal decomposition (POD). TCC-DMD analysis shows that TCC recovers three low-contribution target structures from a dominant background in a synthetic flow. The reconstruction error for the target subfield is 2.15\%. In the cylinder wake, different force observables select structures associated with the vortex-shedding fundamental, slow drag variations, and the second harmonic. In the cavity case, selected modes connect two main frequency bands of the target observable to distinct spatial amplitude and phase distributions. TCC-POD analysis shows that the criterion minimises the weighted truncation error of the correlation operator among equally sized subsets of the fixed orthogonal basis. A comparison with canonical correlation decomposition further reveals differences between fixed-direction selection and correlation-based basis construction in the spatial structure and retention of flow information. Lag aggregation can improve structure identification when zero-lag association is insufficient. The gains depend on the cross-response distribution and window selection. TCC separates target-related ranking from mode-set construction, enabling comparison of spatial structures associated with different observables within the same modal expansion.
\end{abstract}

\section{Introduction}
High-fidelity numerical simulations and time-resolved experiments provide high-dimensional flow-field snapshots; however, engineering diagnostics must also identify dynamical structures associated with a specified response~\citep{taira2017,taira2020,schmid2022}. Structures with large fluctuation energy can dominate low-dimensional representations of the velocity field yet may not be the most relevant to specific engineering observables, such as lift, drag, local pressure, or acoustic responses. Structure extraction for a specific observable must also describe the spatial forms and timescales of the associated structures.

The spectral properties of the Koopman operator provide a theoretical basis for modal decomposition of nonlinear dynamical systems~\citep{mezic2005}. Rowley et al. applied this concept to the structural analysis of nonlinear flows~\citep{rowley2009}. Dynamic mode decomposition (DMD), introduced by Schmid, associates spatial modes with the frequencies and growth rates of an approximate evolution operator~\citep{schmid2010}. Tu et al. generalised its definition through the eigendecomposition of an approximate linear operator and introduced linear consistency~\citep{tu2014}. To improve spectral estimation, DMD developments have addressed optimised fitting, noise bias correction, and residual tests~\citep{chen2012,askham2018,dawson2016,hemati2017,colbrook2023residual}. Subspace DMD uses correlation matrices of past and future snapshots from the same observation sequence to estimate dynamical spectra from noisy data~\citep{takeishi2017}. For compact flow-field representations, sparsity-promoting DMD balances reconstruction error against the number of modes~\citep{jovanovic2014}, whereas recursive DMD progressively extracts structures from the residual~\citep{noack2016}. Time-averaged modal contributions and the Kou--Zhang criterion measure modal contributions to the flow field over the sampling period~\citep{tissot2014,kou2017}. Beyond evaluating modal contributions to the flow-field representation, DMD spectral information also helps to identify and compare oscillatory processes and transient structures~\citep{bagheri2013}. For example, \citet{he2013disturbance} used DMD to analyse boundary-layer disturbances induced by a cylinder wake. The first streamwise growth stage mainly involved amplification at the shedding frequency, whereas the second involved several frequency components.

Proper orthogonal decomposition (POD) provides an energy-optimal orthogonal representation of a given snapshot ensemble~\citep{sirovich1987,berkooz1993}, although individual spatial modes can contain multiple timescales. For statistically stationary flows, spectral proper orthogonal decomposition provides frequency-resolved energy-optimal structures~\citep{towne2018,schmidt2020guide}. \citet{muld2012} used POD and DMD to extract dominant flow structures in a high-speed train wake. For reduced-order modelling, \citet{sipp2020pod} compared the stability, accuracy, and robustness of POD models using different approximations of nonlinear terms. \citet{zong2026couette} compared POD and balanced-truncation modes for Couette flow and showed that reduced-order model performance depends on the flow state. These studies provide different modal tools for flow-structure identification and reduced-order modelling. However, modal spectral characteristics or global flow-field contributions alone cannot establish statistical association with a specified observable.

Engineering responses can enter flow analysis through input--output relationships. DMD with control explicitly separates autonomous dynamics from known control inputs~\citep{proctor2016}. Balanced POD approximately retains directions that are both strongly controllable and observable~\citep{rowley2005}, whereas resolvent analysis uses system operators to examine forcing--response gains~\citep{mckeon2010,taira2017}. In passive measurements, observation signals can directly inform the interpretation or selection of existing DMD modes. Cross-correlations between acoustic signals and the temporal signals of DMD modes have been used to identify acoustically related structures~\citep{assoum2021}. Correlations between DMD spectral amplitudes of pressure and velocity have also been used to analyse statistical associations between flow quantities~\citep{ahani2025}. In underwater acoustics, interfering modes have been rejected using thresholds on correlation coefficients between modes and received data~\citep{remadevi2022}.

Correlation information between a flow field and an observable can also support flow-field estimation or the construction of a new correlation-based basis. Spectral linear stochastic estimation constructs frequency-domain estimates from cross-spectral relationships between observation signals and the flow field~\citep{tinney2006slse}. Lyu's canonical correlation decomposition (CCD) incorporates time-shifted observables and constructs correlation-optimal orthogonal modes from the singular directions of the cross-correlation matrix~\citep{lyu2024}. CCD was subsequently used to identify structures associated with local wall-friction fluctuations in turbulent channel flow~\citep{nie2026}. Principal correlation decomposition has also identified structures associated with far-field sound pressure in rod--airfoil flow~\citep{huang2026rodairfoil}. In this study, we investigate how a specified observable can guide the ranking of existing mode groups while preserving their spatial modes and temporal coefficients. This constraint allows selections for different targets to be compared within the same modal expansion.

We propose a time-lagged cross-correlation (TCC) ranking criterion, which evaluates the time-dependent flow-field contribution of each group within a fixed modal expansion. Spatial modes and temporal coefficients are first determined from the flow field. TCC then ranks groups by the mean squared norm of their cross-responses with the observable over a prescribed lag window. Selected contributions retain the given coefficients; changing the observable changes only the ranking. The criterion applies to expansions with separable spatial modes and temporal coefficients, without requiring orthogonality or exponential temporal evolution. We present the general score, its operator representation, and implementations on DMD and POD mode sets constructed for three cases. These implementations are termed TCC-DMD and TCC-POD, respectively.

Changes in the flow structures and observed responses need not be synchronous. Strong cross-responses may occur at non-zero lags or extend over a finite lag interval. TCC combines squared cross-responses across temporal offsets within the window rather than using only the zero-lag cross-response or a single maximum correlation peak. Whether non-zero lags improve the ranking depends on the actual cross-response distribution and the chosen window. When zero lag already captures the principal association, extending the lag range may provide no additional identification information.

We construct DMD and POD mode sets for a synthetic flow field, a cylinder wake, and cavity experiments to assess target-related structure extraction and reconstruction. The DMD analysis combines eigenvalue frequencies, projection-coordinate spectra, and spatial forms to examine the dynamical processes associated with selected structures. The POD analysis examines target-related selection among weighted orthogonal modes. Using CCD as a reference, it compares structure extraction and reconstruction through fixed-direction selection and correlation-based basis construction. Global-contribution, zero-lag, and peak-aggregation rankings distinguish the effects of the observable and different lag information on structure selection.
\section{TCC on a fixed mode set}
\label{sec:method}
\label{sec:fixed_library_framework}

\begin{figure}[!htbp]
    \centering
    \includegraphics[width=\textwidth]{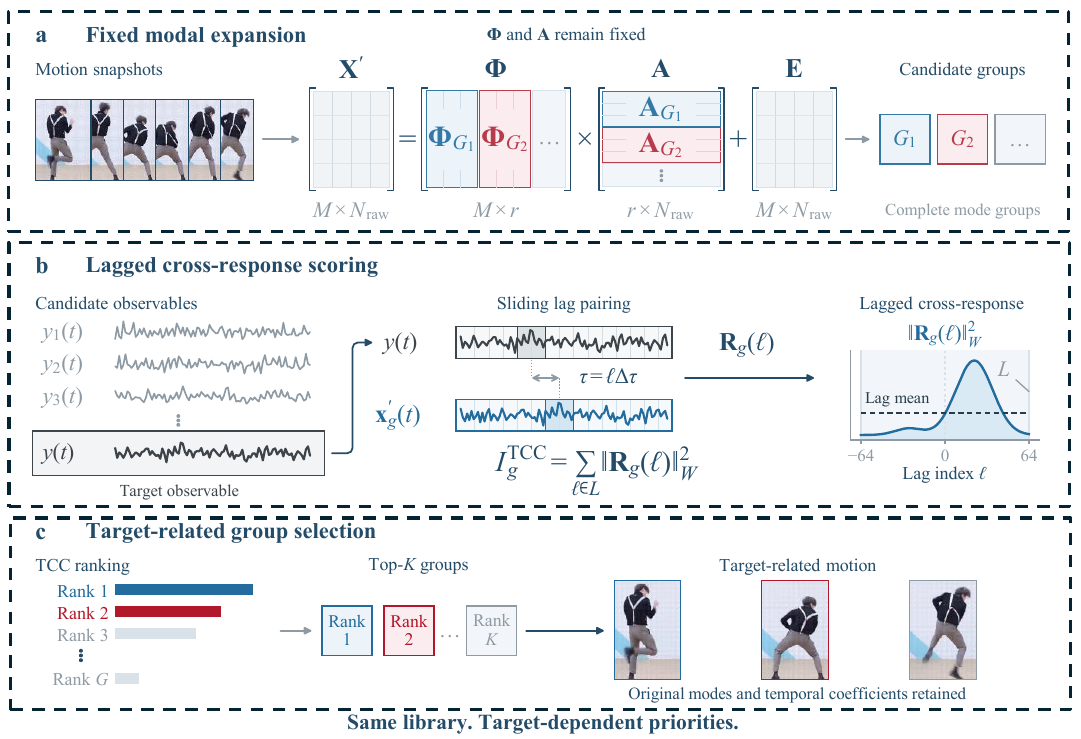}
    \caption{Dance--music analogy illustrating TCC ranking within a fixed modal expansion. (a) Motion snapshots provide spatial modes and temporal coefficients, which remain fixed during target-related ranking. (b) A selected music signal serves as the target observable $y(t)$; mode groups are ranked by the mean squared cross-response norm over a prescribed lag window. (c) The top $K$ mode groups are retained intact, and their contributions are synthesised using the original spatial modes and temporal coefficients. Waveforms and rankings are schematic, and the dance frames illustrate the analogy without representing computed reconstructions.}
    \label{fig:tcc_dmd_overview}
\end{figure}

TCC evaluates flow-field contributions synthesised from fixed spatial modes and their temporal coefficients. The mode set and coefficients are first determined from the flow field. Mode groups are then ranked by their time-lagged cross-responses with a specified observable, and the selected contributions are synthesised (figure~\ref{fig:tcc_dmd_overview}). Within each fixed-mode-set comparison, all ranking methods share spatial modes, temporal coefficients, physical grouping, and spatial weights.

\subsection{Fixed modal expansion and mode-group contributions}
\label{sec:method_library}

Let $\mathbf{x}_k\in\mathbb R^M$ denote the spatially discretised flow state, sampled at $N_{\mathrm{raw}}$ uniformly spaced times $t_k=t_1+(k-1)\Delta t$. Subtracting the mean field $\overline{\mathbf{x}}$ gives the fluctuation state $\mathbf{x}'_k=\mathbf{x}_k-\overline{\mathbf{x}}$ and snapshot matrix $\mathbf{X}'=[\mathbf{x}'_1,\ldots,\mathbf{x}'_{N_{\mathrm{raw}}}]$. Consider an expansion with separable spatial modes and temporal coefficients~\citep{taira2017},
\begin{equation}
    \mathbf X'=\boldsymbol\Phi\mathbf A+\mathbf E,
    \qquad
    \boldsymbol\Phi\mathbf A
    =\sum_{j=1}^{r}\boldsymbol\phi_j\mathbf a_j^T,
    \label{eq:general_modal_expansion}
\end{equation}
where $\boldsymbol{\Phi}=[\boldsymbol\phi_1,\ldots,\boldsymbol\phi_r]\in\mathbb C^{M\times r}$ is the fixed spatial-mode matrix and $\mathbf{A}=[a_j(k)]\in\mathbb C^{r\times N_{\mathrm{raw}}}$ is the temporal-coefficient matrix. Here, $\mathbf{a}_j=[a_j(1),\ldots,a_j(N_{\mathrm{raw}})]^T$, and $\mathbf{E}$ is the representation residual. Spatial modes may be non-orthogonal; coefficients are supplied by the decomposition or obtained by a joint solution over the full mode set. Ranking and truncated reconstruction retain the same coefficients.

Mode-group contributions and cross-responses use the spatially weighted inner product
\begin{equation}
    \langle\mathbf{f},\mathbf{g}\rangle_W
    =\mathbf{f}^{H}\mathbf{W}\mathbf{g},
    \qquad
    \lVert\mathbf{f}\rVert_W^2
    =\mathbf{f}^{H}\mathbf{W}\mathbf{f},
    \label{eq:weighted_inner_product}
\end{equation}
where $\mathbf{W}=\mathbf{W}^{T}\succ0$ is a real, symmetric, positive-definite spatial weighting matrix. Diagonal cell-area or cell-volume weights are used for non-uniform finite-volume meshes; $\mathbf{W}=\mathbf{I}$ is used for uniform meshes without available integration weights. Superscript $H$ denotes the conjugate transpose, and this inner product is used throughout.

Partition the modes into disjoint groups, with index set $\mathcal G_g$ for group $g$. Its flow-field contribution at snapshot $k$ is
\begin{equation}
    \mathbf x'_g(k)
    =\sum_{j\in\mathcal G_g}\boldsymbol\phi_j a_j(k)
    =\boldsymbol\Phi_g\mathbf a_g(k),
    \label{eq:pair_contribution}
\end{equation}
Here, $\boldsymbol\Phi_g$ and $\mathbf{a}_g(k)$ contain the group's spatial modes and corresponding coefficients. Grouping is fixed before ranking so that each group independently synthesises a real-valued flow-field contribution. Real modes can form individual groups, whereas complex-conjugate modes are grouped together; single-mode scoring corresponds to $\mathcal G_g=\{j\}$. Top-$K$ denotes selection of the first $K$ complete mode groups.

\subsection{Observable cross-responses and TCC scoring}
\label{sec:method_scoring}

We measure the global contribution of a group using the root mean square (RMS) of its flow-field contribution. For a non-empty evaluation snapshot index set $\mathcal S\subseteq\{1,\ldots,N_{\mathrm{raw}}\}$, define
\begin{equation}
    I_g^{\mathrm{RMS}}(\mathcal S)
    =\left[
    \frac{1}{|\mathcal S|}
    \sum_{k\in\mathcal S}\lVert\mathbf x'_g(k)\rVert_W^2
    \right]^{1/2}.
    \label{eq:rms_criterion}
\end{equation}
This quantity measures the time-averaged amplitude of the synthesised group contribution, following the time-averaging rationale of existing modal-contribution criteria~\citep{tissot2014}. To identify structures associated with a specified observable, we additionally introduce flow-field--observable cross-responses.

Let $y(t)$ be a scalar observable with a known temporal correspondence to the flow field, such as local velocity, pressure, aerodynamic force, or acoustic pressure. The lag increment $\Delta\tau>0$ can be chosen independently of the snapshot interval $\Delta t$. We define the lag index set and corresponding lags as
\begin{equation}
    \mathcal{L}=\{\ell_0,\ell_0+1,\ldots,\ell_0+Q-1\},
    \qquad \tau_{\ell}=\ell\Delta\tau,
    \label{eq:lag_set}
\end{equation}
where $\ell_0\in\mathbb Z$, and the positive integer $Q=|\mathcal L|$ is the number of lags. Snapshot $\mathbf{x}'_k$ is paired with the observation at $t_k+\tau_\ell$. Negative and positive $\ell$ indicate observation times before and after that snapshot, respectively.

All lags share reference snapshots with paired observations available throughout the window. Their index set is
\begin{equation}
    \mathcal I=\{k_1,\ldots,k_N\},\qquad N=|\mathcal I|>0.
    \label{eq:common_valid_index}
\end{equation}
Writing $\ell_q=\ell_0+q-1$, we centre the paired observations separately at each lag to obtain
\begin{equation}
    \widetilde{y}_{q,n}
    =y(t_{k_n}+\tau_{\ell_q})
    -\frac{1}{N}\sum_{m=1}^{N}y(t_{k_m}+\tau_{\ell_q}),
    \qquad
    \mathbf{Y}_{\mathcal{L}}
    =\left[\widetilde{y}_{q,n}\right]_{
    \substack{q=1,\ldots,Q\\n=1,\ldots,N}}
    \in\mathbb{R}^{Q\times N}.
    \label{eq:lagged_observable_matrix}
\end{equation}
We set $\mathcal S=\mathcal I$ for the RMS such that the global-contribution and target-related scores use the same data interval.

At lag $\tau_{\ell_q}$, define the cross-response between mode $j$'s temporal coefficient and the observable as
\begin{equation}
    c_j(\ell_q)=\frac{1}{N\sqrt{Q}}
    \sum_{n=1}^{N}\widetilde{y}_{q,n}a_j(k_n),
    \label{eq:modal_cross_correlation}
\end{equation}
The factor $1/N$ averages over the common valid samples, while $1/\sqrt Q$ makes the subsequent sum of squared norms a lag average. TCC retains the amplitudes of the flow-field contribution and observable; it is not a normalised correlation coefficient.

Using \eqref{eq:pair_contribution}, the lagged cross-response field between mode group $g$ and the observable is
\begin{equation}
    \begin{aligned}
        \mathbf{R}_g(\ell_q)
        &=\sum_{j\in\mathcal{G}_g}\boldsymbol{\phi}_j c_j(\ell_q)\\
        &=\frac{1}{N\sqrt Q}\sum_{n=1}^{N}
        \widetilde y_{q,n}\mathbf{x}'_g(k_n).
    \end{aligned}
    \label{eq:correlation_field}
\end{equation}
Thus, $\mathbf{R}_g(\ell_q)$ is a cross-covariance-type spatial cross-response between a fixed mode group and the time-shifted observable. Summing its squared norm over the lag window defines the TCC score,
\begin{equation}
    I_g^{\mathrm{TCC}}
    =\sum_{\ell\in\mathcal{L}}
    \left\lVert\mathbf{R}_g(\ell)\right\rVert_W^2.
    \label{eq:tcc_strength}
\end{equation}
Substituting the cross-response into \eqref{eq:tcc_strength} gives
\begin{equation}
    I_g^{\mathrm{TCC}}
    =\frac{1}{Q}\sum_{q=1}^{Q}
    \left\lVert
    \frac{1}{N}\sum_{n=1}^{N}
    \widetilde y_{q,n}\mathbf{x}'_g(k_n)
    \right\rVert_W^2.
    \label{eq:tcc_mean_square_response}
\end{equation}
Normalisation by $1/Q$ gives the mean squared cross-response within the window, avoiding an increase proportional solely to the number of lags.
Let $\mathbf{c}_g(\ell_q)=[c_j(\ell_q)]_{j\in\mathcal G_g}$ be the coefficient cross-response vector and $\mathbf{G}_g=\boldsymbol\Phi_g^H\mathbf{W}\boldsymbol\Phi_g$ the spatial-mode Gram matrix. As $\mathbf{R}_g=\boldsymbol\Phi_g\mathbf{c}_g$,
\begin{equation}
    I_g^{\mathrm{TCC}}
    =\sum_{q=1}^{Q}\mathbf c_g^H(\ell_q)\mathbf G_g\mathbf c_g(\ell_q).
    \label{eq:tcc_general_gram}
\end{equation}
Equation~\eqref{eq:tcc_general_gram} retains cross-terms within non-orthogonal mode groups. For groups whose members are weighted orthonormal, $\mathbf{G}_g=\mathbf{I}$, and each group's score is the sum of its members' scores. All candidate groups are scored using~\eqref{eq:tcc_strength} before the Top-$K$ groups are selected.

Zero-lag cross-correlation ranking (CC$_0$) provides a reference for assessing the contribution of non-zero lags. It evaluates only the zero-lag cross-response; therefore, the common index set $\mathcal I$ must also have observations available at zero lag. Writing $\widetilde y_{0,n}=y(t_{k_n})-N^{-1}\sum_{m=1}^{N}y(t_{k_m})$, define
\begin{equation}
    I_g^{\mathrm{CC0}}
    =\left\lVert\frac{1}{N}\sum_{n=1}^{N}
    \widetilde y_{0,n}\mathbf x'_g(k_n)\right\rVert_W^2.
    \label{eq:cc0_strength}
\end{equation}
On the same common valid samples, CC$_0$ is the special case of TCC with $\mathcal L=\{0\}$. When $0\in\mathcal L$, $I_g^{\mathrm{CC0}}=Q\lVert\mathbf{R}_g(0)\rVert_W^2$.

Peak-aggregation ranking (MaxCC) uses the largest squared cross-response within the same lag window,
\begin{equation}
    I_g^{\mathrm{MaxCC}}=Q\max_{\ell\in\mathcal L}
    \|\mathbf R_g(\ell)\|_W^2.
    \label{eq:maxcc_strength}
\end{equation}
On the same common valid samples, CC$_0$ measures the zero-lag cross-response, MaxCC the strongest cross-response, and TCC the mean squared cross-response over the window. Therefore, broad cross-responses and peaks confined to a few lags can yield different rankings. The conjugate-group score expansion is given in Appendix~\ref{app:group_algebra}.

\subsection{Selected-mode flow-field reconstruction and operator representation}
\label{sec:method_extraction}

Let $\mathcal S_K$ contain the mode indices in the first $K$ ranked groups. Retaining the temporal coefficients of the full expansion, we synthesise the selected contributions and restore the mean field,
\begin{equation}
    \widehat{\mathbf x}_k
    =\overline{\mathbf x}
    +\operatorname{Re}\left[
    \sum_{j\in\mathcal S_K}\boldsymbol\phi_j a_j(k)
    \right].
    \label{eq:ranked_reconstruction}
\end{equation}
where $\widehat{\mathbf{x}}'_k=\widehat{\mathbf{x}}_k-\overline{\mathbf{x}}$ is the fluctuation contribution of the selected groups.

To relate group scores to the full flow-field cross-response, let $\mathbf{X}_{\mathcal I}=[\mathbf{x}'_{k_1},\ldots,\mathbf{x}'_{k_N}]$ and define the time-lagged flow-field--observable correlation operator,
\begin{equation}
    \mathbf C_{xy}
    =\frac{1}{N\sqrt Q}\mathbf X_{\mathcal I}\mathbf Y_{\mathcal L}^{H}.
    \label{eq:full_correlation_operator}
\end{equation}
When coefficients follow a fixed linear mapping $\mathbf{a}(k)=\mathbf{B}\mathbf{x}'_k$, let the diagonal matrix $\mathbf{S}_g$ select group $g$'s coordinates. The group reconstruction operator is
\begin{equation}
    \mathcal P_g=\boldsymbol\Phi\mathbf S_g\mathbf B,
    \qquad
    \mathbf x'_g(k)=\mathcal P_g\mathbf x'_k.
    \label{eq:dmd_group_reconstruction_operator}
\end{equation}
TCC scores are determined by the lagged cross-responses extracted by this operator. Cross-terms between non-orthogonal group cross-responses distinguish individual-group ranking from joint subset optimisation. Matrix forms, cross-response decompositions, and invariance to within-group representations are given in Appendix~\ref{app:operator_theory}. Next, we  describe fixed-mode-set construction and TCC implementation for DMD and POD.

\subsection{DMD mode-set construction and projection coordinates}
\label{sec:method_dmd}

We construct a fixed mode set using exact DMD~\citep{schmid2010,tu2014} and denote TCC applied to this set by TCC-DMD. Adjacent fluctuation snapshots form
\begin{equation}
    \mathbf{X}_1=
    \begin{bmatrix}
        \mathbf{x}'_1&\mathbf{x}'_2&\cdots&\mathbf{x}'_{N_{\mathrm{raw}}-1}
    \end{bmatrix},\qquad
    \mathbf{X}_2=
    \begin{bmatrix}
        \mathbf{x}'_2&\mathbf{x}'_3&\cdots&\mathbf{x}'_{N_{\mathrm{raw}}}
    \end{bmatrix}.
    \label{eq:snapshot_matrices}
\end{equation}
Define the weighted snapshot matrices
\begin{equation}
    \mathbf{X}_{1,W}=\mathbf{W}^{1/2}\mathbf{X}_1,
    \qquad
    \mathbf{X}_{2,W}=\mathbf{W}^{1/2}\mathbf{X}_2,
    \label{eq:weighted_snapshot_matrices}
\end{equation}
and a rank-$r$ truncated singular value decomposition (SVD) is applied directly to $\mathbf{X}_{1,W}$,
\begin{equation}
    \mathbf{X}_{1,W}
    \approx\mathbf{U}_r\boldsymbol{\Sigma}_r\mathbf{V}_r^{H},
    \qquad
    \mathbf{U}_r^{H}\mathbf{U}_r=\mathbf{I}.
    \label{eq:svd}
\end{equation}
After preprocessing, the truncation rank $r$ is determined solely from the numerical rank of the flow snapshots or a prescribed singular-value truncation criterion. Neither target-related scores nor target-signal reconstruction errors were used to tune $r$. Once the mode set is established, every ranking criterion uses the same $r$, eigenvalue matrix $\boldsymbol{\Lambda}$, and spatial mode matrix $\boldsymbol{\Phi}$.
The weighted snapshot matrices give the reduced evolution operator and its eigendecomposition,
\begin{equation}
    \widetilde{\mathbf{A}}_W
    =\mathbf{U}_r^{H}\mathbf{X}_{2,W}
    \mathbf{V}_r\boldsymbol{\Sigma}_r^{-1},
    \qquad
    \widetilde{\mathbf{A}}_W\mathbf{Y}
    =\mathbf{Y}\boldsymbol{\Lambda}.
    \label{eq:reduced_operator}
\end{equation}
Mapping the resulting reduced eigenvectors to the physical space yields the exact DMD modes,
\begin{equation}
    \boldsymbol{\Phi}
    =\mathbf{W}^{-1/2}\mathbf{X}_{2,W}
    \mathbf{V}_r\boldsymbol{\Sigma}_r^{-1}\mathbf{Y}
    =\mathbf{X}_2\mathbf{V}_r\boldsymbol{\Sigma}_r^{-1}\mathbf{Y},
    \label{eq:exact_dmd_modes}
\end{equation}
DMD eigenvectors can be multiplied by an arbitrary non-zero complex number. To establish a common scale for comparing projection coordinates, each physical-space mode is normalised before calculation as
\begin{equation}
    \boldsymbol{\phi}_j
    \leftarrow
    \frac{\boldsymbol{\phi}_j}
    {\sqrt{\boldsymbol{\phi}_j^{H}\mathbf{W}\boldsymbol{\phi}_j}},
    \qquad
    \lVert\boldsymbol{\phi}_j\rVert_W=1,
    \qquad j=1,\ldots,r.
    \label{eq:dmd_mode_normalization}
\end{equation}
Hereafter, $\boldsymbol{\Phi}$ denotes the normalised mode matrix.
The eigenvalues provide continuous-time spectral quantities
\begin{equation}
    \omega_j=\frac{\log\lambda_j}{\Delta t},\qquad
    f_j=\frac{\operatorname{Im}(\omega_j)}{2\pi},\qquad
    \sigma_j=\operatorname{Re}(\omega_j),
    \label{eq:dmd_frequency_growth}
\end{equation}
Here, $\lambda_j$, $f_j$, and $\sigma_j$ denote the discrete eigenvalue, DMD frequency, and growth rate, respectively. These spectral quantities and spatial modes depend on the flow field and weights. Snapshot coefficients are then computed with the mode set held fixed.

For each snapshot, all modal coefficients are jointly determined by spatially weighted least squares over the full DMD mode set $\boldsymbol{\Phi}$,
\begin{equation}
    \mathbf{a}^{\mathrm{full}}_k
    =\underset{\mathbf{c}}{\operatorname{arg\,min}}
    \left\lVert\mathbf{x}'_k-\boldsymbol{\Phi}\mathbf{c}\right\rVert_W^2
    =\left(\boldsymbol{\Phi}^{H}\mathbf{W}\boldsymbol{\Phi}\right)^{+}
    \boldsymbol{\Phi}^{H}\mathbf{W}\mathbf{x}'_k,
    \qquad k=1,\ldots,N_{\mathrm{raw}}.
    \label{eq:projected_coefficient}
\end{equation}
Here, superscript $+$ denotes the Moore--Penrose pseudoinverse, and $a_j(k)$ is the $j$th snapshot-wise projection coordinate, abbreviated to projection coordinate~\citep{tu2014,priebe2016}. All ranking methods share the coordinates computed over the full mode set. These coordinates need not follow the exponential evolution $a_j(k)=b_j\exp[\omega_j(t_k-t_1)]$ implied by initial amplitude $b_j$ and the eigenvalue. Whereas $f_j$ and $\sigma_j$ describe the operator spectrum, the power spectrum of $a_j(k)$ describes the frequency distribution of the actual snapshot contribution. Numerical rank, conditioning, and conjugate-pairing checks are given in Appendix~\ref{app:numerical_audit}.

Each complex-conjugate DMD pair forms one group, while each mode with a real eigenvalue forms its own group. The ranking comparisons here consider only the oscillatory groups. Every valid conjugate group participates and is identified by its positive-frequency member satisfying $0<f_j<f_N=1/(2\Delta t)$. Modes with real eigenvalues remain in the full-mode-set projection but are excluded from this oscillatory-group comparison. Both conjugate members contribute to the cross-response in~\eqref{eq:correlation_field}, retaining within-group phase coupling in the score; the expansion is given in Appendix~\ref{app:group_algebra}.

\subsection{POD mode-set construction and temporal coefficients}
\label{sec:method_pod}

POD constructs an energy-optimal low-dimensional representation through an orthogonal expansion of the snapshot ensemble~\citep{sirovich1987,berkooz1993}. Using the spatial weights in~\eqref{eq:weighted_inner_product}, apply a rank-$r$ truncated SVD to the weighted matrix of all fluctuation snapshots,
\begin{equation}
    \mathbf W^{1/2}\mathbf X'
    \approx\mathbf U_r^{\mathrm P}\boldsymbol\Sigma_r^{\mathrm P}
    (\mathbf V_r^{\mathrm P})^H,
    \qquad
    (\mathbf U_r^{\mathrm P})^H\mathbf U_r^{\mathrm P}=\mathbf I.
    \label{eq:pod_weighted_svd}
\end{equation}
Superscript $\mathrm P$ distinguishes the POD matrices from those obtained using adjacent snapshots in DMD construction. The physical-space POD modes and temporal coefficients are
\begin{equation}
    \begin{aligned}
        \boldsymbol\Phi^{\mathrm P}
        &=\mathbf W^{-1/2}\mathbf U_r^{\mathrm P},\\
        \mathbf A^{\mathrm P}
        &=\boldsymbol\Sigma_r^{\mathrm P}(\mathbf V_r^{\mathrm P})^H
        =(\boldsymbol\Phi^{\mathrm P})^H\mathbf W\mathbf X'.
    \end{aligned}
    \label{eq:pod_modes_coefficients}
\end{equation}
Thus, $(\boldsymbol{\Phi}^{\mathrm P})^H\mathbf{W}\boldsymbol{\Phi}^{\mathrm P}=\mathbf{I}$, and POD temporal coefficients are snapshot projection coordinates on the weighted orthogonal modes. Mode-set construction and the truncation rank $r$ depend only on flow snapshots, without using the target observable; singular values determine the retained energy. Writing the $j$th singular value as $s_j^{\mathrm P}$, the mean squared modal contribution and cumulative energy fraction are
\begin{equation}
    \mu_j=\frac{(s_j^{\mathrm P})^2}{N_{\mathrm{raw}}}
    =\frac{1}{N_{\mathrm{raw}}}\sum_{k=1}^{N_{\mathrm{raw}}}|a_j^{\mathrm P}(k)|^2,
    \qquad
    \eta_r=\frac{\sum_{j=1}^{r}(s_j^{\mathrm P})^2}
    {\sum_{j=1}^{r_{\mathrm{full}}}(s_j^{\mathrm P})^2},
    \label{eq:pod_energy}
\end{equation}
where $r_{\mathrm{full}}$ is the effective rank of the weighted snapshot matrix. POD energy ranking retains modes in descending order of $\mu_j$. On evaluation index set $\mathcal S$, the squared contribution corresponding to the RMS is $|\mathcal S|^{-1}\sum_{k\in\mathcal S}|a_j^{\mathrm P}(k)|^2$. The RMS and POD energy rankings coincide when all snapshots are used; their orders may differ when the evaluation interval changes.

TCC reorders priorities within this fixed POD mode set, and this implementation is denoted TCC-POD. Each real POD mode is a selection unit, with $\mathcal G_j=\{j\}$. Substituting~\eqref{eq:pod_modes_coefficients} into~\eqref{eq:modal_cross_correlation} gives
\begin{equation}
    c_j^{\mathrm P}(\ell_q)
    =\frac{1}{N\sqrt Q}\sum_{n=1}^{N}
    \widetilde y_{q,n}a_j^{\mathrm P}(k_n),
    \qquad
    I_j^{\mathrm{TCC}}
    =\sum_{q=1}^{Q}|c_j^{\mathrm P}(\ell_q)|^2.
    \label{eq:tcc_pod_specialization}
\end{equation}
POD energy ranking measures global squared contributions, whereas TCC measures target cross-responses; both retain the same spatial modes and temporal coefficients. Weighted orthogonality makes the squared cross-responses of different modes additive. Among all $K$-mode subsets of this fixed POD set, the first $K$ TCC modes minimise the weighted truncated-approximation error of the correlation operator. The orthogonal-projection error decomposition is given in~\eqref{eq:tcc_orthogonal_selection_error} of the appendix.

After selecting the first $K$ real modes,~\eqref{eq:ranked_reconstruction} becomes
\begin{equation}
    \widehat{\mathbf x}_k
    =\overline{\mathbf x}
    +\sum_{j\in\mathcal S_K}\boldsymbol\phi_j^{\mathrm P}a_j^{\mathrm P}(k).
    \label{eq:pod_ranked_reconstruction}
\end{equation}
The fluctuation part is the weighted orthogonal projection of the original fluctuations onto the selected POD subspace; the mean field is then added back. Retaining $K$ real modes gives a $K$-dimensional real fluctuation subspace.

\subsection{Comparison methods and evaluation settings}
\label{sec:method_evaluation}

We evaluate ranking criteria through selected structures, target-signal reconstruction, and flow-field representation. Known target structures and the target subfield provide validation in the synthetic case. For the cylinder and cavity cases, spatial forms, DMD spectra, and actual projection-coordinate spectra relate selected structures to the observable. Within each mode set, all criteria share spatial modes, temporal coefficients, selection units, weights, and retention budgets. DMD comparisons retain equal numbers of complete groups; POD comparisons retain equal numbers of real modes.

RMS, CC$_0$, MaxCC, and TCC are defined in \S\ref{sec:method_scoring}; table~\ref{tab:baseline_hierarchy} summarises their comparison purposes and additional references. All correlation criteria use the same common valid samples, as does the DMD RMS. The POD energy ranking uses $\mu_j$ computed over all snapshots. Window sweeps keep $\Delta\tau$ fixed and define one common valid index set $\mathcal I_{\max}$ from the largest candidate window. All smaller windows retain this index set to separate lag-range effects from sample-interval effects.

\begin{table}[htbp]
    \centering
    \small
    \caption{Comparison purposes for fixed-mode ranking and reference methods for DMD and POD.}
    \label{tab:baseline_hierarchy}
    \renewcommand{\arraystretch}{1.15}
    \begin{tabular*}{\textwidth}{@{\extracolsep{\fill}}p{0.23\textwidth}p{0.23\textwidth}p{0.48\textwidth}@{}}
        \toprule
        Comparison role & Method & Purpose \\
        \midrule
        Global contribution & RMS; POD energy ranking & Compare selections based on global contribution and target relevance \\
        Zero-lag comparison & CC$_0$ & Assess additional selection information from nonzero lags \\
        Lag-aggregation comparison & MaxCC & Compare peak and mean squared cross-response aggregation \\
        DMD ranking reference & Kou--Zhang criterion & Compare existing full-record modal-contribution ranking \\
        POD correlation-basis reference & CCD & Compare fixed-direction selection with construction of correlation-based directions \\
        \bottomrule
    \end{tabular*}
\end{table}

For DMD, we use the Kou--Zhang full-record criterion as an existing ranking reference~\citep{kou2017}. Let $\widehat{\boldsymbol\phi}_j=\boldsymbol\phi_j/\lVert\boldsymbol\phi_j\rVert_F$ and $\widehat b_j=b_j\lVert\boldsymbol\phi_j\rVert_F$, where $\mathbf{b}=\mathbf{a}_1^{\mathrm{full}}$. The group score is
\begin{equation}
    I_g^{\mathrm{KZ}}
    =\sum_{j\in\mathcal G_g}
    \sum_{k=1}^{N_{\mathrm{raw}}}
    \left|\widehat b_j\lambda_j^{k-1}\right|\Delta t.
    \label{eq:kou_zhang_criterion}
\end{equation}
This criterion uses exponential temporal sequences constructed from initial amplitudes and eigenvalues, with all samples and Frobenius-normalised modes.

The POD comparison includes CCD, which constructs spatial directions from flow-field--observable correlations~\citep{lyu2024}. CCD and TCC-POD share the flow field, spatial weights, observable, lag window, and common valid samples. A weighted SVD of the full correlation operator gives
\begin{equation}
    \begin{aligned}
        \mathbf W^{1/2}\mathbf C_{xy}&=\mathbf U_C\boldsymbol\Sigma_C\mathbf V_C^H,\\
        \boldsymbol\Psi_K^{\mathrm C}&=\mathbf W^{-1/2}\mathbf U_{C,K},
        \qquad
        \mathbf A_K^{\mathrm C}=(\boldsymbol\Psi_K^{\mathrm C})^H\mathbf W\mathbf X'.
    \end{aligned}
    \label{eq:ccd_comparison_basis}
\end{equation}
CCD constructs new spatial directions from the full correlation operator. The POD energy ranking, TCC-POD, and CCD each retain $K$ real modes and reconstruct the flow field using their respective weighted projection coordinates.

Target-signal reconstruction consistently uses real-valued features from the selected temporal coefficients. Real DMD groups use real coordinates; conjugate groups use the real and imaginary parts of the coordinate for positive-frequency representative $j_g$,
\begin{equation}
    \mathbf z_K(k)=
    \left[
    \left\{\operatorname{Re}a_j(k)\right\}_{\mathcal G_g=\{j\}\subset\mathcal S_K};
    \left\{\operatorname{Re}a_{j_g}(k),
    \operatorname{Im}a_{j_g}(k)\right\}_{
    \mathcal G_g=\{j_g,j_g^*\}\subset\mathcal S_K}
    \right]^T
    \in\mathbb R^{d_K}.
    \label{eq:observable_features}
\end{equation}
Each POD mode directly supplies its real coefficient.
Let $\mathcal E=\{e_1,\ldots,e_{N_E}\}$ be the fitting and evaluation index set, with $N_E$ samples. Each feature of $\mathbf{z}_K$ is centred and scaled by its standard deviation on this set, giving $\widetilde{\mathbf{z}}_K$. Including an intercept gives the design matrix
\begin{equation}
    \mathbf H_K=
    \begin{bmatrix}
        1&\widetilde{\mathbf z}_K(e_1)^T\\
        \vdots&\vdots\\
        1&\widetilde{\mathbf z}_K(e_{N_E})^T
    \end{bmatrix},
    \qquad
    \mathbf y_{\mathcal E}=
    \begin{bmatrix}y_{e_1}&\cdots&y_{e_{N_E}}\end{bmatrix}^T.
    \label{eq:observable_design_matrix}
\end{equation}
Ordinary least squares determines the regression coefficients,
\begin{equation}
    \widehat{\boldsymbol\theta}_K
    =\underset{\boldsymbol\theta}{\operatorname{arg\,min}}
    \left\lVert
    \mathbf y_{\mathcal E}-\mathbf H_K\boldsymbol\theta
    \right\rVert_2^2
    =\mathbf H_K^+\mathbf y_{\mathcal E},
    \qquad
    \widehat{\mathbf y}_{\mathcal E}
    =\mathbf H_K\widehat{\boldsymbol\theta}_K,
    \label{eq:observable_readout}
\end{equation}
where $\widehat{\boldsymbol\theta}_K=[\widehat\beta_0;\widehat{\boldsymbol\beta}]$. The main comparisons for all three cases fit and evaluate on all snapshots; therefore, $\widehat{\mathbf{y}}$ is an in-sample reconstruction. The target-signal reconstruction error $E_y$ and Pearson correlation coefficient $\rho_y$ are defined as
\begin{equation}
    E_y=100\,
    \frac{\lVert\mathbf y-\widehat{\mathbf y}\rVert_2}
    {\lVert\mathbf y-\overline y_{\mathrm{ref}}\mathbf 1\rVert_2},
    \qquad
    \rho_y=
    \frac{(\mathbf y-\overline y\mathbf 1)^T
    (\widehat{\mathbf y}-\overline{\widehat y}\mathbf 1)}
    {\lVert\mathbf y-\overline y\mathbf 1\rVert_2
    \lVert\widehat{\mathbf y}-\overline{\widehat y}\mathbf 1\rVert_2},
    \label{eq:observable_metrics}
\end{equation}
The main comparisons use $\overline y_{\mathrm{ref}}=\overline y$. The primary target-signal metric is $E_y$, supplemented by $\rho_y$ to quantify the linear correlation between the reconstructed and target signals.

The flow-field reconstruction error $E_x$ is evaluated over all snapshots relative to the fluctuation field, using the spatially weighted relative Frobenius error,
\begin{equation}
    E_x=100\,
    \frac{\lVert\mathbf W^{1/2}(\mathbf X'-\widehat{\mathbf X}')\rVert_F}
    {\lVert\mathbf W^{1/2}\mathbf X'\rVert_F}.
    \label{eq:field_error}
\end{equation}
Truncated reconstruction retains the coefficients of the full mode set and assesses how much of the original field the selected contributions preserve. The synthetic case additionally uses analytical target-subspace coverage and target-subfield reconstruction error to assess structure recovery, as defined in Appendix~\ref{app:pod_protocol}.

To assess how closely a selected DMD group set $\mathcal T$ approximates the full correlation operator, define the correlation-operator residual,
\begin{equation}
 E_C(\mathcal T)=100\,
 \frac{\|\mathbf W^{1/2}(\mathbf C_{xy}-\sum_{g\in\mathcal T}
 \mathcal P_g\mathbf C_{xy})\|_F}
 {\|\mathbf W^{1/2}\mathbf C_{xy}\|_F}.
 \label{eq:correlation_residual}
\end{equation}
For the weighted orthogonal bases $\boldsymbol\Psi_K$ of POD, TCC-POD and CCD, we report the retained cross-response fraction,
\begin{equation}
    \eta_C(K)=100\,
    \frac{\|\boldsymbol\Psi_K^H\mathbf W\mathbf C_{xy}\|_F^2}
    {\|\mathbf W^{1/2}\mathbf C_{xy}\|_F^2}.
    \label{eq:retained_correlation_fraction}
\end{equation}
$E_C$ measures correlation-operator approximation residuals, whereas $\eta_C$ measures the squared cross-response retained in an orthogonal subspace. Together with target-signal and flow-field reconstruction errors, these quantities characterise the selected structures.

\FloatBarrier
\section{Validation cases}
\label{sec:validation}

TCC is applied to DMD and POD mode sets for a synthetic flow field, a cylinder wake, and broadband cavity flow. These cases examine low-contribution target recovery, selection for different observables, and correspondence between target bands and spatial organisation in complex experimental data. Known synthetic oscillations, cylinder-force spectra, and cavity target bands provide frequency references. Therefore, Sections 3.1--3.3 use fixed DMD sets, combining eigenvalue frequencies, projection-coordinate spectra, and spatial forms to identify target-related structures.

Section~\ref{sec:pod_validation} tests single-real-mode TCC ranking on fixed POD sets and compares structure selection and reconstruction using TCC-POD, POD energy ranking, and CCD. POD supplies orthogonal directions determined by flow snapshots, whereas CCD constructs orthogonal directions from correlations between the flow field and lagged observables. This comparison examines how retaining existing directions or constructing correlation-based directions affects target-structure concentration and flow-field representation.

\subsection{Synthetic flow field: extraction and recovery of target structures with low global contributions}

A two-dimensional synthetic flow field is constructed by superimposing five known oscillatory structures to test whether target structures with low global contributions can be extracted from a strong background. Two structures with high global contributions form an unrelated background, while three structures with lower contributions jointly generate the target observable. First, the analytical ground truth is used to verify that the target components are present in the candidate mode set before assessing whether each criterion selects them when the same number of groups is retained. Subsequently, local amplitude modulation is introduced into the target mode to examine whether non-zero lags improve its identification when the observable response is temporally delayed.

First, we consider a two-dimensional incompressible analytical flow field with sustained oscillations, defined by the stream function
\begin{equation}
    \psi(x,y,t)=U_{\infty}y+
    \sum_{j=1}^{5}A_jG_j(x,y)
    \cos(k_jx-\Omega_jt+\varphi_j),
    \label{eq:synthetic_streamfunction}
\end{equation}
where
\begin{equation}
    G_j(x,y)=\exp\left[-\left(\frac{x-x_{c,j}}{L_{x,j}}\right)^2
    -\left(\frac{y-y_{c,j}}{L_{y,j}}\right)^2\right].
    \label{eq:synthetic_envelope}
\end{equation}
The velocity components $u=\partial\psi/\partial y$ and $v=-\partial\psi/\partial x$ satisfy two-dimensional incompressibility. Table~\ref{tab:synthetic_modes} lists the five components; spatial discretisation, sampling, and noise settings are given in Appendix~\ref{app:synthetic_stationary_details}.

\begin{table}[htbp]
    \centering
    \caption{Five oscillatory components of the two-dimensional synthetic analytical flow field.}
    \label{tab:synthetic_modes}
    \begin{tabular}{@{}cccccccccc@{}}
        \toprule
        $j$ & $A_j$ & $k_j$ & $\Omega_j$ & $\varphi_j$ & $x_{c,j}$ & $y_{c,j}$ & $L_{x,j}$ & $L_{y,j}$ & Target-related \\
        \midrule
        1 & 3.00 & 0.70 & 0.80 & 0.10  & 3.00 & 0.00  & 2.60 & 1.05 & No \\
        2 & 2.20 & 2.20 & 3.60 & 0.55  & 5.00 & 1.05  & 1.70 & 0.42 & No \\
        3 & 1.00 & 1.40 & 1.40 & 0.25  & 4.70 & 0.00  & 2.20 & 0.42 & Yes \\
        4 & 0.80 & 2.60 & 2.20 & $-0.35$ & 3.80 & $-0.58$ & 1.80 & 0.28 & Yes \\
        5 & 0.65 & 3.50 & 2.90 & 0.70  & 5.70 & 0.35  & 1.30 & 0.30 & Yes \\
        \bottomrule
    \end{tabular}
\end{table}

\begin{figure}[!t]
    \centering
    \includegraphics[width=0.90\textwidth]{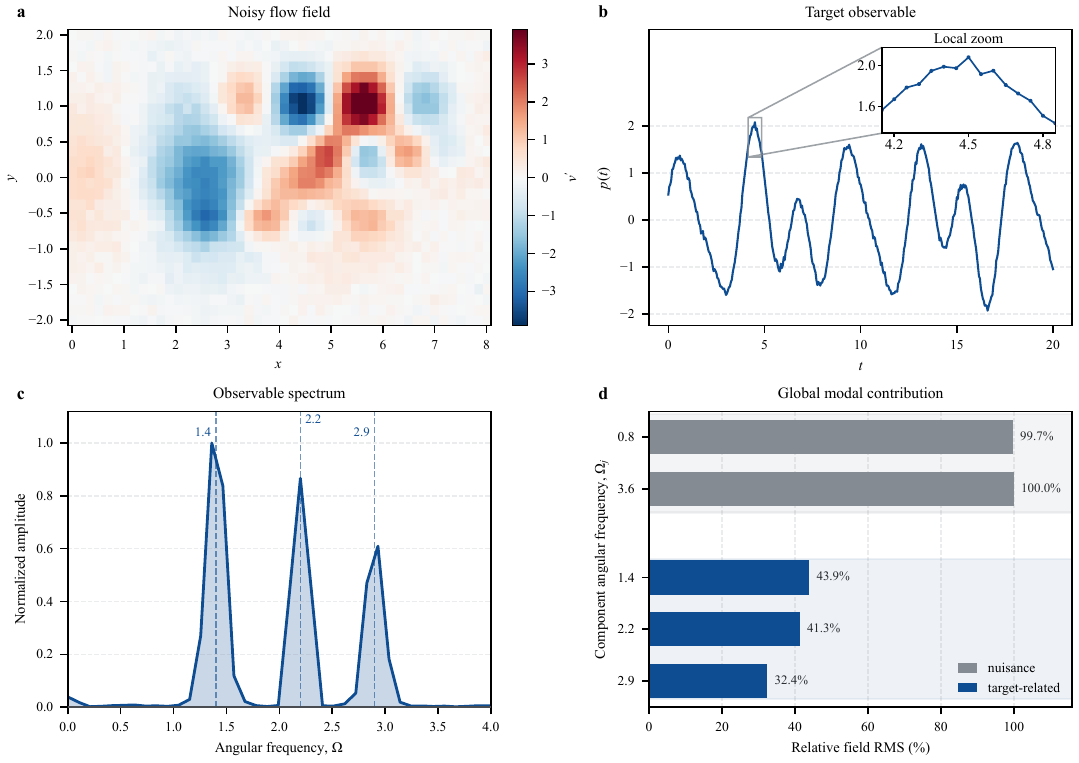}
    \caption{Two-dimensional synthetic analytical flow field and target observable. (a) A noisy snapshot of the transverse velocity fluctuations. (b) The noisy observable $p(t)$ and a local enlargement; the inset shows signal details over $t\in[4.15,4.85]$. (c) The amplitude spectrum of the observable, normalised by the maximum amplitude over the full spectrum; blue dashed lines mark the three target frequencies. (d) The RMS contributions of the analytical components to the velocity field, normalised such that the maximum among the five components is 100\%; grey and blue denote unrelated and target-related components, respectively.}
    \label{fig:synthetic_setup}
\end{figure}

The components at $\Omega=0.80$ and 3.60 have large global mean squared contributions but do not enter the target observable. The other three lower-contribution components jointly generate the target,
\begin{equation}
    p(t)=\cos[1.40(t-0.35)]
    +0.8\sin[2.20(t-0.55)]
    +0.6\cos[2.90(t-0.25)]+\varepsilon(t),
    \label{eq:synthetic_probe}
\end{equation}
where $\varepsilon(t)$ is Gaussian noise. Figure~\ref{fig:synthetic_setup} shows the instantaneous velocity field, the target observable, and the global contribution of each component. The three target carrier frequencies are distinct from those of the background components with high global contributions.

\begin{figure}[!t]
    \centering
    \includegraphics[width=0.94\textwidth]{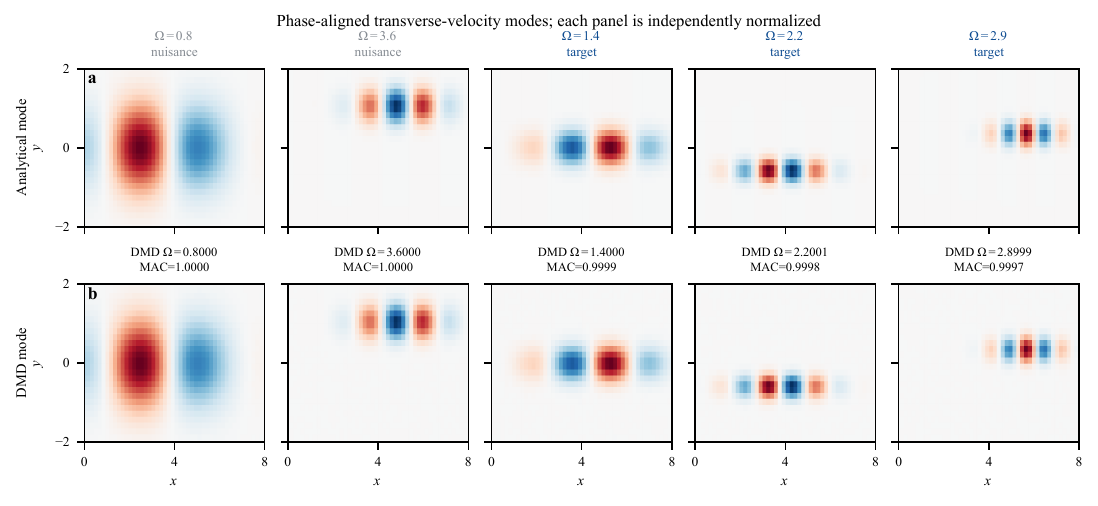}
    \caption{Analytical and DMD modes of the two-dimensional synthetic flow field. (a) Analytical transverse velocity modes. (b) DMD modes after frequency matching.}
    \label{fig:synthetic_mode_recovery}
\end{figure}

For this synthetic field, exact DMD with truncation rank $r=10$ uses only velocity-fluctuation snapshots to construct a fixed candidate mode set. The target observable does not enter its construction. Five real oscillatory components yield five conjugate groups, such that $r=10$ is the known ground-truth rank and separates mode-identification errors from ranking errors. Under sustained oscillations, RMS, CC$_0$, and TCC share full-mode-set snapshot-wise projection coordinates and common valid samples. Target-signal and flow-field reconstruction are evaluated on all samples. The modal assurance criterion (MAC) measures similarity between DMD spatial modes and their analytical counterparts; the definitions and settings are given in Appendix~\ref{app:synthetic_stationary_details}.
DMD recovered all five prescribed frequencies, with a maximum angular-frequency error of $1.21\times10^{-4}$. The five spatial modes had MAC values of 0.99974--0.99997 relative to the analytical ground truth (figure~\ref{fig:synthetic_mode_recovery}). Therefore, the low-contribution target structures are present in the candidate set; subsequent comparisons test whether ranking gives them priority. Increasing the truncation rank to 12, 14, and 16 leaves the TCC Top-3 selection unchanged (Appendix~\ref{app:synthetic_stationary_details}).

\FloatBarrier

\begin{figure}[!htbp]
    \centering
    \includegraphics[width=0.98\textwidth]{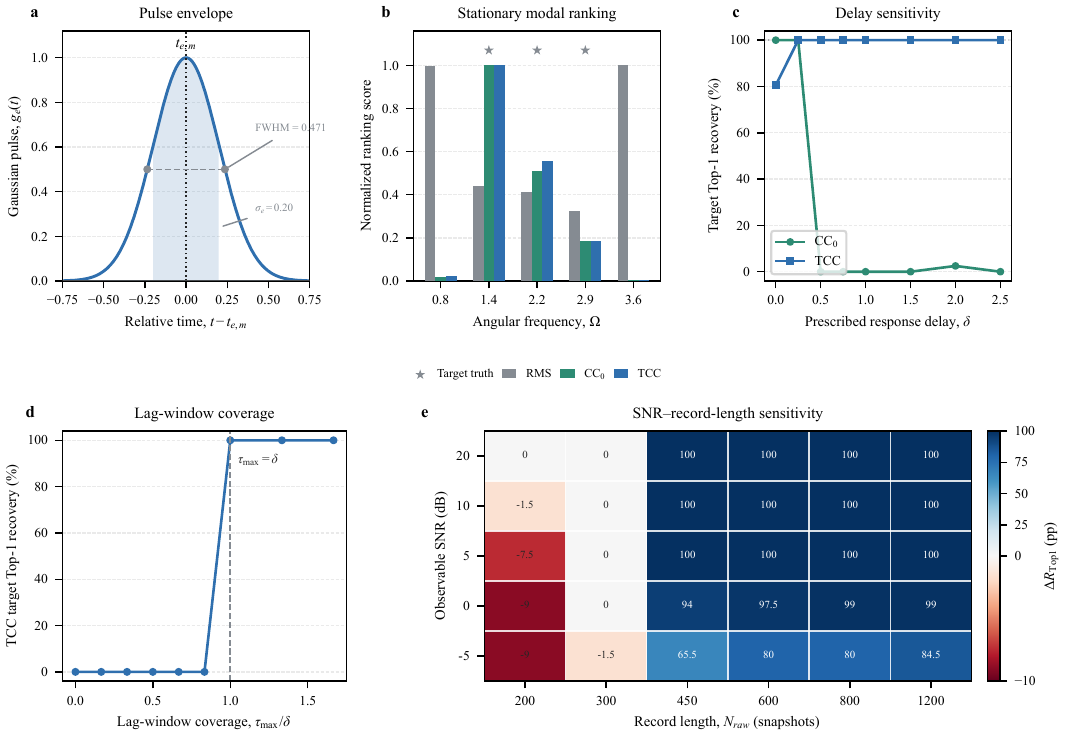}
    \caption{Rankings for the synthetic flow field and identification under amplitude modulation. (a) Pulse envelope. (b) Sustained-oscillation rankings, each normalised by its own maximum candidate score; vertical-axis magnitudes cannot be compared across criteria. (c) Response-delay sensitivity. (d) Lag-window coverage, with the dashed line marking $\tau_{\max}=\delta$. (e) Sensitivity to signal-to-noise ratio (SNR) and sample count. Colours and cell values show Top-1 recall differences $\Delta R_{\mathrm{Top1}}=R_{\mathrm{TCC}}-R_{\mathrm{CC}_0}$ at discrete parameter nodes, in percentage points (pp). Each node uses 200 independent noise trials; sampling and sweep settings are given in Appendix~\ref{app:synthetic_event_details}. Three additional aggregation comparisons and Wilson 95\% confidence intervals are reported in Appendix~\ref{app:synthetic_event_checks}.}
    \label{fig:synthetic_controlled}
\end{figure}

When each criterion selected three groups from the same candidate mode set, the RMS selected $\Omega=3.60$, 0.80, and 1.40 in that order, retaining only one target frequency. CC$_0$ and TCC both selected $\Omega=1.40$, 2.20, and 2.90 in that order, recovering all three target groups (figure~\ref{fig:synthetic_controlled}b). As the candidate modes, projection coordinates, and number of retained groups were identical, the difference arose from the observable changing the ranking priorities of the existing structures.
In this case, CC$_0$ and TCC produced the same Top-3 selection and target-signal reconstruction, indicating that inclusion of non-zero lags did not provide additional discrimination among the candidate groups. Although the target signal in Equation~\eqref{eq:synthetic_probe} includes fixed time offsets relative to the individual harmonics, these offsets primarily appear as phase differences under sustained oscillations, and the target groups remain distinguishable from the unrelated background at zero lag.

A subfield is constructed from the spatial modes and projection coordinates of the selected groups and compared with the sum of the three noise-free target components. CC$_0$ and TCC each achieved a Top-3 target-subfield reconstruction error of 2.15\%. Their target-signal reconstruction errors were 4.11\%, close to the relative RMS level of the added probe noise, 4.03\% (table~\ref{tab:synthetic_results}). The RMS retained two high-amplitude background components, giving a reconstruction error of 219.78\% relative to the target subfield.

\begin{figure}[!htbp]
    \centering
    \includegraphics[width=0.70\textwidth]{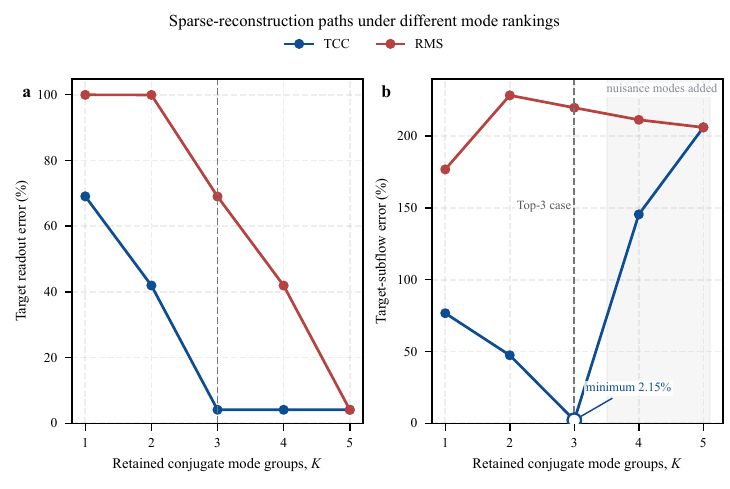}
    \caption{Target reconstruction error versus retained group count $K$ for the synthetic flow field. (a) Noisy target-signal reconstruction error. (b) Target-subfield reconstruction error relative to the sum of the three noise-free target components. The grey dashed line marks $K=3$. All curves share the fixed DMD mode set and its full-set snapshot-wise projection coordinates.}
    \label{fig:synthetic_tradeoff}
\end{figure}

Varying the number of selected groups shows the order in which target structures are extracted (figure~\ref{fig:synthetic_tradeoff}). Along the path $K=1$--5, TCC includes the three target groups first, whereas the RMS first includes background components with high global contributions. When $K>3$, TCC begins to add background components. This raises the reconstruction error relative to the target subfield, even though the projection coordinates and spatial contributions of the original target groups remain unchanged.

\begin{table}[htbp]
    \centering
    \caption{Top-3 reconstruction of the two-dimensional synthetic flow with sustained oscillations, using snapshot-wise projection coordinates.}
    \label{tab:synthetic_results}
    \begin{tabular}{lcc}
        \toprule
        Metric & TCC / CC$_0$ & RMS \\
        \midrule
        Noisy target-signal reconstruction error (\%) & 4.11 & 69.02 \\
        Noise-free target-subfield reconstruction error (\%) & 2.15 & 219.78 \\
        \bottomrule
    \end{tabular}
\end{table}

Local amplitude modulation in the same synthetic field tests how temporal offsets between structures and observed responses affect identification. The DMD spatial modes and prescribed carrier frequencies remain fixed. The target component receives a local pulse envelope, while the other components receive fixed weak amplitude modulation. These flow-field construction parameters remain unchanged throughout the parameter sweeps. At target carrier frequency $\Omega_{\mathrm{tar}}=1.40$, the temporal coefficient uses a Gaussian pulse envelope,
\begin{equation}
    g_e(t;t_{e,m})=\exp\!\left[-\frac{(t-t_{e,m})^2}{2\sigma_e^2}\right],
    \label{eq:synthetic_event_envelope}
\end{equation}
Multiple pulses form a normalised envelope $g(t)$. An observable delayed by $\delta$ relative to the amplitude modulation of the target mode in the flow field is then constructed,
\begin{equation}
    y_{\delta}(t)=g(t-\delta)\cos\!\left[\Omega_{\mathrm{tar}}(t-\delta)\right]
    +0.10\cos(2.20t+0.35)+\varepsilon_y(t).
    \label{eq:synthetic_delayed_observable}
\end{equation}
Here, the $\Omega=2.20$ component serves as a weak competitor and is not shifted with the delayed target response. The noisy synthetic flow field with amplitude modulation is projected onto the fixed DMD mode set to obtain the projection coordinates of each mode. All criteria rank the groups using the same common valid samples, with 200 independent noise trials at each parameter point. Top-1 recall is defined as the fraction of trials in which the target group ranks first. The amplitude-modulation construction, noise settings, and values used in each scan are given in Appendix~\ref{app:synthetic_event_details}.

With the spatial structures and prescribed carrier frequencies unchanged, the target recall of CC$_0$ decreased rapidly as the response delay $\delta$ increased. At the tested nodes with $\delta\geq0.50$, its recall was already close to zero, whereas TCC continued to identify the target group reliably when its window covered the main cross-response (figure~\ref{fig:synthetic_controlled}c). Thus, even when the target structure itself is unchanged, temporal misalignment between the structure and the observable can affect its identification.

To examine how lag coverage recovers this identification information, the response delay is fixed at $\delta=1.5$ and the window is expanded using identical common valid samples. TCC cannot reliably identify the target before the main correlation peak is covered; Top-1 recall reaches 100\% once it is included (figure~\ref{fig:synthetic_controlled}d). At $(\delta,\tau_{\max})=(1.5,2.5)$, MaxCC and TCC both achieve 100\% recall, identifying peak coverage as their shared recovery condition here (Appendix~\ref{app:synthetic_event_checks}). The prescribed delay $\delta$ does not alone determine the actual peak location and width, which also depend on the envelope, carrier phase, noise, and sampling. Therefore, window coverage should be assessed against the actual cross-response.

Once the window covers the correlation peak, further widening does not necessarily improve identification. For example, at $\delta=0$ and $\tau_{\max}=3.0$, CC$_0$ retained 100\% recall, whereas TCC recall decreased to 80.5\%. An overly broad lag window changed the relative scores of the target and competing groups. Therefore, window selection must account both for coverage of the effective cross-response and for the effect of additional lags on the mean-square cross-response.

When the window covers the main peak, TCC consistently exceeds CC$_0$ in target-group recall for records containing more pulses and observables with moderate SNR. For shorter records or stronger noise, the recall difference narrows, and TCC yields lower recall under some conditions (figure~\ref{fig:synthetic_controlled}e).

To examine the combined effects of identification and ranking, DMD is reidentified from independent noisy fields after amplitude modulation. Projection coordinates and correlation rankings are computed on each resulting mode set. TCC consistently ranks first a group closely matching the target spatial form; however, its DMD angular frequency is approximately 12\% above the prescribed value (Appendix~\ref{app:synthetic_event_checks}).

The synthetic case demonstrates extraction of low-contribution targets from an existing mode set and reveals the effects of response misalignment, window coverage, and finite samples. Next, we consider physical flow without an analytical target-subfield reference, using frequency and spatial form to assess selection for different engineering observables.

\FloatBarrier
\subsection{Cylinder wake: extraction of structures related to different aerodynamic forces}
\label{sec:cylinder_results}

The cylinder case examines whether changing the aerodynamic-force observable and its frequency band changes the timescales and spatial forms of extracted structures. We use public three-dimensional cylinder-wake data at $Re=3900$~\citep{wallian2025re3900}, analysing mid-plane transverse velocity fluctuations $v'$ over $x/D\in[3.5,13.0]$, $y/D\in[-2,2]$. The 310 snapshots span $77.25D/U_{\infty}$, approximately 17 lift-fluctuation cycles. Lift and drag are obtained by integrating pressure and viscous stresses over the cylinder wall. Mean removal gives lift fluctuations $C_L'$ and drag fluctuations $C_D'$; the latter are termed raw drag hereafter. Lift fluctuations, raw drag, and bandpass-filtered drag are used as targets, with identical velocity snapshots, DMD candidates, and projection coordinates. Figure~\ref{fig:cylinder_setup}(a,b) shows the flow field; extraction and processing details are given in Appendix~\ref{app:cylinder_protocol}.

The two aerodynamic forces differ primarily in their timescales. Frequency is expressed as the Strouhal number $St=fD/U_\infty$, where $D$ is the cylinder diameter and $U_\infty$ is the free-stream velocity. Lift fluctuations $C_L'$ are dominated by shedding near $St=0.2194$, whereas raw drag $C_D'$ mainly varies slowly. The second harmonic near $St=0.4645$ contributes only a small fraction of drag power (figure~\ref{fig:cylinder_setup}(c,d); additional diagnostics are given in Appendix~\ref{app:cylinder_sensitivity}).

\begin{figure}[!htbp]
    \centering
    \includegraphics[width=0.98\textwidth]{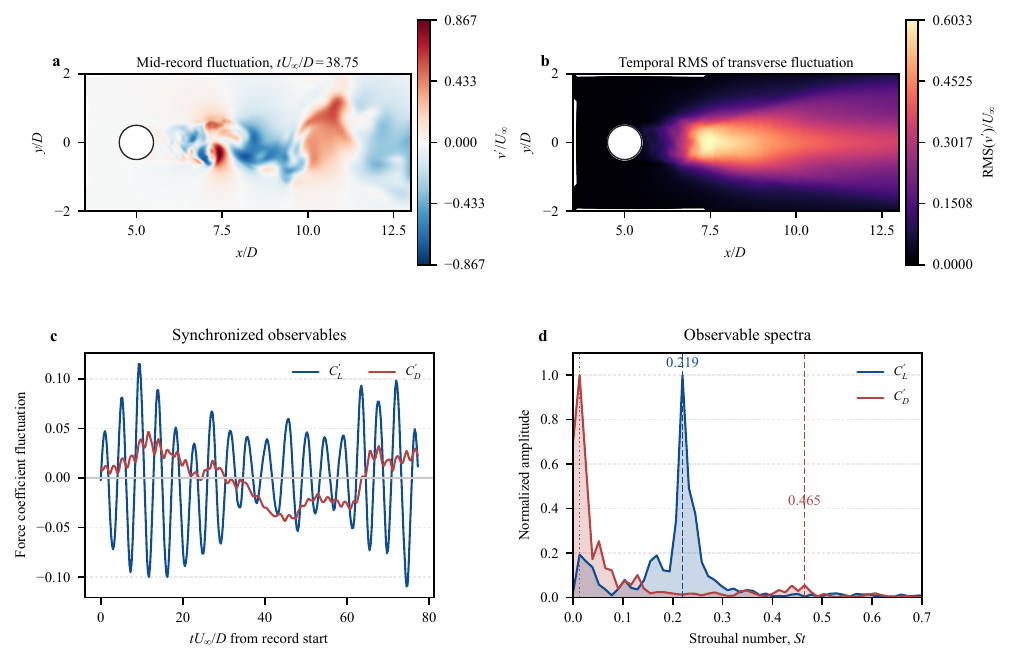}
    \caption{Cylinder wake and aerodynamic-force timescales. (a) Mid-plane transverse velocity fluctuation snapshot. (b) RMS transverse velocity fluctuations. (c) Aerodynamic-force fluctuations. (d) Amplitude spectra of aerodynamic-force fluctuations. Time is measured from the start of the analysed interval.}
    \label{fig:cylinder_setup}
\end{figure}

Lift fluctuations $C_L'$ and raw drag $C_D'$ are the primary target-signal reconstruction cases. Linearly detrended drag $C_D^{\mathrm{det}}$ and bandpass-filtered drag $C_D^{\mathrm{band}}$ test sensitivity to the drag-target definition; the latter extracts $0.35\leq St\leq0.55$ after detrending. All four targets share a rank-252 area-weighted DMD set and its full-set snapshot-wise projection coordinates, with all 123 valid oscillatory conjugate groups ranked. A zero-centred two-sided window, $-5\leq\tau U_{\infty}/D\leq5$ ($Q=41$), captures force signals both leading and lagging the flow snapshots. Every criterion uses the same 270 common valid samples. Retaining $K=1,\ldots,12$ groups tests the effects of group count on extraction and reconstruction; details are given in Appendix~\ref{app:cylinder_protocol}.

\begin{figure}[!htbp]
    \centering
    \includegraphics[width=0.98\textwidth]{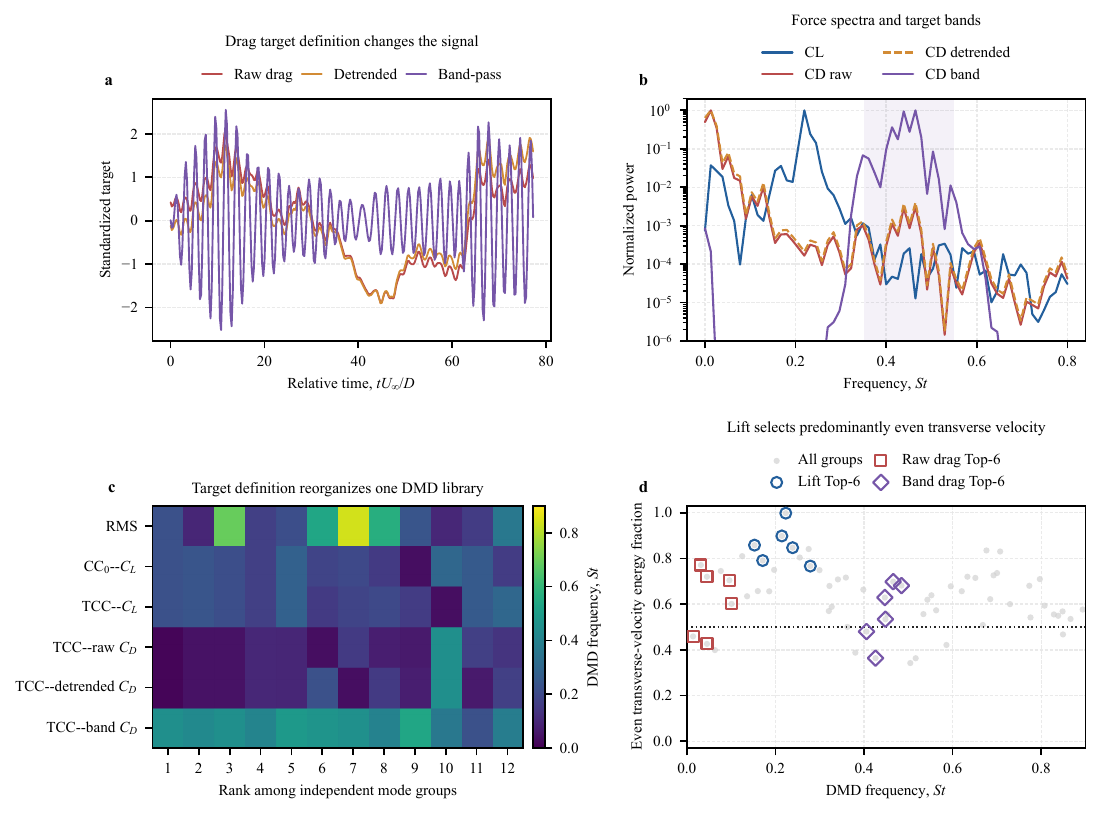}
    \caption{Aerodynamic-force targets and selection within a fixed DMD mode set. (a) Three drag targets. (b) Force power spectra. (c) Top-12 frequencies for each ranking; colour indicates only $St$. (d) Even-symmetry energy fraction of candidate transverse-velocity modes, with open symbols marking three TCC Top-6 sets. All rankings share 123 oscillatory conjugate groups from the rank-252 set and 270 common valid samples.}
    \label{fig:cylinder_conditioning}
\end{figure}

Under this common mode set and evaluation protocol, different force targets select distinct frequency clusters (figure~\ref{fig:cylinder_conditioning}). TCC Top-6 frequencies lie in $St=0.153$--0.279 for lift fluctuations, $St=0.014$--0.101 for raw drag, and $St=0.406$--0.484 for bandpass-filtered drag. These DMD frequencies lie near the shedding fundamental, in the low-frequency range, and in the second-harmonic band, respectively. For linearly detrended drag, five Top-6 groups remain in the low-frequency range, indicating that low-frequency power still dominates after removing the full-record linear trend. Subsequent spatial comparisons therefore use lift fluctuations and bandpass-filtered drag to examine fundamental and second-harmonic structures.

Comparison with the RMS reveals agreement at the shedding fundamental. The first RMS group, $St=0.2233$, also ranks first under lift-targeted TCC, combining a large transverse-velocity contribution with strong lift-fluctuation association. This agreement does not extend to the entire selected set. Beyond the first group, the RMS spans a wider frequency range; lift-targeted TCC remains near the fundamental, while drag targets emphasise low-frequency or second-harmonic candidates. A dominant flow frequency can therefore coincide with a target's main associated frequency without uniquely determining the full mode set required for that engineering task.

Selected modes also differ in spatial symmetry across force targets. Mid-plane transverse-velocity modes about $y=0$ are split into odd and even components. Their centreline parity is quantified by the area-weighted even-symmetry energy fraction $S_j^{(v)}$, defined in Appendix~\ref{app:cylinder_protocol}. The lift-fluctuation Top-6 has mean $S_j^{(v)}=0.860$, with 0.998 for the leading group (figure~\ref{fig:cylinder_conditioning}(d)). Its nearly even transverse-velocity structure and fundamental shedding frequency support an alternating-vortex-street interpretation. Raw-drag and bandpass-filtered-drag Top-6 means are 0.613 and 0.564, respectively, indicating mixed parity in their selected structures.

\begin{figure}[!htbp]
    \centering
    \includegraphics[width=0.98\textwidth]{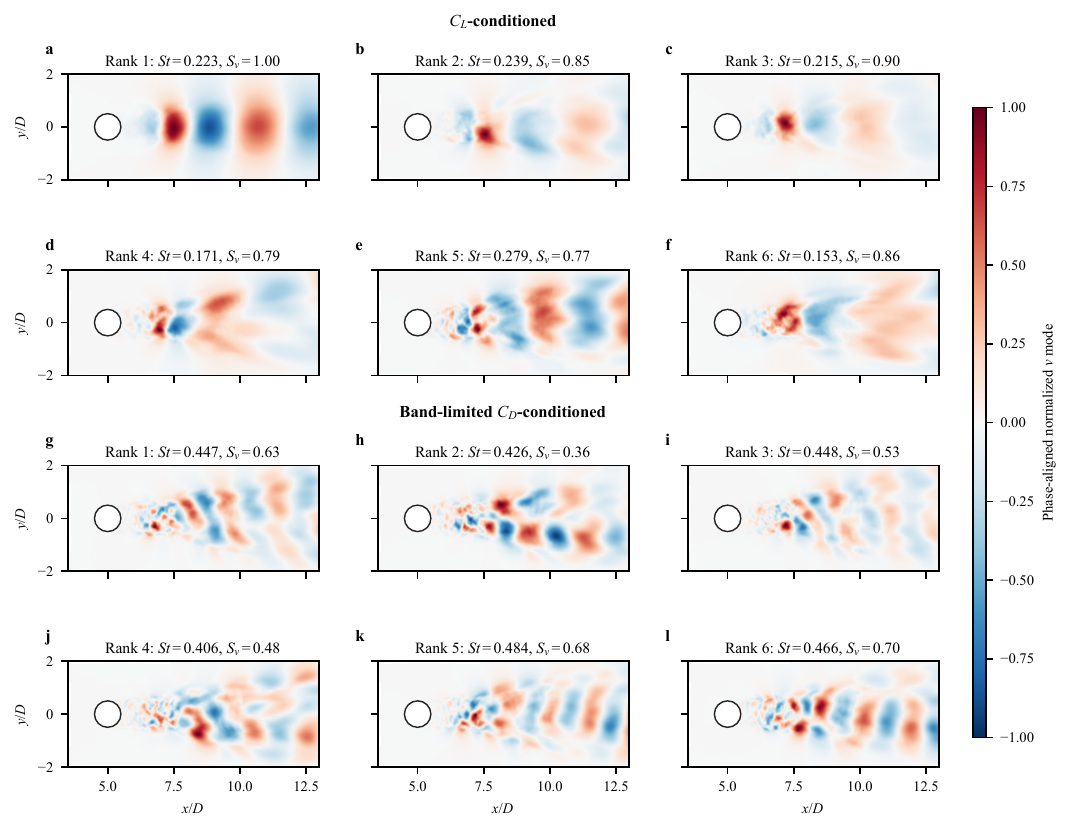}
    \caption{Representative transverse-velocity DMD modes selected for lift fluctuations and bandpass-filtered drag. (a--f, upper two rows) Lift-targeted TCC Top-6. (g--l, lower two rows) Bandpass-filtered-drag TCC Top-6. Each set follows ranking order from left to right and top to bottom. Each panel independently aligns the overall complex phase and normalises by the maximum absolute real-mode value; colours do not compare physical amplitudes across modes. Triangulation is used only for display.}
    \label{fig:cylinder_modes}
\end{figure}

Figure~\ref{fig:cylinder_modes} compares TCC Top-6 spatial modes for lift fluctuations and bandpass-filtered drag. The leading lift group forms large alternating structures downstream; the second and third groups also have predominantly large-scale distributions. The first three bandpass-filtered-drag groups contain denser positive and negative patches in the near wake and smaller downstream variation scales. The $St=0.426$ group has $S_j^{(v)}=0.36$, indicating odd-component dominance, whereas the other two have mixed parity. Together, frequency and spatial form associate lift fluctuations with large-scale organisation near the shedding fundamental, and bandpass-filtered drag with faster oscillations and denser spatial variations. Differences between raw and bandpass-filtered drag further show that slow variations and second-harmonic fluctuations of the same force relate to different wake structures. Therefore, physical interpretation requires both the observable and its frequency band to be specified.

\begin{figure}[!htbp]
    \centering
    \includegraphics[width=0.70\textwidth]{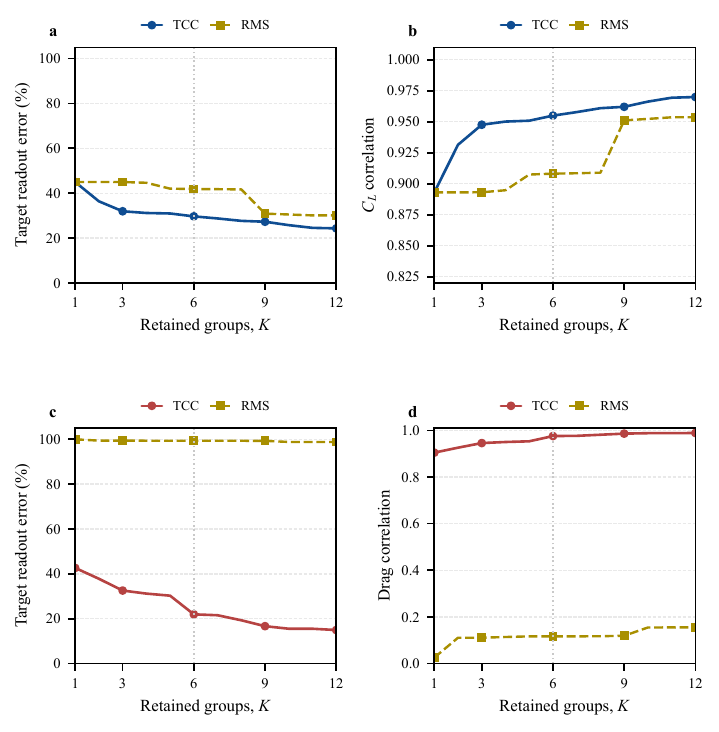}
    \caption{Effect of retained group count on force-signal reconstruction. (a,b) Lift-fluctuation signal reconstruction error and correlation coefficient. (c,d) Raw-drag signal reconstruction error and correlation coefficient. Rankings use 270 common valid samples; all errors are evaluated on the same 310 samples.}
    \label{fig:cylinder_tradeoff}
\end{figure}

With six groups retained, TCC reduces target-signal reconstruction error relative to the RMS from 41.86\% to 29.68\% for lift fluctuations and from 99.32\% to 21.96\% for raw drag (figure~\ref{fig:cylinder_tradeoff}). Zero-lag ranking already provides most of this improvement. For lift fluctuations, CC$_0$ gives 29.92\%, only 0.24 percentage points above TCC. For raw and linearly detrended drag, CC$_0$ and TCC select identical Top-6 sets within each target, giving errors of 21.96\% and 35.92\%, respectively. Therefore, non-zero lags add limited improvement to six-group reconstruction for these three targets.

\begin{figure}[!htbp]
    \centering
    \includegraphics[width=0.98\textwidth]{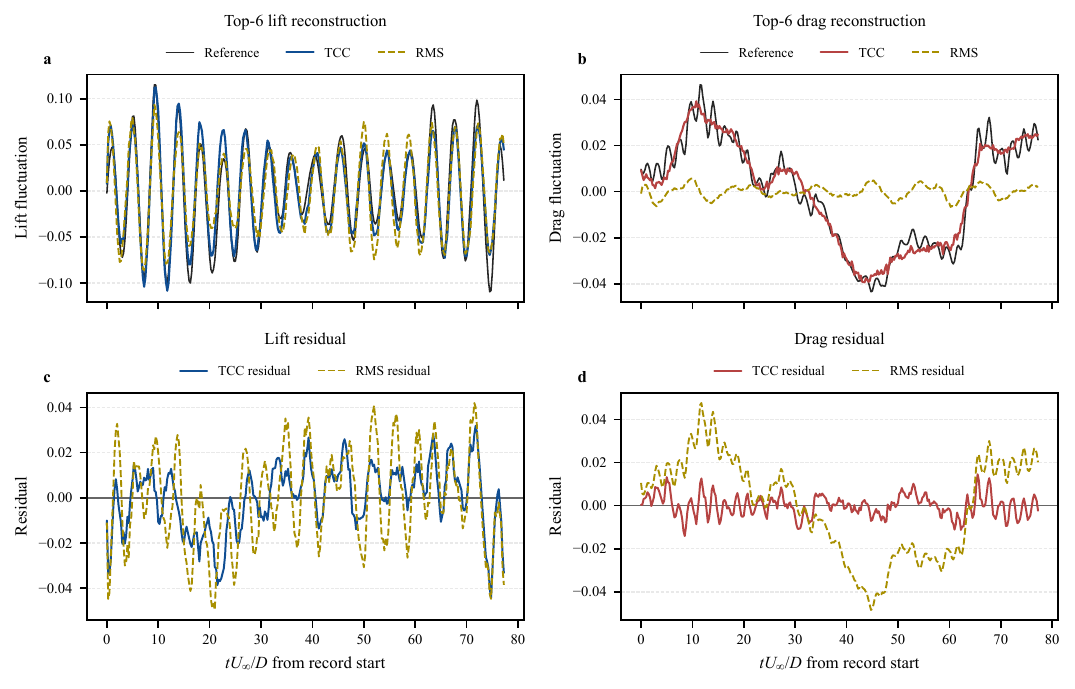}
    \caption{In-sample force-signal reconstruction using Top-6 mode groups. (a,b) Lift fluctuations and raw drag. (c,d) Corresponding residuals. All methods use the same 310 samples and linear regression form.}
    \label{fig:cylinder_observable_reconstruction}
\end{figure}

Figure~\ref{fig:cylinder_observable_reconstruction} shows that TCC-based lift reconstruction follows shedding oscillations and their amplitude modulation. Raw-drag reconstruction recovers the initial positive peak, middle negative plateau, and final increase. Although the RMS includes the leading shedding group, its other selections differ, leaving some lift-amplitude variations and slow drag variations unresolved.

Lag aggregation improves bandpass-filtered-drag reconstruction more clearly. Across all samples, TCC gives a Top-6 target-signal error of 51.19\%, compared with 60.51\% for CC$_0$. After trimming 30 points from each end and repeating ranking and reconstruction, the errors are 43.47\% and 49.46\%, respectively (Appendix~\ref{app:cylinder_sensitivity}). TCC gives lower errors on both sample sets, supporting additional selection information from lags for this target.

To test temporal sensitivity of lift-targeted DMD ranking, the 310 snapshots are divided chronologically into three approximately equal contiguous segments. The DMD set identified from all snapshots and its projection coordinates remain fixed, while TCC is recalculated using each segment's common valid samples. Regression fitted on the selected coordinates within one segment is then applied to reconstruct lift fluctuations in the other segments. The shedding-fundamental group ranks first in all three segments, whereas subsequent selections and cross-segment target-signal errors vary (figure~\ref{fig:cylinder_revision}(a,b)). Fitting and evaluation details are given in Appendix~\ref{app:cylinder_sensitivity}.

To separate window-width effects from sample-range effects, the window sweep uses the 250 common valid samples defined by the largest window, $Q=61$. Windows with $Q=1$, 11, 21, 41, and 61 give the same Top-6 set and 29.92\% target-signal error (figure~\ref{fig:cylinder_revision}(c)). Only the fifth and sixth groups exchange order at $Q=61$ (Appendix~\ref{app:cylinder_sensitivity}). Thus, expanding the symmetric two-sided window neither changes the selected set nor improves lift-fluctuation reconstruction on these common samples. Direction comparisons with equal lag count, total width, and common valid samples are given in Appendix~\ref{app:cylinder_sensitivity}.

\begin{figure}[!htbp]
    \centering
    \includegraphics[width=0.98\textwidth]{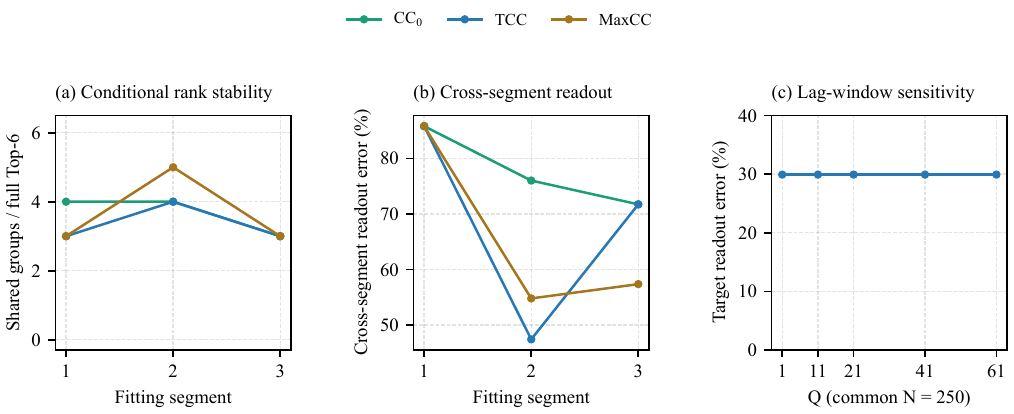}
    \caption{Fixed-mode-set ranking stability for the cylinder lift-fluctuation target. (a) Number of shared groups between segment-specific and full-record Top-6 sets. (b) Cross-segment target-signal reconstruction error after fitting on one segment and evaluating jointly on the others. (c) Lag-count sweep for symmetric two-sided windows using the same 250 common valid samples. All segments share the rank-252 mode set identified from the complete flow field.}
    \label{fig:cylinder_revision}
\end{figure}

\FloatBarrier

The cylinder case shows that force targets guide selection near the shedding fundamental, in the low-frequency range, and in the second-harmonic band within one fixed DMD set. Lag aggregation adds little to lift-fluctuation structure selection but improves bandpass-filtered-drag reconstruction more clearly, demonstrating dependence on the target. Next, we examine multi-band target-related extraction alongside its response to lag-window changes in broadband cavity flow.

\subsection{Broadband cavity flow: identification and spatial organisation of multi-band target structures}
\label{sec:cavity_results}

Shear-layer oscillations and their control in open cavities have been studied previously~\citep{rowley2002cavity,rowley2006cavity}. The cavity experiment extends target-related extraction to broadband particle image velocimetry (PIV) data, where a local observable contains several principal frequency bands. We examine whether structures associated with these bands can be extracted simultaneously from one fixed DMD set and how their spatial organisation differs. Without target-subfield ground truth, identification combines DMD frequencies and growth rates, actual projection-coordinate spectra, spatial forms near the probe, and local mode-group signals. We then assess target-signal reconstruction and sensitivity to analysis settings.

\begin{figure}[!htbp]
    \centering
    \includegraphics[width=0.98\textwidth]{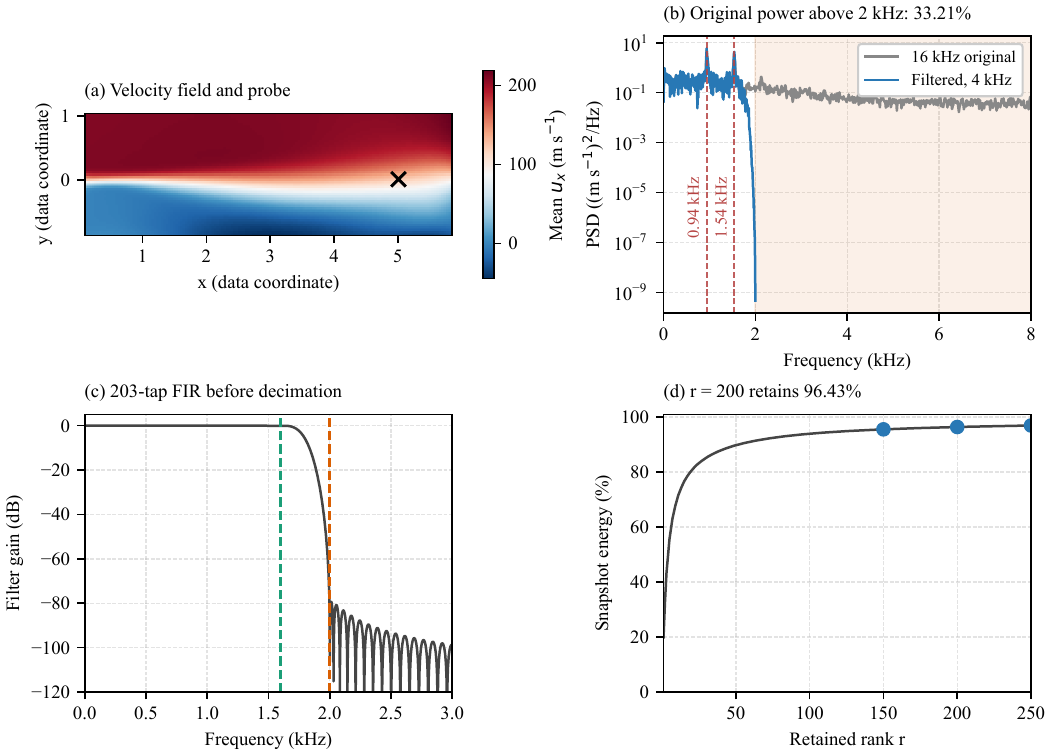}
    \caption{Cavity preprocessing and inputs for constructing the DMD mode set. (a) Mean streamwise velocity and observation point. (b) Raw and anti-aliased probe power spectra; shading marks the original frequency range above the Nyquist frequency after downsampling. (c) Frequency response of the finite impulse response (FIR) low-pass filter before downsampling. (d) Cumulative singular-value energy of the filtered and downsampled velocity snapshots.}
    \label{fig:cavity_setup}
\end{figure}

The flow field is the open-cavity PIV dataset from Challenge 2.1 of the American Institute of Aeronautics and Astronautics Reduced-Complexity Modelling Community Challenge~\citep{schmidt2026community}. The free-stream Mach number is 0.6, the Reynolds number is $3.3\times10^5$, and the cavity length-to-depth ratio is 6. A fixed downstream PIV point supplies the transverse velocity fluctuation target $u_y'$ (figure~\ref{fig:cavity_setup}(a)). Its two strongest local spectral peaks occur at 940 and 1540\,Hz (figure~\ref{fig:cavity_setup}(b)). The original sampling rate is 16\,kHz; downsampling to 4\,kHz gives a Nyquist frequency of 2000\,Hz. Before downsampling, identical antialiasing low-pass filters are applied to the flow field and target, with passband and stopband edges at 1600 and 2000\,Hz, respectively. Both target bands remain within the passband, with a 400\,Hz transition band for high-frequency attenuation.
After filtering and downsampling, a fixed rank-$r=200$ DMD set is constructed from the flow field, and all 94 valid oscillatory conjugate groups are ranked. Every criterion shares the full-set snapshot-wise projection coordinates and selects six conjugate groups. TCC and MaxCC use the main two-sided window $[-16,16]$\,ms; all correlation criteria use the same 3020 common valid samples. Selected timescales are checked against the original probe spectrum. Preprocessing and comparison settings are given in Appendix~\ref{app:cavity_protocol}.

\begin{figure}[!htbp]
    \centering
    \includegraphics[width=0.98\textwidth]{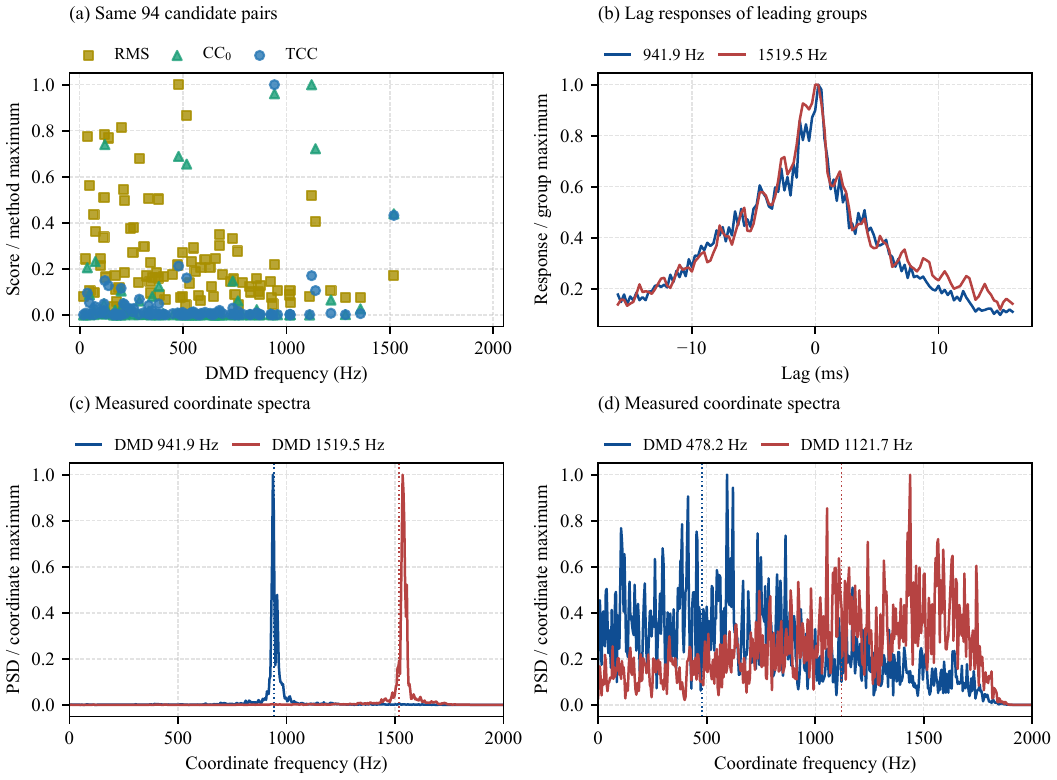}
    \caption{Cavity candidate rankings and DMD-frequency checks. (a) Scores on the same 94 candidates, each criterion normalised by its own maximum for ranking comparisons only. (b) Independently normalised lagged cross-response distributions of the first two TCC groups. (c,d) Normalised one-sided power spectra of the real projection coordinates for the first four TCC groups; vertical dashed lines mark their DMD frequencies.}
    \label{fig:cavity_ranking}
\end{figure}

With six groups selected from the same set, the RMS misses candidates near both principal target peaks, while CC$_0$ retains only the lower-frequency group. TCC ranks these groups first and second, at DMD frequencies of 941.87 and 1519.53\,Hz, respectively (hereafter the first two TCC groups; figure~\ref{fig:cavity_ranking}(a)). Their cross-responses peak near zero lag and retain finite width on both sides (figure~\ref{fig:cavity_ranking}(b)). MaxCC, using the same window, still excludes the higher-frequency group from its Top-6 (Appendix~\ref{app:cavity_spectral_details}). The additional selection information therefore comes mainly from finite-width cross-responses around the peaks. Mean squared aggregation includes these cross-responses and changes the relative group priorities.

\begin{figure}[!htbp]
    \centering
    \includegraphics[width=0.98\textwidth]{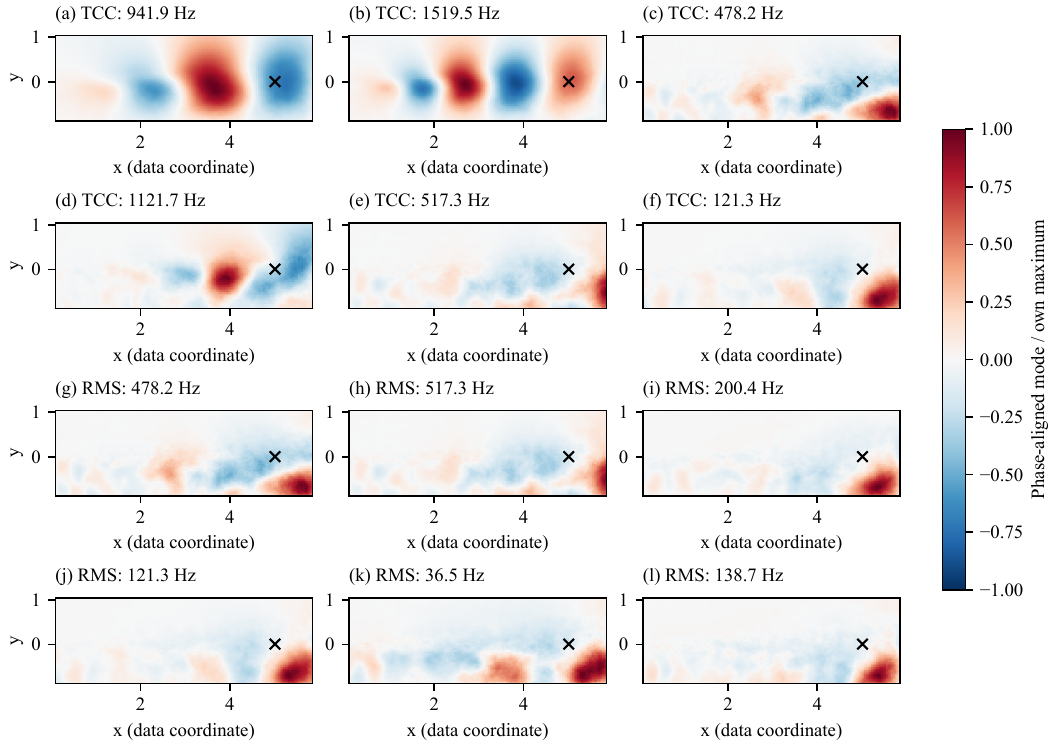}
    \caption{Transverse velocity spatial modes of the first six groups selected by TCC and RMS in the cavity. (a--f, upper two rows) TCC Top-6. (g--l, lower two rows) RMS Top-6. Each set is arranged in rank order from left to right and from top to bottom. The overall complex phase is aligned independently in each panel, and the real part is normalised by its maximum absolute value. Crosses mark the observation point, and colours do not indicate physical amplitudes across modes.}
    \label{fig:cavity_modes}
\end{figure}

Actual projection-coordinate spectra further link the first two TCC groups to the two target bands. Their principal spectral peaks lie within the respective target bands (figure~\ref{fig:cavity_ranking}(c)). Approximately 80\% of each complex-coordinate spectral power lies within $\pm50$\,Hz of its DMD frequency (Appendix~\ref{app:cavity_spectral_details}). For the third and fourth groups, less than 10\% lies within the corresponding neighbourhoods, and their coordinate peaks differ substantially from their DMD frequencies (figure~\ref{fig:cavity_ranking}(d)). Results for other neighbourhood widths are given in Appendix~\ref{app:cavity_spectral_details}. DMD frequencies follow the identified operator's eigenvalues, whereas projection coordinates follow snapshot-wise fitting. Their spectra can contain multiple components and need not peak at the corresponding DMD frequency~\eqref{eq:projected_coefficient}.
Spatially, both of the first two TCC groups show transverse velocity patterns that alternate along the streamwise direction near the probe height and extend into the downstream region containing the probe (figure~\ref{fig:cavity_modes}). The group near 942\,Hz varies more gradually along the streamwise direction, whereas the group near 1520\,Hz shows more closely spaced alternating positive and negative regions. The latter has a streamwise variation scale of 0.3672, approximately 27\% smaller than the 0.5017 scale of the former (Appendix~\ref{app:cavity_spectral_details}). By contrast, the stronger normalised fluctuations of the first RMS group (478.19 Hz, also the third TCC group) are concentrated mainly in the lower part of the downstream region.

To connect selected structures with the local target, we examine velocity-fluctuation signals reconstructed directly from each group at the probe. The first two TCC local signals are dominated by the respective principal target bands, consistent with their projection-coordinate spectra. Although the first RMS group's local amplitude is approximately 1.45 times the target amplitude, it peaks near 594\,Hz. Only 8.36\% of its power lies in the two target-band neighbourhoods combined (Appendix~\ref{app:cavity_spectral_details}). Therefore, the TCC groups exhibit the relevant principal bands at the observation point; a large local amplitude alone does not ensure this correspondence.\par

\begin{figure}[!htbp]
    \centering
    \includegraphics[width=0.98\textwidth]{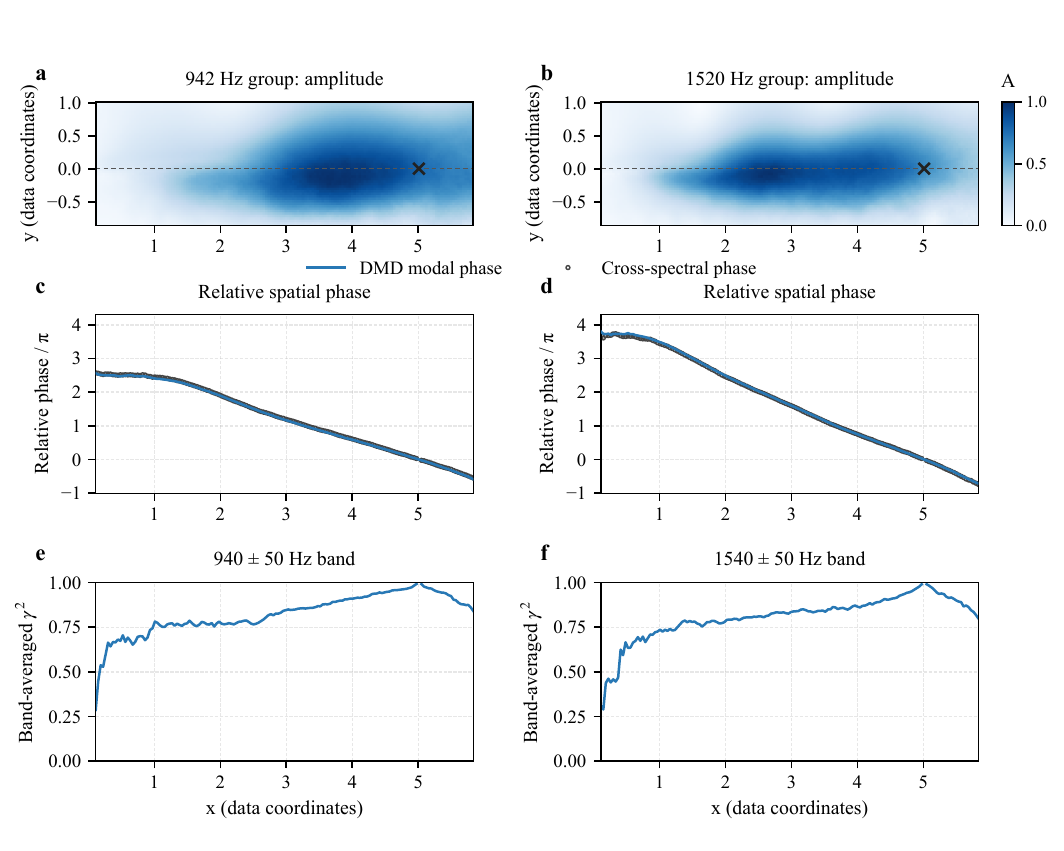}
    \caption{Spatial organisation of the two target frequency bands in the cavity. (a,b) Normalised modal amplitudes. (c,d) Comparison of spatial phases. (e,f) Band-averaged magnitude-squared coherence.}
    \label{fig:cavity_spatial_organization}
\end{figure}

Complex-mode amplitude and phase further reveal the spatial organisation of these two structures. Amplitudes in figure~\ref{fig:cavity_spatial_organization}a,b are normalised by each group's maximum; dashed lines and crosses indicate the observation height and target probe. Both groups concentrate near the observation height and extend into the downstream observation region; the group near 1520\,Hz lies relatively farther upstream. Their streamwise centres, weighted by squared modal amplitude, are $x=3.91$ and 3.40 in data coordinates. Relative to the original probe, both unwrapped phases generally decrease with $x$ along the observation height. The higher-frequency group accumulates more phase change over the same interval, consistent with its shorter streamwise variation scale.

Along the observation height, the cross-spectral phase measures the local velocity's phase difference from the target probe at a specified frequency. It is calculated from pre-processed transverse-velocity and probe signals. At the respective probe peaks, 937.50 and 1539.06\,Hz, the cross-spectral phases follow the corresponding DMD relative spatial phases (figure~\ref{fig:cavity_spatial_organization}c,d, circles and blue curves). Excluding the reference point, the mean magnitude-squared coherences over 153 locations are 0.824 and 0.806 for the two target bands. At each location, band averaging is weighted by target spectral power (figure~\ref{fig:cavity_spatial_organization}e,f; definitions in Appendix~\ref{app:cavity_spatial_organization}). Phase agreement and coherence beyond the probe support association of the two target bands with spatial structures differing in streamwise scale, amplitude distribution, and phase variation. Therefore, the selected modes connect multi-band features in a single-point signal to spatial organisation, resolving the distributions and phase relationships associated with each band.\par

\begin{figure}[!htbp]
    \centering
    \includegraphics[width=0.64\textwidth]{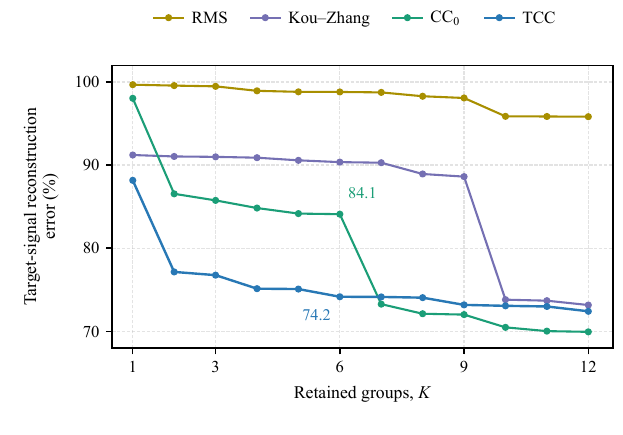}
    \caption{Target-signal reconstruction error versus retained group count for all cavity samples. The RMS error varies by less than four percentage points over $K=1$--12; only its key value at $K=6$ is labelled.}
    \label{fig:cavity_errors}
\end{figure}

Next, we test whether the selected combinations reconstruct the target signal. On all 3148 pre-processed samples, each criterion fits the same linear regression form using its selected coordinates. The TCC set containing both target-related groups gives a target-signal error of 74.16\%, below the 84.09\% of CC$_0$ (figure~\ref{fig:cavity_errors}). The same set gives a fixed-coefficient truncated flow-field error of 185.44\% (Appendix~\ref{app:cavity_expression}), illustrating the different objectives of target-signal and full-flow reconstruction. Non-orthogonal-superposition effects and additional reconstruction comparisons are discussed in Appendix~\ref{app:cavity_expression}.

Changes in analysis settings also distinguish structural identification from signal reconstruction by selected combinations. Identification of the two principal target bands is relatively stable across the tested DMD truncation ranks and lag directions. At $r=150$, 200, and 250, the first two groups consistently correspond to the bands near 940 and 1540\,Hz, respectively. Negative, positive, and two-sided windows preserve this correspondence. However, stable band identification does not imply consistently lower target-signal error. At $r=150$ and 250, CC$_0$ gives a lower error than TCC at the same rank. The same trend appears for $K=7$--12 in the main mode set, and further window expansion does not produce sustained error reduction. Therefore, principal bands can remain identifiable while reconstruction depends on truncation rank, group count, and lag window.

To test temporal sensitivity, the 3148 pre-processed samples are divided into four contiguous segments. The DMD set identified from all snapshots and its projection coordinates remain fixed. TCC is recalculated on each segment's common valid samples; regression fitted on that segment's selected coordinates is applied to the other three segments. Both principal target bands remain identifiable in every segment's probe signal, but only one segment's TCC Top-6 retains both groups. Cross-segment target-signal errors also vary. Thus, the principal bands persist while the full Top-$K$ set and reconstruction performance depend on the evaluation segment. Complete sensitivity results are given in Appendix~\ref{app:cavity_sensitivity}.

The broadband cavity case shows that TCC can extract structures associated with a local multi-band target from an existing DMD set. Lag aggregation supplements zero-lag identification under the main settings. Structural identification and reconstruction by selected combinations remain distinct, with reconstruction affected by non-orthogonal combinations, analysis settings, and the analysed interval.

\FloatBarrier
\begingroup
\raggedbottom
\setlength{\textfloatsep}{12pt}
\setlength{\intextsep}{12pt}
\subsection{Structure extraction by TCC on a fixed POD mode set}
\label{sec:pod_validation}

We compare TCC-POD, POD energy ranking, and CCD in all three cases to examine target-related extraction from fixed POD modes. The comparison distinguishes fixed-direction selection from correlation-based direction construction in structure extraction and flow-field reconstruction. Each case retains its flow field and observable; the cylinder uses lift fluctuations only. Truncation ranks, common valid samples, and lag windows are summarised in Appendix~\ref{app:pod_protocol}. All three methods retain six real modes for the quantitative comparisons, with the results listed in table~\ref{tab:pod_ccd_summary}; CC$_0$ and MaxCC comparisons are given in Appendix~\ref{app:pod_synthetic}.

POD modes are labelled P1, P2,\ldots\ in full-record energy order, and retain these labels after TCC reordering. CCD directions are labelled C1, C2,\ldots\ in correlation singular-value order. All spatial modes are normalised to unit weighted norm, with consistent arbitrary signs for display. Colours show their transverse velocity components, using a shared scale across panels within each case.

\begin{table}[!htbp]
    \centering
    \caption{Cross-response fraction, flow-field reconstruction error, and retained energy with six real modes in three cases. All values are percentages. Cross-responses and flow-field energy are normalised by their respective total squared norms.}
    \label{tab:pod_ccd_summary}
    \begin{tabular}{llrrr}
        \toprule
        Case & Method & $\eta_C$ & $E_x$ & Retained energy\\
        \midrule
        Synthetic & POD energy ranking & 72.16 & 32.43 & 89.48\\
         & TCC-POD & 90.45 & 67.16 & 54.90\\
         & CCD & 100.00 & 85.87 & 26.27\\
        Cylinder & POD energy ranking & 99.61 & 50.92 & 74.07\\
         & TCC-POD & 99.65 & 53.01 & 71.90\\
         & CCD & 99.99 & 54.42 & 70.38\\
        Cavity & POD energy ranking & 87.33 & 63.69 & 59.43\\
         & TCC-POD & 87.51 & 65.29 & 57.37\\
         & CCD & 97.45 & 68.18 & 53.52\\
        \bottomrule
    \end{tabular}
\end{table}

The synthetic case directly reveals the spatial content of ranking changes. POD energy ranking prioritises broad central modes P1 and P4 and narrower upper modes P2 and P3 (figure~\ref{fig:pod_synthetic_modes}). These distributions correspond to the two background components in figure~\ref{fig:synthetic_mode_recovery}. TCC promotes P5, P6, P8, and P7 to the first four positions and retains P4 and P1. This emphasises alternating patches below the centreline and central fluctuations, spatially consistent with the analytical targets and DMD-extracted target structures. The first two CCD modes cover these regions; the later modes also retain short-wave patches near $y\approx0.35$. Both methods emphasise target-related regions; however, POD ranking preserves mixtures of analytical components within existing directions, whereas CCD reorganises cross-responses into new orthogonal directions.

Known synthetic structures allow direct assessment of the selected subspace's coverage of each target component (table~\ref{tab:pod_synthetic_ccd_metrics}(a)). With six real modes, TCC-POD increases target-subspace coverage from 38.67\% under POD energy ranking to 66.20\%. Coverage is 97.54\% and 98.49\% for angular frequencies 1.40 and 2.20, respectively, but only 3.57\% for 2.90. CCD reaches 99.07\% overall, exceeding 98\% for each component and reaching 99.03\% for the 2.90 component. TCC-POD thus approaches CCD coverage for the first two targets; the main gap concerns the third. TCC-POD selects existing directions and retains their mixed background content. CCD constructs new directions from lagged flow-field--observable correlations, allowing all three targets to be well covered within six dimensions in this case.

These different target--background distributions also affect reconstruction. Retained modes P4 and P1 carry strong background contributions, leaving TCC-POD's target-subfield error at 153.18\%, above CCD's 61.43\%. The same contributions lower TCC-POD's full-flow error to 67.16\%, compared with 85.87\% for CCD (table~\ref{tab:pod_synthetic_ccd_metrics}(b)). TCC-POD reduces target-signal error from 63.20\% under POD energy ranking to 39.09\%, also slightly below CCD's 43.93\%. CCD optimises concentration of lagged cross-responses; target regression evaluates the selected coordinates' linear representation of simultaneous observations. Accordingly, CCD concentrates analytical targets and recovers their subfield more fully, whereas TCC-POD improves target-signal representation while preserving more original flow contribution. CC$_0$ gives still lower target-signal error under the same six-dimensional budget; detailed rankings and errors are given in Appendix~\ref{app:pod_synthetic}.

\begin{figure}[!htbp]
    \centering
    \includegraphics[width=\textwidth]{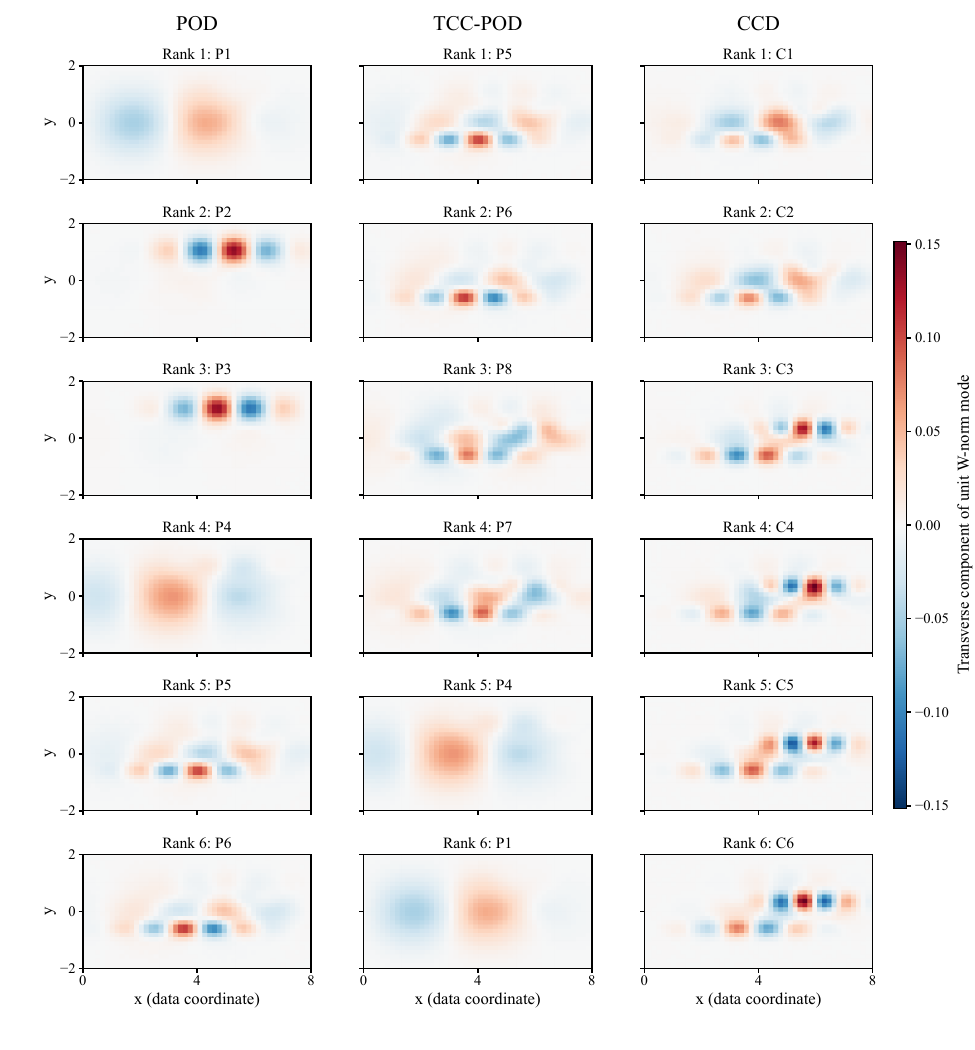}
    \caption{First six modes selected by POD energy ranking, TCC-POD, and CCD for the synthetic flow.}
    \label{fig:pod_synthetic_modes}
\end{figure}

Cylinder lift-fluctuation structures are already concentrated in dominant POD directions. POD energy ranking and TCC-POD both retain P1 and P2 first. Their broad alternating patches along the central wake resemble CCD's C1 and C2 and the leading lift-targeted DMD group (figures~\ref{fig:pod_cylinder_modes} and~\ref{fig:cylinder_modes}). Together with the spectra in \S\ref{sec:cylinder_results}, this shared organisation supports dominance of shedding oscillations in lift-related structures. TCC-POD retains P1, P2, P11, P6, P12, and P3. Relative to the first six energy-ranked modes, P11 and P12 replace P4 and P5, retaining additional near-wake and downstream fluctuations (complete six-mode comparison in Appendix~\ref{app:pod_full_modes}). All three methods retain over 99\% of the cross-response in six dimensions (table~\ref{tab:pod_ccd_summary}). Therefore, TCC-POD and CCD closely capture the principal structures and cross-responses. POD reordering yields a modest full-sample lift-signal error reduction, from 42.47\% to 41.42\% (Appendix~\ref{app:pod_reconstruction}).

\begin{figure}[!htbp]
    \centering
    \includegraphics[width=\textwidth]{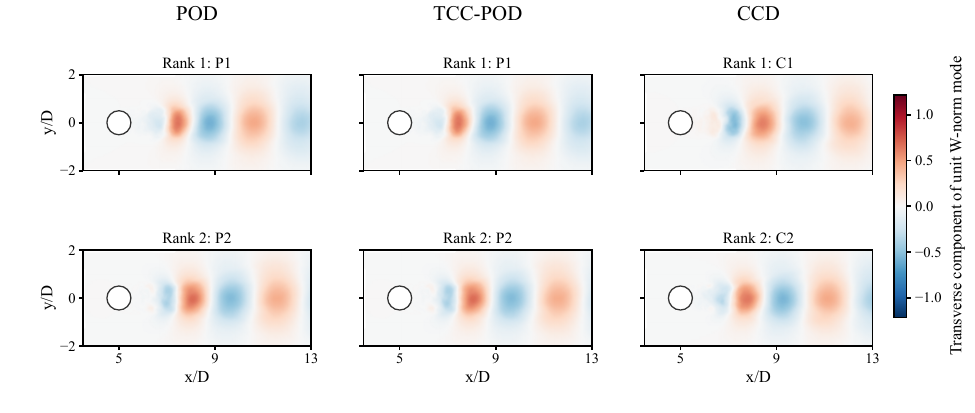}
    \caption{First two modes selected by each method for cylinder lift fluctuations.}
    \label{fig:pod_cylinder_modes}
\end{figure}

For the cavity, TCC-POD first retains P2 and P3, whose streamwise-alternating patches near the observation height extend downstream to the probe. The first two modes of CCD show similar organisation (figure~\ref{fig:pod_cavity_modes}). These patterns occupy the same principal regions as the two target-band structures identified through DMD and projection-coordinate spectra in \S\ref{sec:cavity_results}. The first two TCC-POD modes retain 55.14\% of the cross-response, above 47.55\% for POD energy ranking (Appendix~\ref{app:pod_reconstruction}). Reordering thus increases target-related cross-response retention by the first two POD directions. Its six-mode order is P2, P3, P1, P4, P7, and P6. It retains P1, with stronger fluctuations in the lower downstream region, and replaces energy-ranked P5 with P7 (complete six-mode comparison in Appendix~\ref{app:pod_full_modes}). The six-dimensional TCC-POD and energy-ranked spaces share five directions and retain similar cross-response fractions, whereas CCD retains more through newly constructed directions (table~\ref{tab:pod_ccd_summary}).

\begin{figure}[!htbp]
    \centering
    \includegraphics[width=\textwidth]{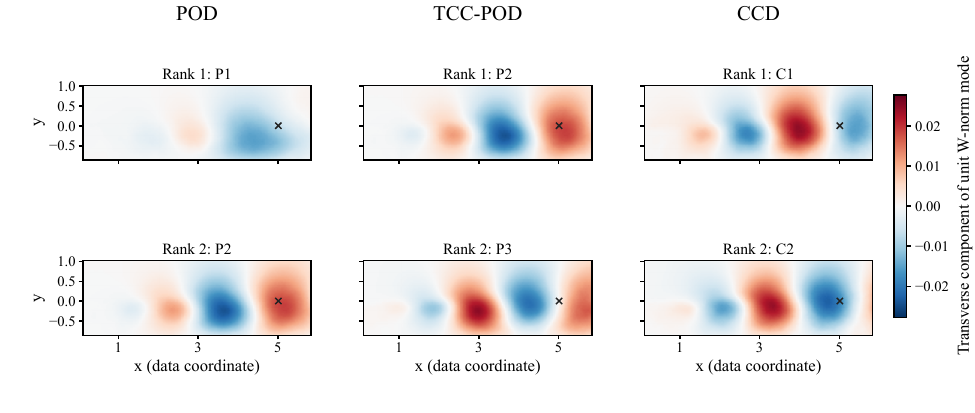}
    \caption{First two modes selected by each method for the cavity probe target. Crosses mark the target probe position.}
    \label{fig:pod_cavity_modes}
\end{figure}

With six real modes, CCD retains more cross-response in all three cases, whereas TCC-POD gives lower flow-field reconstruction errors (table~\ref{tab:pod_ccd_summary}). In the synthetic field, the error difference mainly reflects background contributions retained through P4 and P1; cylinder and cavity errors are closer. The flow-field advantage varies with retention budget. Curves across mode counts and instantaneous six-mode reconstruction maps are given in Appendix~\ref{app:pod_reconstruction}.

These cases show that TCC-POD extracts observable-related structures by reprioritising existing POD directions. It preserves spatial forms, energy meanings, and projection coordinates, tying selections to established POD modes and retaining their global flow contributions. Mixtures of target and background components within those directions are also preserved. Therefore, extraction depends on how targets are distributed across candidate directions and which combination fits the retention budget. CCD instead constructs directions from flow-field--observable correlations to concentrate cross-responses. The two approaches represent target-related selection among fixed directions and construction of new correlation-based directions, respectively.

\FloatBarrier
\endgroup
\section{Discussion and conclusions}
\label{sec:cross_case_discussion}

TCC introduces observables into selection within a fixed modal expansion, allowing structures from the same flow field to be reprioritised for different engineering targets. Relative to flow-contribution ranking~\citep{tissot2014,kou2017}, it adds statistical association with a specified response. Relative to CCD, which constructs a basis from the correlation operator~\citep{lyu2024}, it retains existing spatial modes and temporal coefficients. For DMD, this preserves connections among frequency, growth rate, and spatial form, providing a common dynamical reference for selections across targets.

The three principal cases show that global flow contribution and association with a specified observable provide distinct criteria for evaluating structures. Cylinder and cavity results connect observable timescales with spatial organisation, interpreting differences among force targets and multi-band local responses within their respective fixed expansions. DMD frequencies provide candidate spectral information. Snapshot-wise projection-coordinate spectra and local group signals describe how these structures contribute to the actual flow sequence. Together, they support dynamical interpretation.

The effect of lag aggregation depends on the temporal distribution of the cross-response. Zero lag already selects all three synthetic targets under sustained oscillations; after amplitude modulation introduces a response offset, covering the main correlation peak becomes decisive. In the cavity, finite-width cross-responses around the peaks allow mean squared aggregation to retain a higher-frequency group omitted by MaxCC. Cylinder lift-fluctuation gains mainly arise from introducing the observable, with limited additional benefit from non-zero lags. Wider windows can alter relative scores through newly included lags, as reflected by reduced recall under synchronous synthetic modulation and window-dependent cavity errors. Therefore, windows should reflect actual cross-response distributions, effective sampling duration, and relevant timescales.

Extraction is also constrained by how target structures are represented in the existing modes. For non-orthogonal DMD, joint group cross-responses contain cross-terms, such that weighted-orthogonal subset optimality does not directly extend to this setting. Individual TCC scores measure each group's cross-response, whereas reconstruction error also depends on the group combination. Therefore, ranking and selected-set reconstruction should be assessed together. Target-signal and flow-field reconstruction have different objectives: the former evaluates the coordinates' representation of the observable, while the latter evaluates retention of original flow contributions.

For weighted orthogonal POD, TCC minimises the weighted truncated-approximation error of the correlation operator among all $K$-mode subsets of the fixed set. CCD constructs new directions from flow-field--observable correlations and therefore has a broader subspace selection domain. POD analysis shows that fixed-direction selection preserves original spatial forms together with their mixed target and background contributions; CCD can further concentrate cross-responses. Their different target-subspace coverage and flow-field errors show why target concentration and retention of original flow information require separate evaluation.

\medskip
We propose TCC to separate mode-set construction from target-related selection, enabling comparisons across observables within one modal expansion. The three cases show that observable association provides a selection criterion distinct from global contribution. DMD analysis connects principal response timescales to selected spatial organisation. POD analysis shows that the criterion also applies to fixed orthogonal expansions. Extraction is jointly constrained by lagged cross-response distributions and the spatial representation of targets within candidate modes.

The spatial distributions and timescales of target-related modes may also inform design and flow control for specified physical quantities. Future work could combine configuration changes or active control with tests of how interventions on associated structures alter the target response. Such tests would assess whether modal analysis can guide design and regulation of the target quantity.

\FloatBarrier

\begin{appen}
\renewcommand{\theHsection}{appendix.\Alph{section}}
\renewcommand{\theHsubsection}{\theHsection.\arabic{subsection}}
\renewcommand{\theHsubsubsection}{\theHsubsection.\arabic{subsubsection}}
\renewcommand{\theHequation}{\theHsection.\arabic{equation}}
\section{Numerical implementation and theoretical derivations}

\subsection{Numerical validity and conjugate-pairing checks for the mode set}
\label{app:numerical_audit}

Before ranking, the numerical suitability of the fixed mode set is assessed using the singular values of $\mathbf{W}^{1/2}\boldsymbol{\Phi}$, with the threshold
\begin{equation}
    \varepsilon_{\mathrm{rank}}
    =\epsilon_{\mathrm{mach}}\max(M,r)s_1
    \label{eq:rank_tolerance}
\end{equation}
Here, $\epsilon_{\mathrm{mach}}$ is double-precision machine precision, and $s_1$ and $s_r$ are the largest and smallest singular values. A zero-norm mode or $s_r\leq\varepsilon_{\mathrm{rank}}$ indicates numerical rank deficiency and requires reassessment of the truncation rank and nearly linearly dependent modes. The condition number $\kappa_2(\mathbf{W}^{1/2}\boldsymbol{\Phi})=s_1/s_r$ characterises potential amplification of coordinate perturbations. The main synthetic, cylinder, and cavity mode sets have full numerical rank ($r=10,252,200$), with condition numbers 1.63, 81.06, and 1266.70, respectively.

Independent numerical checks for the cavity confirm that the left singular vectors remain orthogonal to within numerical precision. For sampled snapshots, reconstruction errors computed directly and through the modal inner-product matrix differ only at the level of double-precision roundoff. The imaginary residual of the field synthesised from conjugate members is negligible.

The criteria for identifying real modes and pairing conjugate modes are, respectively,
\begin{equation}
    \frac{|\operatorname{Im}(\lambda_j)|}
    {\max(1,|\lambda_j|)}<\varepsilon_{\mathrm{real}},
    \qquad
    \frac{|\lambda_{j^*}-\lambda_j^*|}
    {\max(1,|\lambda_j|)}<\varepsilon_{\mathrm{pair}},
    \label{eq:conjugate_tolerances}
\end{equation}
For the cylinder and cavity, $\varepsilon_{\mathrm{real}}=\varepsilon_{\mathrm{pair}}=64\epsilon_{\mathrm{mach}}$, whereas the synthetic case uses $\varepsilon_{\mathrm{real}}\approx10^{-8}$ and $\varepsilon_{\mathrm{pair}}=10^{-6}$. Non-finite or zero eigenvalues cannot provide the continuous-time frequencies and growth rates defined in this study. If a complex eigenvalue has no unpaired conjugate partner satisfying the tolerance, the mode set does not proceed to grouping and ranking. Each complex-conjugate group is counted once, with both members participating in scoring and reconstruction. For the field synthesised from each group, the ratio of the imaginary-part norm to the real-part norm must be below $10^{-10}$ before the real part is taken to remove roundoff residuals.

\subsection{RMS criterion and conjugate-group expansion}
\label{app:group_algebra}

The time-averaged modal contribution of \citet{tissot2014} can be written as
\begin{equation}
    E_j^{\mathrm{Tissot}}
    =\frac{1}{T}\int_0^T
    \left\lVert
    \widetilde{\boldsymbol\phi}_j
    \lambda_j^{t/\Delta t}
    \right\rVert_W^2\,\mathrm dt,
    \label{eq:tissot_energy}
\end{equation}
where $\widetilde{\boldsymbol\phi}_j$ includes the initial amplitude. Equation~\eqref{eq:rms_criterion} in the main text likewise averages modal contributions over the sampling period but uses snapshot-wise projection coordinates over the full mode set in place of prescribed exponential time series. The field is first synthesised within each mode group, and its norm is then evaluated.

For an ideal conjugate group, write $c=c_{j_g}(\ell)$ and $\boldsymbol\phi=\boldsymbol\phi_{j_g}$. The lagged cross-response field is $\mathbf{R}_g=c\boldsymbol\phi+c^*\boldsymbol\phi^*=2\operatorname{Re}(c\boldsymbol\phi)$. Expanding $\|\mathbf{R}_g\|_W^2$ for real symmetric $\mathbf{W}$ gives
\begin{equation}
    I_g^{\mathrm{TCC}}
    =2\sum_{\ell\in\mathcal L}
    \left[
    |c_{j_g}(\ell)|^2
    \boldsymbol\phi_{j_g}^H\mathbf W\boldsymbol\phi_{j_g}
    +\operatorname{Re}\!\left(
    c_{j_g}^2(\ell)
    \boldsymbol\phi_{j_g}^T\mathbf W\boldsymbol\phi_{j_g}
    \right)
    \right].
    \label{eq:tcc_pair_expansion}
\end{equation}
The second term retains the phase coupling between conjugate members. The numerical implementation retains the full quadratic form of the combined cross-response without assuming exact conjugacy. Real single modes also use Equation~\eqref{eq:tcc_strength} directly, without introducing a factor of 2.

The phase term in~\eqref{eq:tcc_pair_expansion} shows that zero-lag and lag-averaged scores can yield different rankings, even for purely harmonic temporal coefficients. The phase term vanishes when the spatial mode satisfies $\boldsymbol\phi^{T}\mathbf{W}\boldsymbol\phi=0$. This condition is equivalent to the mode's real and imaginary parts being orthogonal and having equal norms in the $W$ inner product.

\subsection{Matrix cross-response forms, selection optimality, and representation invariance}
\label{app:operator_theory}

Let $\mathbf{A}_{\mathcal I}=[\mathbf{a}(k_1),\ldots,\mathbf{a}(k_N)]$, where $\mathbf{a}(k)=[a_1(k),\ldots,a_r(k)]^T$, and let diagonal matrix $\mathbf{S}_g$ select group $g$'s coefficients. The coefficient--observable correlation and group cross-response matrices are
\begin{equation}
    \mathbf C_{ay}=\frac{1}{N\sqrt Q}\mathbf A_{\mathcal I}\mathbf Y_{\mathcal L}^{H},
    \qquad
    \mathbf C_{gy}=\boldsymbol\Phi\mathbf S_g\mathbf C_{ay}.
    \label{eq:general_group_correlation}
\end{equation}
Column $q$ of $\mathbf{C}_{gy}$ is $\mathbf{R}_g(\ell_q)$ in~\eqref{eq:correlation_field}, giving the matrix form
\begin{equation}
    I_g^{\mathrm{TCC}}
    =\|\mathbf W^{1/2}\boldsymbol\Phi\mathbf S_g\mathbf C_{ay}\|_F^2
    =\frac{1}{N^2Q}
    \|\mathbf W^{1/2}\boldsymbol\Phi\mathbf S_g
    \mathbf A_{\mathcal I}\mathbf Y_{\mathcal L}^{H}\|_F^2.
    \label{eq:tcc_general_matrix}
\end{equation}
This expression applies directly to the fixed expansion in~\eqref{eq:general_modal_expansion}. For a single mode,~\eqref{eq:tcc_general_gram} reduces to
\begin{equation}
    I_j^{\mathrm{TCC}}
    =\|\boldsymbol\phi_j\|_W^2\sum_{q=1}^{Q}|c_j(\ell_q)|^2.
    \label{eq:tcc_single_mode}
\end{equation}

When $\mathbf{a}(k)=\mathbf{B}\mathbf{x}'_k$, the identity $\mathbf{C}_{ay}=\mathbf{B}\mathbf{C}_{xy}$ gives
\begin{equation}
    \mathbf C_{gy}=\mathcal P_g\mathbf C_{xy},
    \qquad
    I_g^{\mathrm{TCC}}
    =\|\mathbf W^{1/2}\mathcal P_g\mathbf C_{xy}\|_F^2.
    \label{eq:restricted_correlation_energy}
\end{equation}
The same group reconstruction operator gives an equivalent RMS expression,
\begin{equation}
    I_g^{\mathrm{RMS}}(\mathcal S)
    =\left[\frac{1}{|\mathcal S|}
    \|\mathbf W^{1/2}\mathcal P_g\mathbf X_{\mathcal S}\|_F^2\right]^{1/2},
    \qquad
    \mathbf X_{\mathcal S}=[\mathbf x'_k]_{k\in\mathcal S}.
    \label{eq:rms_operator_form}
\end{equation}
Weighted least squares over the full mode set corresponds to
\begin{equation}
    \mathbf B=\mathbf B_\Phi
    =(\mathbf W^{1/2}\boldsymbol\Phi)^+\mathbf W^{1/2},
    \label{eq:least_squares_analysis_map}
\end{equation}
where superscript $+$ denotes the Moore--Penrose pseudoinverse. For weighted orthonormal modes, $\boldsymbol{\Phi}^H\mathbf{W}\boldsymbol{\Phi}=\mathbf{I}$, and $\mathbf{B}_\Phi=\boldsymbol{\Phi}^H\mathbf{W}$. For non-orthogonal modes, coefficients are solved jointly over the full set, and $\mathcal P_g$ is generally not the group's weighted orthogonal projector. If a decomposition's coefficients lack a fixed linear mapping,~\eqref{eq:tcc_general_matrix} is used directly.

For the partition $\mathcal J$ containing all groups, $\sum_{g\in\mathcal J}\mathbf{S}_g=\mathbf{I}_r$. Equations~\eqref{eq:general_modal_expansion} and~\eqref{eq:general_group_correlation} give
\begin{equation}
    \sum_{g\in\mathcal J}\mathbf C_{gy}
    =\boldsymbol\Phi\mathbf C_{ay}
    =\mathbf C_{xy}-\mathbf C_{ey},
    \qquad
    \mathbf C_{ey}=\frac{1}{N\sqrt Q}\mathbf E_{\mathcal I}\mathbf Y_{\mathcal L}^H,
    \label{eq:general_correlation_resolution}
\end{equation}
where $\mathbf{E}_{\mathcal I}$ contains representation residuals on the common valid samples. Therefore, the summed group cross-responses give the cross-response of the retained expansion, while $\mathbf{C}_{ey}$ measures the unrepresented part's correlation with the target.

Writing the weighted matrix inner product as $\langle\mathbf{G},\mathbf{H}\rangle_{W,F}=\operatorname{tr}(\mathbf{G}^H\mathbf{W}\mathbf{H})$ gives
\begin{equation}
    \left\|\mathbf W^{1/2}\sum_{g\in\mathcal J}\mathbf C_{gy}\right\|_F^2
    =\sum_{g\in\mathcal J}I_g^{\mathrm{TCC}}
    +2\operatorname{Re}\sum_{g<h}
    \langle\mathbf C_{gy},\mathbf C_{hy}\rangle_{W,F}.
    \label{eq:tcc_group_cross_terms}
\end{equation}
For any group set $\mathcal T$, the squared norm of the synthesised cross-response is
\begin{equation}
    J(\mathcal T)
    =\left\|\mathbf W^{1/2}\sum_{g\in\mathcal T}\mathbf C_{gy}\right\|_F^2.
    \label{eq:tcc_joint_subset_objective}
\end{equation}
Its difference from $\sum_{g\in\mathcal T}I_g^{\mathrm{TCC}}$ is given by the cross-group terms in~\eqref{eq:tcc_group_cross_terms}. These vanish when $\boldsymbol\Phi_g^H\mathbf{W}\boldsymbol\Phi_h=\mathbf0$ for different groups. This condition does not require orthogonality within each group.

For the weighted least-squares coefficients in~\eqref{eq:least_squares_analysis_map},
\begin{equation}
    \sum_{g\in\mathcal J}\mathcal P_g
    =\boldsymbol\Phi\mathbf B_\Phi
    \equiv\boldsymbol\Pi_\Phi^W,
    \qquad
    \sum_{g\in\mathcal J}\mathbf C_{gy}
    =\boldsymbol\Pi_\Phi^W\mathbf C_{xy}.
    \label{eq:dmd_group_resolution}
\end{equation}
As $\boldsymbol\Pi_\Phi^W$ is the weighted orthogonal projector onto $\operatorname{span}(\boldsymbol{\Phi})$,
\begin{equation}
    \|\mathbf W^{1/2}\mathbf C_{xy}\|_F^2
    =\|\mathbf W^{1/2}\boldsymbol\Pi_\Phi^W\mathbf C_{xy}\|_F^2
    +\|\mathbf W^{1/2}(\mathbf I-\boldsymbol\Pi_\Phi^W)\mathbf C_{xy}\|_F^2.
    \label{eq:dmd_library_residual}
\end{equation}
If $\boldsymbol{\Phi}$ additionally has full column rank, $\mathbf{B}_\Phi\boldsymbol{\Phi}=\mathbf{I}_r$, implying $\mathcal P_g^2=\mathcal P_g$ and $\mathcal P_g\mathcal P_h=\mathbf0$ for $g\neq h$. Different group operators retain disjoint modal coordinates, but generally do not satisfy $\mathcal P_g^H\mathbf{W}=\mathbf{W}\mathcal P_g$.

Consider a fixed weighted orthogonal set, $\boldsymbol{\Phi}^H\mathbf{W}\boldsymbol{\Phi}=\mathbf{I}$, with coefficients $\mathbf{a}(k)=\boldsymbol{\Phi}^H\mathbf{W}\mathbf{x}'_k$ and individual modes as selection units. Then, $\mathcal P_j=\boldsymbol\phi_j\boldsymbol\phi_j^H\mathbf{W}$ satisfies $\mathcal P_j^H\mathbf{W}=\mathbf{W}\mathcal P_j$ and $\mathcal P_i\mathcal P_j=\delta_{ij}\mathcal P_j$, where $\delta_{ij}$ is the Kronecker delta. For any mode-index subset $\mathcal S\subseteq\{1,\ldots,r\}$, $\sum_{j\in\mathcal S}\mathcal P_j$ is the weighted orthogonal projector onto the selected subspace. Its cross-response is orthogonal to the residual cross-response, and individual-mode cross-responses are mutually orthogonal. Therefore, Equation~\eqref{eq:restricted_correlation_energy} gives
\begin{equation}
    \left\|\mathbf W^{1/2}\left(\mathbf C_{xy}
    -\sum_{j\in\mathcal S}\mathcal P_j\mathbf C_{xy}\right)\right\|_F^2
    =\left\|\mathbf W^{1/2}\mathbf C_{xy}\right\|_F^2
    -\sum_{j\in\mathcal S}I_j^{\mathrm{TCC}}.
    \label{eq:tcc_orthogonal_selection_error}
\end{equation}
The first term on the right is independent of $\mathcal S$. For $|\mathcal S|=K$, minimising the weighted truncated-approximation error therefore maximises $\sum_{j\in\mathcal S}I_j^{\mathrm{TCC}}$. The solution consists of the $K$ highest-scoring TCC modes.

TCC is invariant to within-group representation changes that preserve group contributions. For a non-singular matrix $\mathbf{T}_g$, simultaneously transform the group's modes and coefficients,
\begin{equation}
    \widetilde{\boldsymbol\Phi}_g=\boldsymbol\Phi_g\mathbf T_g,
    \qquad
    \widetilde{\mathbf A}_g=\mathbf T_g^{-1}\mathbf A_g.
    \label{eq:group_basis_transform}
\end{equation}
As $\widetilde{\boldsymbol{\Phi}}_g\widetilde{\mathbf{A}}_g=\boldsymbol\Phi_g\mathbf{A}_g$, both $\widetilde{\mathbf{C}}_{gy}=\mathbf{C}_{gy}$ and $\widetilde I_g^{\mathrm{TCC}}=I_g^{\mathrm{TCC}}$ follow. Non-zero scaling of a mode with inverse scaling of its coefficients is a special case. Full-column-rank least-squares representations automatically obey this coefficient transformation after a within-group basis change; general representations require modes and coefficients to be transformed together. Mixing across groups changes their individual contributions and does not preserve individual scores in this way.

\FloatBarrier
\section{Construction parameters and test results for the synthetic validation}
\label{app:synthetic_details}

\subsection{Synthetic flow field with sustained oscillations: analytical modes and evaluation settings}
\label{app:synthetic_stationary_details}

The sustained-oscillation synthetic domain is $x\in[0,8]$, $y\in[-2,2]$, with a $48\times28$ grid, $U_{\infty}=1$, $\Delta t=0.05$, and 1200 snapshots. All synthetic cases in this appendix use $\Delta\tau=\Delta t=0.05$, reading observations directly at integer lag indices. Independent Gaussian velocity noise has a standard deviation equal to 8\% of the noise-free fluctuation-field RMS. The probe noise in~\eqref{eq:synthetic_probe} has standard deviation 0.04.

To specify the analytical ground truth used for MAC, the $j$th fluctuating component is written as
\begin{equation}
    \psi'_j(x,y,t)=\operatorname{Re}\!\left\{
    \widehat{\psi}_j(x,y)e^{-\mathrm{i}\Omega_jt}\right\},
    \qquad
    \widehat{\psi}_j=A_jG_j(x,y)e^{\mathrm{i}(k_jx+\varphi_j)},
    \label{eq:synthetic_complex_streamfunction}
\end{equation}
and its complex velocity mode is defined as
\begin{equation}
    \boldsymbol{\phi}^{\mathrm{true}}_j=
    \begin{bmatrix}
        \partial_y\widehat{\psi}_j \\
        -\partial_x\widehat{\psi}_j
    \end{bmatrix}.
    \label{eq:synthetic_true_mode}
\end{equation}
The analytical temporal oscillation is $e^{-\mathrm i\Omega_jt}$, and the physical field is the real part of the complex expression. Therefore, its positive-frequency representative is $\boldsymbol\phi^{\mathrm{true},+}_j=(\boldsymbol\phi^{\mathrm{true}}_j)^*$. Each positive-frequency DMD mode is matched to the nearest analytical angular frequency, and MAC is computed using the same frequency-sign convention,
\begin{equation}
    \mathrm{MAC}_j=
    \frac{\left|\left(\boldsymbol{\phi}^{\mathrm{true},+}_j\right)^H
    \boldsymbol{\phi}^{\mathrm{DMD}}_j\right|^2}
    {\left\|\boldsymbol{\phi}^{\mathrm{true},+}_j\right\|_2^2
    \left\|\boldsymbol{\phi}^{\mathrm{DMD}}_j\right\|_2^2}.
    \label{eq:synthetic_mac}
\end{equation}
The global phase and amplitude are aligned before plotting figure~\ref{fig:synthetic_mode_recovery}, although the MAC is invariant to both transformations. Across the five DMD groups, the maximum deviation from the unit circle is $\max_j\left|\lvert\lambda_j\rvert-1\right|=1.21\times10^{-4}$, and the maximum absolute continuous growth rate is $\max_j|\sigma_j|=2.43\times10^{-3}$.

RMS, CC$_0$, and TCC share the rank-10 set and common indices $\mathcal I=\{1,\ldots,1121\}$ determined by $Q=80$ and $0\leq\tau\leq3.95$. Coordinates are obtained for all 1200 times using~\eqref{eq:projected_coefficient}; in-sample target-signal and truncated flow-field reconstructions are evaluated on all samples. The main comparison uses $K=3$, supplemented by $K=1,\ldots,5$. Target-subfield error uses the three noise-free analytical targets as reference, while flow-field error uses the noisy fluctuation field; both include both velocity components. With three groups, CC$_0$ and TCC have lower target-signal errors but higher flow-field errors than the RMS. Table~\ref{tab:synthetic_equal_dimension} reports the target-signal comparison.

\begin{table}[!htbp]
    \centering
    \caption{Fixed-mode-set Top-3 target-signal reconstruction for the synthetic field over all 1200 samples. Every method retains three conjugate groups.}
    \label{tab:synthetic_equal_dimension}
    \begin{tabular}{lc}
        \toprule
        Method & Target-signal reconstruction error (\%)\\
        \midrule
        RMS & 69.02\\
        Kou--Zhang & 69.02\\
        CC$_0$ & 4.11\\
        TCC & 4.11\\
        \bottomrule
    \end{tabular}
\end{table}

At truncation ranks $r=10$, 12, 14, and 16, TCC Top-3 angular frequencies remain $1.400$, $2.200$, and $2.900$ in that order, to three decimal places. Target groups and their order are unchanged across these tested ranks.

\subsection{Synthetic flow fields with local amplitude modulation: construction, noise, and parameter sweeps}
\label{app:synthetic_event_details}

Equation~\eqref{eq:synthetic_event_envelope} uses $\sigma_e=0.20$ and pulse centres $t_{e,m}=6.0+7.5m$, where $m=0,1,\ldots$. Only pulses satisfying $t_{e,m}<t_{\mathrm{end}}-2.0$ are retained, with $t_{\mathrm{end}}$ the final snapshot time. A pulse has full width at half maximum $2\sqrt{2\ln 2}\,\sigma_e\approx0.471$, and the normalised envelope is
\begin{equation}
    g(t)=\frac{0.025+\sum_m g_e(t;t_{e,m})}
    {\displaystyle\max_s\left[0.025+\sum_m g_e(s;t_{e,m})\right]}.
    \label{eq:synthetic_event_train}
\end{equation}
Let $\mathcal J^+$ be the index set of the five positive-frequency representative modes, and set $\Omega_{\mathrm{tar}}=1.40$. The synthesis coordinates associated with the unit-norm DMD modes are
\begin{align}
    a_{\mathrm{tar}}^{(e)}(t)
    &=4.0\,g(t)e^{-\mathrm{i}\Omega_{\mathrm{tar}}t}, \\
    a_j^{(e)}(t)
    &=A_j^{(e)}\left[0.75+0.25\cos(0.11t+0.7\Omega_j)\right]
      e^{-\mathrm{i}\Omega_jt},\qquad j\neq\mathrm{tar},
    \label{eq:synthetic_event_coefficients}
\end{align}
For $\Omega_j=(0.80,1.40,2.20,2.90,3.60)$, the amplitudes are $A_j^{(e)}=(3.0,4.0,0.8,0.65,2.2)$, and these parameters remain fixed in all sweeps. Before constructing the synthetic field, the phase of each unit-norm mode's largest-magnitude component is set to zero to establish a consistent synthesis convention. The baseline value of 0.025 keeps the projection coordinates non-zero between successive pulses. The other components undergo weak modulation with a mean of 0.75, an amplitude of 0.25, and an angular frequency of 0.11, while $0.7\Omega_j$ offsets their modulation phases. Carrier frequencies are labelled by their absolute values here, and the negative exponent specifies the temporal direction of the synthetic coordinates. Amplitude modulation causes the components to deviate from the exponential evolution implied by the original DMD eigenvalues.

The synthetic flow field with amplitude modulation is
\begin{equation}
    \mathbf x'_e(t)=2\operatorname{Re}\!\left[\sum_{j\in\mathcal J^+}
    \boldsymbol\phi_j a_j^{(e)}(t)\right]+\boldsymbol\eta_x(t),
    \label{eq:synthetic_event_field}
\end{equation}
Here, $\boldsymbol\eta_x(t)$ is flow-field noise with RMS equal to 3\% of the amplitude-modulated noise-free field RMS. The noisy field is projected snapshot-by-snapshot onto the fixed mode set. In~\eqref{eq:synthetic_delayed_observable}, the weak competitor at angular frequency 2.20 has fixed amplitude 0.10. Linear interpolation constructs $g(t-\delta)$, with zero values outside the sampled interval. The variance of $\varepsilon_y$ follows the mean squared power of the first two terms in~\eqref{eq:synthetic_delayed_observable} and the specified SNR. The delay $\delta$ shifts the synthetic response in time and differs from the correlation-analysis lag increment $\Delta\tau$.

Each parameter point uses 200 independent noise trials, giving a counting increment of 0.5 percentage points for Top-1 recall. The three parameter sweeps are configured as follows, with all criteria using the same common index set $\mathcal I$ at each point:
\begin{itemize}
    \item The delay sweep uses $\delta=0$, 0.25, 0.50, 0.75, 1.00, 1.50, 2.00, and 2.50, with $N_{\mathrm{raw}}=800$, SNR of 8\,dB, and $0\leq\tau\leq3.0$. Thus, $Q=61$ and the common sample count is $N=740$.
    \item The window sweep uses $\tau_{\max}=0$, 0.25, 0.50, 0.75, 1.00, 1.25, 1.50, 2.00, and 2.50, with $\delta=1.5$, $N_{\mathrm{raw}}=800$, and SNR of 8\,dB. The common maximum window $0\leq\tau\leq2.5$ fixes $N=750$ common valid samples.
    \item The sample-count--SNR sweep combines $N_{\mathrm{raw}}=200$, 300, 450, 600, 800, and 1200 with SNR levels of $-5$, 0, 5, 10, and 20\,dB. With $\delta=1.5$ and $\tau_{\max}=2.5$, $Q=51$ and $N=N_{\mathrm{raw}}-50$. Sample count changes both observation duration and the number of pulses within the sampled interval.
\end{itemize}

\subsection{Time-lag aggregation comparisons and DMD mode reidentification after amplitude modulation}
\label{app:synthetic_event_checks}

Figure~\ref{fig:event_revision}(a) compares CC$_0$, TCC, and MaxCC using 800 snapshots, SNR of 8\,dB, and 750 common valid samples fixed by the maximum lag of 2.5. Projection-coordinate noise is sampled directly from the covariance induced by projecting independent real Gaussian field noise through the fixed mode matrix's pseudoinverse. This is equivalent to adding independent real Gaussian field noise before projection. All methods use the same noise within each trial. Wilson 95\% confidence intervals quantify recall uncertainty over 200 trials.

At $(\delta,\tau_{\max})=(0,2.5)$, the recall rates of CC$_0$, TCC, and MaxCC are 100\%, 96.5\%, and 100\%, respectively. The corresponding rates are 0\%, 0\%, and 21.5\% at $(1.5,1.0)$, and 0\%, 100\%, and 100\% at $(1.5,2.5)$. These comparisons show that both TCC and MaxCC recover the target once the main correlation peak is covered. MaxCC retains partial recall when the lag window does not encompass the main cross-response peak. Whether its upper limit reaches the prescribed response delay is therefore insufficient to determine whether the target can be identified.

Twenty independent noisy amplitude-modulated fields are separately centred, decomposed by rank-10 DMD and projected over their full mode sets, with $\delta=1.5$ and $\tau_{\max}=2.5$. Target matching uses the largest MAC within each conjugate group. Joint spatial-mode and carrier-frequency matching requires MAC at least 0.8 and relative carrier-frequency error no greater than 10\%. Matched MAC values are 0.99492--0.99498. TCC and MaxCC rank the matched group first, whereas CC$_0$ ranks it fourth or fifth. Reidentified angular frequencies are 1.5702--1.5728, approximately 12.2\%--12.3\% above the imposed 1.40 carrier, giving a joint pass rate of 0/20 (figure~\ref{fig:event_revision}b). This test distinguishes DMD identification accuracy from ranking performance on a fixed mode set.

\begin{figure}[!htbp]
    \centering
    \includegraphics[width=0.98\textwidth]{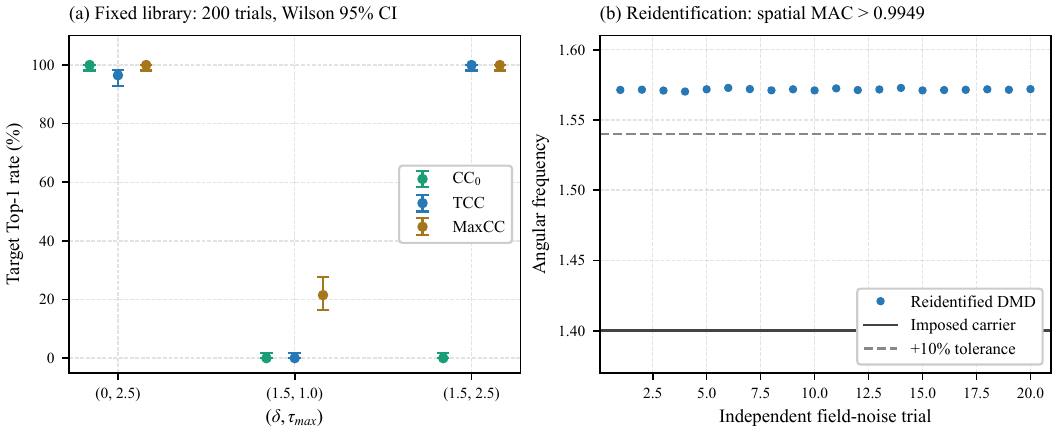}
    \caption{Time-lag aggregation and DMD mode identification in synthetic flow fields with amplitude modulation. (a) Top-1 recall over 200 trials with a fixed mode set and Wilson 95\% confidence intervals; the horizontal axis specifies the true delay and the upper limit of the window. (b) Target angular frequencies from 20 independent DMD identifications of noisy synthetic flow fields after amplitude modulation. The dashed line marks the upper bound for a 10\% relative carrier-frequency error.}
    \label{fig:event_revision}
\end{figure}
\FloatBarrier

\section{Evaluation settings and sensitivity results for the cylinder wake}
\label{app:cylinder_details}

\subsection{Data extraction, aerodynamic forces, and spatial parity}
\label{app:cylinder_protocol}

The cylinder data have $Re=U_{\infty}D/\nu=3900$, with $U_{\infty}=1$, $D=1$, $\nu=2.5641\times10^{-4}$, and span $L\approx\pi D$. Extracting fields from $t=599.75$ to 677.00 at $\Delta t=0.25$ gives 310 snapshots. All lag analyses use $\Delta\tau=\Delta t$, reading force values directly from the same time grid. The mid-plane lies at $z\approx\pi D/2$, and the analysis region contains 14600 cells. Coordinates, areas, and velocities are extracted and checked using original volume-mesh cell indices. DMD inner products and flow-field errors use original cell centres and area weights, without spatial interpolation. Figure~\ref{fig:cylinder_setup}(a) uses snapshot 156, labelled at relative time $tU_{\infty}/D=38.75$; triangulation is only for display.

Aerodynamic force divided by density follows the surface integral of pressure and viscous stresses,
\begin{equation}
    \frac{\mathbf{F}}{\rho}
    =\int_{S}\left[-p\mathbf{n}
    +(\nu+\nu_t)(\nabla\mathbf{u}+\nabla\mathbf{u}^{T})\mathbf{n}\right]\,\mathrm{d}S,
    \label{eq:cylinder_force}
\end{equation}
Here, $\rho$ is density, $S$ is the cylinder wall, and $p=p_{\mathrm{phys}}/\rho$, with $p_{\mathrm{phys}}$ the physical pressure. The turbulent kinematic viscosity is $\nu_t$, and $\mathbf{n}$ is the unit normal pointing from the cylinder into the fluid. Force coefficients are normalised by $\tfrac{1}{2}U_{\infty}^{2}DL$. Mean drag is 1.00694 and raw-drag RMS is 0.02291; mean lift is $-0.00557$ and lift-fluctuation RMS is 0.04895. Separate mean removal gives $C_L'$ and $C_D'$. Bandpass-filtered drag is first linearly detrended by least squares, then filtered over $0.35\leq St\leq0.55$ using a fourth-order zero-phase Butterworth filter.

Exact DMD uses area-weighted transverse velocities after pointwise mean removal. A 99.9\% singular-value energy threshold gives $r=252$. Full-set projection coordinates are computed for all 310 snapshots using~\eqref{eq:projected_coefficient}. All 123 finite, non-zero oscillatory groups with valid conjugates participate in ranking. Six real modes with negative eigenvalues lie at the Nyquist frequency and remain in the full-mode-set projection. Real-mode and conjugate-pairing checks are given in Appendix~\ref{app:numerical_audit}. The main $Q=41$ window spans $-5\leq\tau U_{\infty}/D\leq5$, giving 270 common indices $\mathcal I=\{21,\ldots,290\}$. RMS, CC$_0$, TCC, and MaxCC use this interval, with $K=6$ yielding 12 real-valued projection-coordinate features. For each selected set, coordinate real and imaginary parts form features for least-squares force-signal regression over all 310 samples.

The centreline-even and centreline-odd components of a transverse velocity mode are defined as
\begin{equation}
    \boldsymbol{\phi}^{+}_{j}(x,y)=
    \frac{\boldsymbol{\phi}_{j}(x,y)+\boldsymbol{\phi}_{j}(x,-y)}{2},
    \qquad
    \boldsymbol{\phi}^{-}_{j}(x,y)=
    \frac{\boldsymbol{\phi}_{j}(x,y)-\boldsymbol{\phi}_{j}(x,-y)}{2},
    \label{eq:cylinder_parity_components}
\end{equation}
The corresponding energy fraction is
\begin{equation}
    S_j^{(v)}=\frac{\lVert\boldsymbol{\phi}^{+}_{j}\rVert_W^2}
    {\lVert\boldsymbol{\phi}_{j}\rVert_W^2},
    \qquad
    A_j^{(v)}=\frac{\lVert\boldsymbol{\phi}^{-}_{j}\rVert_W^2}
    {\lVert\boldsymbol{\phi}_{j}\rVert_W^2}.
    \label{eq:cylinder_parity_fraction}
\end{equation}
Cells in the unstructured mesh are paired with their nearest mirror counterparts, with a maximum centre-to-centre distance of $5.35\times10^{-3}D$; the weight is the mean area of each cell pair. These energy fractions describe the centreline parity of the scalar transverse velocity.

\subsection{Diagnostics of slow variations in drag and ranking sensitivity}
\label{app:cylinder_sensitivity}

Slow drag variations appear in both the full-record trend and changes in segment means. We fit $C_D(t)=\beta_0+\beta_1t+\varepsilon(t)$ to all drag samples and measure the end-to-end linear change as $\Delta C_D^{\mathrm{lin}}=\beta_1(t_N-t_1)$. Dividing the 310 points chronologically into three approximately equal segments, define
\begin{equation}
    R_{\mu,D}=\frac{\max_s\overline{C_D}^{(s)}-\min_s\overline{C_D}^{(s)}}{\sigma(C_D')},
    \qquad s=1,2,3,
    \label{eq:cylinder_drift_metric}
\end{equation}
to measure the range of segment means. These means first decrease and then recover (table~\ref{tab:cylinder_drift_summary}). Local slopes also change from negative to positive (figure~\ref{fig:cylinder_cd_stability}), such that the interval cannot be described as a monotonic full-record drift. The moving window spans approximately $19.8D/U_{\infty}$. Window-mean deviations and end-to-end changes of window-fitted lines are normalised by $\sigma(C_D')$. Hann-windowed one-sided periodograms are computed for uncentred, mean-removed and linearly detrended drag. The lowest non-zero-frequency peak is compared with the second-harmonic peak within $0.3\leq St\leq0.6$. Their power ratio remains far below unity after linear detrending (table~\ref{tab:cylinder_drift_summary}); therefore, removing the linear trend does not remove low-frequency dominance.

\begin{table}[!htbp]
    \centering
    \caption{Diagnostics of slow cylinder-drag variations. Peak power ratios compare the second-harmonic peak at $St=0.4645$ with the lowest non-zero-frequency peak at $St=0.0129$.}
    \label{tab:cylinder_drift_summary}
    \begin{tabular*}{\textwidth}{@{\extracolsep{\fill}}llr@{}}
        \toprule
        Diagnostic quantity & Evaluation range or processing & Value\\
        \midrule
        Linear change $\Delta C_D^{\mathrm{lin}}$ & All samples & $-0.02452$\\
        Normalised linear change $\Delta C_D^{\mathrm{lin}}/\sigma(C_D')$ & All samples & $-1.07$\\
        Mean drag $\overline{C_D}^{(s)}$ & Segment 1 & 1.02523\\
        & Segment 2 & 0.98897\\
        & Segment 3 & 1.00644\\
        Segment-mean range $R_{\mu,D}$ & Three segments & 1.58\\
        \midrule
        Second-harmonic/lowest-frequency peak power ratio & Mean removal & $3.11\times10^{-3}$\\
        & Linear detrending & $3.99\times10^{-3}$\\
        \bottomrule
    \end{tabular*}
\end{table}

\begin{figure}[!htbp]
    \centering
    \includegraphics[width=0.98\textwidth]{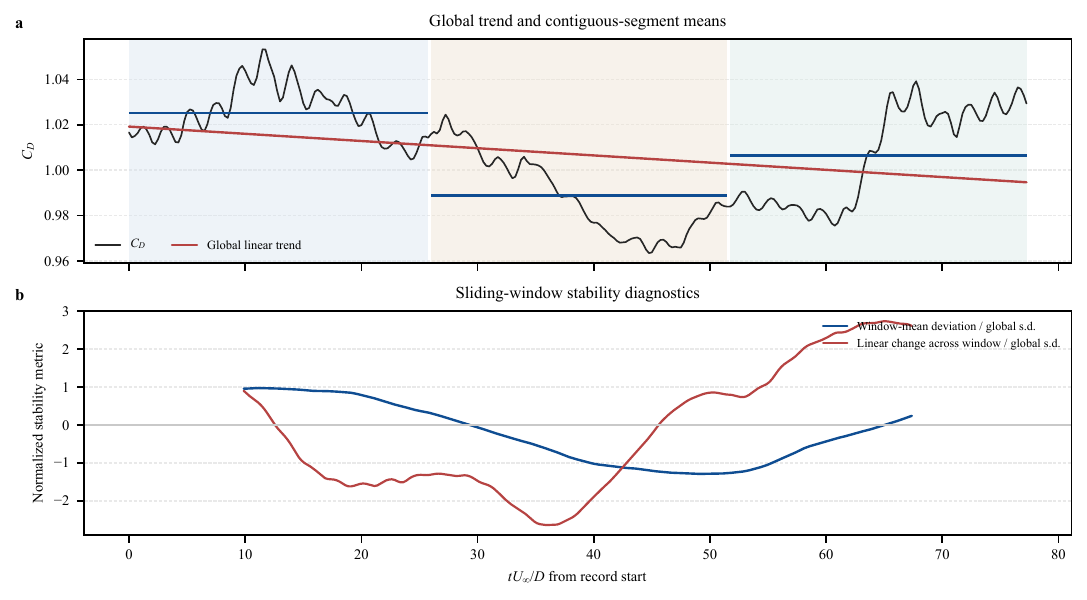}
    \caption{Diagnostics of slow cylinder-drag variations. (a) Drag signal, linear trend, and segment means. (b) Moving-window means and slopes.}
    \label{fig:cylinder_cd_stability}
\end{figure}

\begin{table}[!htbp]
    \centering
    \renewcommand{\arraystretch}{0.94}
    \caption{Frequency rankings and reconstruction metrics for the main cylinder comparison.}
    \label{tab:cylinder_comparison_details}
    \begin{tabular*}{\textwidth}{@{\extracolsep{\fill}}lrrrrrr@{}}
        \toprule
        \multicolumn{7}{@{}l}{(a) Selected frequencies $St$ ordered by each ranking criterion}\\[2pt]
        Ranking & Group 1 & Group 2 & Group 3 & Group 4 & Group 5 & Group 6\\
        \midrule
        RMS & 0.223 & 0.096 & 0.692 & 0.171 & 0.215 & 0.534\\
        Bandpass-filtered drag TCC & 0.447 & 0.426 & 0.448 & 0.406 & 0.484 & 0.466\\
        \bottomrule
    \end{tabular*}
    \par\vspace{5pt}
    \begin{tabular*}{\textwidth}{@{\extracolsep{\fill}}lcc@{}}
        \toprule
        \multicolumn{3}{@{}l}{(b) Target-signal reconstruction correlation coefficients for the main rankings}\\[2pt]
        Ranking & $K=6$ lift fluctuations & $K=6$ raw drag\\
        \midrule
        Lift-targeted TCC & 0.955 & --\\
        Raw drag TCC & -- & 0.976\\
        RMS & 0.908 & 0.117\\
        \bottomrule
    \end{tabular*}
    \par\vspace{5pt}
    \begin{tabular*}{\textwidth}{@{\extracolsep{\fill}}llcc@{}}
        \toprule
        \multicolumn{4}{@{}l}{(c) Top-6 reconstruction for cylinder lift fluctuations}\\[2pt]
        Method & Groups & \shortstack{Target-signal\\error (\%)} & \shortstack{Flow-field\\error (\%)}\\
        \midrule
        RMS & 6 conjugate groups & 41.86 & 77.26\\
        Kou--Zhang & 6 conjugate groups & 38.23 & 63.72\\
        CC$_0$ & 6 conjugate groups & 29.92 & 63.30\\
        TCC & 6 conjugate groups & 29.68 & 64.21\\
        \bottomrule
    \end{tabular*}
    \par\vspace{5pt}
    \parbox{\textwidth}{\footnotesize Note: -- denotes an unlisted entry. Lift-fluctuation and raw-drag TCC use their respective target-specific selections.}
\end{table}

Table~\ref{tab:cylinder_comparison_details} lists frequency orders and reconstruction metrics for the main comparison. The TCC--RMS difference in target-signal correlation is larger for raw drag than for lift fluctuations. For linearly detrended drag, TCC and CC$_0$ have equal Top-6 target-signal errors. Panel (c) compares Top-6 reconstruction over all 310 samples, where lift-targeted TCC improves both target-signal and transverse-velocity-field reconstruction relative to the RMS. Raw-drag-targeted TCC gives a six-group flow-field error of 97.47\%. Therefore, these groups, associated with slow drag variations, represent the transverse-velocity field only to a limited extent.

The segment test in \S\ref{sec:cylinder_results} measures selection stability by overlap with the full-record Top-6. Feature centring, standard-deviation scaling, regression intercept, and coefficients are determined within the fitting segment and then applied to the remaining segments. Cross-segment $E_y$ uses the target mean over the combined evaluation segments as $\overline y_{\mathrm{ref}}$.\par

\begin{table}[!htbp]
    \centering
    \caption{Sensitivity of cylinder lift-fluctuation rankings and bandpass-filtered-drag reconstruction to the mode set, time segment, and sample interval.}
    \label{tab:cylinder_sensitivity_details}
    \renewcommand{\arraystretch}{0.9}
    \begin{tabular*}{\textwidth}{@{\extracolsep{\fill}}lrrr@{}}
        \toprule
        \multicolumn{4}{@{}l}{(a) DMD reidentified at different truncation ranks}\\[2pt]
        Truncation rank $r$ & 200 & 252 & 280\\
        \midrule
        Leading TCC group frequency $St$ & 0.2248 & 0.2233 & 0.2228\\
        Top-6 target-signal error (\%) & 33.70 & 29.68 & 30.44\\
        \bottomrule
    \end{tabular*}
    \par\vspace{5pt}
    \begin{tabular*}{\textwidth}{@{\extracolsep{\fill}}lrrr@{}}
        \toprule
        \multicolumn{4}{@{}l}{(b) Consecutive-segment test with the fixed rank-252 mode set}\\[2pt]
        Segment & 1 & 2 & 3\\
        \midrule
        Groups shared with full-record Top-6 & 3 & 4 & 3\\
        Leading group frequency $St$ & 0.2233 & 0.2233 & 0.2233\\
        Cross-segment target-signal error (\%) & 85.81 & 47.42 & 71.73\\
        \bottomrule
    \end{tabular*}
    \par\vspace{5pt}
    \begin{tabular*}{\textwidth}{@{\extracolsep{\fill}}llr@{}}
        \toprule
        \multicolumn{3}{@{}l}{(c) Symmetric two-sided window sweep on 250 common valid samples}\\[2pt]
        Lag count $Q$ & Top-6 $St$ (priority order) & \shortstack{Target-signal\\error (\%)}\\
        \midrule
        1, 11, 21, 41 & 0.2233, 0.2145, 0.2388, 0.1713, 0.2788, 0.1860 & 29.92\\
        61 & 0.2233, 0.2145, 0.2388, 0.1713, 0.1860, 0.2788 & 29.92\\
        \bottomrule
    \end{tabular*}
    \par\vspace{5pt}
    \begin{tabular*}{\textwidth}{@{\extracolsep{\fill}}lrr@{}}
        \toprule
        \multicolumn{3}{@{}l}{(d) Top-6 target-signal error for bandpass-filtered drag (\%)}\\[2pt]
        Fitting and evaluation samples & CC$_0$ & TCC\\
        \midrule
        Complete 310-point record & 60.51 & 51.19\\
        250 points after trimming 30 points from each end & 49.46 & 43.47\\
        \bottomrule
    \end{tabular*}
    \par\vspace{5pt}
    \parbox{\textwidth}{\footnotesize Note: (b) Rankings and regression are fitted within each segment and applied to the remaining segments. (c) The maximum symmetric window, $Q_{\max}=61$ over $[-7.5,7.5]D/U_{\infty}$, fixes 250 common valid samples; $Q=1$ gives CC$_0$. The main comparison ranks on 270 points; both evaluations reconstruct over all 310 samples. (d) Common valid samples, rankings, and regressions are recalculated within each retained sample set; 250 points denotes the sample count after trimming.}
\end{table}

Leading-group frequency stability and stability of the full truncated set require separate assessment. Across DMD truncation ranks, lift-targeted TCC consistently ranks a near-fundamental group first, but Top-6 target-signal error does not decrease monotonically (table~\ref{tab:cylinder_sensitivity_details}(a)). With the mode set fixed, the fundamental group leads in all three contiguous segments, while remaining selections and cross-segment errors vary substantially (panel b). In the main lift comparison, CC$_0$ and TCC share their Top-5 and give similar target-signal errors. MaxCC selects a different Top-6, with 29.19\% error.

All tested symmetric two-sided windows give the same Top-6 on the shared 250 ranking samples; non-zero lags do not reduce lift-fluctuation error (table~\ref{tab:cylinder_sensitivity_details}(c)). For bandpass-filtered drag, TCC gives lower errors both before and after trimming (panel d). After trimming, rankings and regression are refitted on the retained samples.

Direction comparisons fix $Q=41$ and total width $10D/U_{\infty}$, using windows $[-10,0]$, $[0,10]$, and $[-5,5]D/U_{\infty}$. All share 230 common valid samples covering the union $[-10,10]D/U_{\infty}$. Negative, positive, and two-sided windows give TCC Top-6 lift-signal errors of 28.90\%, 29.56\%, and 29.36\%, respectively; CC$_0$ gives 29.56\% on the same samples. Direction affects secondary-group selection over this finite interval, but all three windows retain the leading group at $St\approx0.223$.

\section{Cavity data preprocessing and sensitivity results}
\label{app:cavity_details}

\subsection{Data preprocessing and selection of common valid samples}
\label{app:cavity_protocol}

Power spectra are estimated using Welch's method~\citep{welch1967}; window lengths, overlap lengths, and frequency spacings are specified for each analysis.

Challenge 2.1 training data are two-dimensional, two-component HDF5 records containing 12800 snapshots over 0.79994\,s on a $154\times51$ grid. The public description gives a cavity width-to-depth ratio of 3.85 and velocity units of $\mathrm{m\,s^{-1}}$. File coordinates span $0.1104\leq x\leq5.8343$ and $-0.8535\leq y\leq1.0171$. Spatial scales are reported in data coordinates; their physical-length conversion is undetermined, such that no Strouhal conversion is made. The observable is local $u_y'$ at array index $(22,23)$ and coordinates $(5.0113,0.00697)$.

The original probe spectrum uses Welch estimation with 1600-point Hann windows and 800-point overlap, giving 10\,Hz spacing. Power above the post-downsampling Nyquist frequency of 2000\,Hz accounts for 33.21\%. Before downsampling, the same Kaiser finite impulse response low-pass filter is applied to both velocity components at every grid point (table~\ref{tab:cavity_filter_parameters}). It retains near-unit gain across both main target bands and suppresses components above the new Nyquist frequency. Centred polyphase filtering uses linear boundary extension. End samples affected by extension are removed according to the filter half-width, leaving 3148 synchronised snapshots over 0.78675\,s. Pointwise temporal-mean removal gives $\mathbf{x}'_k=[\mathbf{u}'_{x,k};\mathbf{u}'_{y,k}]\in\mathbb R^{15708}$. The target is the pre-processed probe's transverse velocity fluctuation, identical to independently filtering and decimating the original probe signal in the same way.

\begin{table}[!htbp]
    \centering
    \caption{Anti-aliasing and downsampling parameters for the cavity data.}
    \label{tab:cavity_filter_parameters}
    \begin{tabular}{ll}
        \toprule
        Parameter & Value\\
        \midrule
        Original/downsampled sampling rate & 16/4\,kHz\\
        Passband/stopband edge & 1600/2000\,Hz\\
        FIR tap count/cutoff frequency & 203 taps; 1800\,Hz\\
        Design/measured stopband attenuation & 80/approximately 79\,dB\\
        Kaiser parameter $\beta$ & 7.85726\\
        Passband ripple & $<0.002$\,dB\\
        Trimming after downsampling & 26 samples at each end\\
        Retained snapshots/snapshot interval $\Delta t$ & 3148; $2.5\times10^{-4}$\,s\\
        \bottomrule
    \end{tabular}
\end{table}

Equal weights are used at all grid points, such that $\mathbf{W}=\mathbf{I}$. After preprocessing, exact DMD is constructed through the snapshot Gram matrix, with $r=200$ retaining 96.43\% of snapshot singular-value energy. After mode normalisation, all 3148 snapshots are projected by least squares over the full set; these coordinates remain fixed throughout ranking. Numerical rank and conditioning are reported in Appendix~\ref{app:numerical_audit}. All lag analyses use $\Delta\tau=\Delta t=2.5\times10^{-4}$\,s and directly read pre-processed observations. The main window is $[-16,16]$\,ms ($Q=129$), with $\mathcal L=\{-64,\ldots,64\}$ and 3020 common indices $\mathcal I=\{65,\ldots,3084\}$. RMS, CC$_0$, TCC, and MaxCC use these samples. Kou--Zhang integrates the exponential sequence defined by initial amplitudes and eigenvalues over all 3148 times, following its original definition. All fixed-set rankings share 94 oscillatory candidates and full-set projection coordinates. Sensitivity sweeps supplement, but do not replace, the main $[-16,16]$\,ms comparison (Appendix~\ref{app:cavity_sensitivity}).

\subsection{Mode rankings and projection-coordinate spectra}
\label{app:cavity_spectral_details}

The first two TCC groups have lagged cross-response peaks at 0.25 and 0\,ms, respectively. CC$_0$ and MaxCC share five Top-6 groups; MaxCC replaces the 517.30\,Hz group with the 138.74\,Hz group. Neither includes the 1519.53\,Hz group; complete rankings are given in table~\ref{tab:cavity_rank_spectrum}.

\begin{table}[!htbp]
    \centering
    \caption{Top-6 cavity rankings, dominant projection-coordinate spectral peaks, and power fractions in neighbourhoods of the DMD characteristic frequencies.}
    \label{tab:cavity_rank_spectrum}
    \renewcommand{\arraystretch}{0.9}
    \setlength{\tabcolsep}{4pt}
    \begin{tabular*}{\textwidth}{@{\extracolsep{\fill}}lrrrrrr@{}}
        \toprule
        Rank & 1 & 2 & 3 & 4 & 5 & 6\\
        \midrule
        \multicolumn{7}{@{}l}{(a) DMD frequencies of selected groups (Hz)}\\[2pt]
        TCC & 941.87 & 1519.53 & 478.19 & 1121.75 & 517.30 & 121.32\\
        RMS & 478.19 & 517.30 & 200.42 & 121.32 & 36.53 & 138.74\\
        CC$_0$ & 1121.75 & 941.87 & 121.32 & 1140.48 & 478.19 & 517.30\\
        MaxCC & 1121.75 & 941.87 & 121.32 & 138.74 & 478.19 & 1140.48\\
        \midrule
        \multicolumn{7}{@{}l}{(b) TCC coordinate spectra and DMD-frequency neighbourhood power (TCC order in a)}\\[2pt]
        \shortstack[l]{DMD frequency\\(neighbourhood centre, Hz)} & 941.87 & 1519.53 & 478.19 & 1121.75 & 517.30 & 121.32\\
        Dominant coordinate peak (Hz) & 937.50 & 1535.16 & 593.75 & 1437.50 & 593.75 & 50.78\\
        Power within $\pm25$\,Hz of DMD frequency (\%) & 68.41 & 57.31 & 3.93 & 4.39 & 3.37 & 8.02\\
        Power within $\pm50$\,Hz of DMD frequency (\%) & 78.80 & 80.91 & 8.21 & 9.25 & 7.18 & 13.67\\
        Power within $\pm75$\,Hz of DMD frequency (\%) & 83.02 & 85.84 & 11.92 & 13.44 & 11.51 & 21.76\\
        \bottomrule
    \end{tabular*}
    \par\vspace{5pt}
    \parbox{\textwidth}{\footnotesize Note: Two-sided Welch spectra use the positive-frequency representative's complex projection coordinate, 1024-point Hann windows and 512-point overlap, giving 3.90625\,Hz spacing. Peak frequencies are absolute values. Percentages measure coordinate power near each group's DMD frequency relative to its own total power, combining positive and negative frequency neighbourhoods. Figure~\ref{fig:cavity_ranking}c,d shows one-sided spectra of the real coordinates for the first four groups; both spectral calculations give the same principal peaks for these groups.}
\end{table}

Coordinate power fractions are measured in neighbourhoods centred on each DMD frequency $f_j$, with half-widths $\Delta f=25$, 50 and 75\,Hz. The main results use $\Delta f=50$\,Hz. For the two-sided complex-coordinate Welch spectrum $P_j(\nu)$, the fraction is $\sum_{\lvert |\nu|-f_j\rvert<\Delta f}P_j(\nu)/\sum_\nu P_j(\nu)$. The numerator combines neighbourhoods at positive and negative frequencies; the denominator is the coordinate's total spectral power. The first two TCC groups retain 68.41\% and 57.31\%, respectively, within $\pm25$\,Hz. The third and fourth coordinate spectra peak away from their DMD frequencies; even at $\pm75$\,Hz, fractions are only 11.92\% and 13.44\%, respectively (table~\ref{tab:cavity_rank_spectrum}).

The first two TCC groups (941.87 and 1519.53\,Hz) have $|\lambda|$ values of 0.9501 and 0.9583, respectively. These values describe the decay prescribed by the identified operator, but the snapshot-wise projection coordinates need not follow that decay law.
\FloatBarrier

\begingroup
\raggedbottom
\setlength{\intextsep}{7pt}
\setlength{\abovedisplayskip}{4pt}
\setlength{\belowdisplayskip}{4pt}
\setlength{\textfloatsep}{7pt plus 2pt minus 2pt}
\paragraph{Mode-group signals at the observation point and spatial variation scales.}
To connect the spatial modes directly to the local target, let $\mathbf{e}_p$ extract the transverse velocity component at the observation point, such that $y(k)=\mathbf{e}_p^T\mathbf{x}'_k$. The signal of mode group $g$ at this location is
\begin{equation}
    y_g(k)=\mathbf e_p^T\mathbf x'_g(k)
    =\operatorname{Re}\left[
    \sum_{j\in\mathcal G_g}\phi_{j,p}a_j(k)\right].
    \label{eq:cavity_local_group_signal}
\end{equation}
This signal is synthesised directly from the spatial modes and projection coordinates over the full mode set, without fitting the regression weights used in target-signal reconstruction. Its one-sided Welch spectrum uses a 1024-point Hann window and a 512-point overlap. For neighbourhoods centred on the original target spectral peaks at 940 and 1540\,Hz, the fraction of the group's total local spectral power within each $\pm50$\,Hz neighbourhood is calculated. The local amplitude ratio in table~\ref{tab:cavity_local_group_signal} is $\operatorname{std}(y_g)/\operatorname{std}(y)$. Cross-terms between mode groups prevent these quantities from forming an additive partition of target energy.

Length scales can be constructed from the ratio of mean-square fluctuating velocity to its mean-square spatial derivative~\citep{zhu2022jet}. Following this form, we define a streamwise variation scale for the positive-frequency representative mode's complex transverse velocity profile at the observation height $y_p$:
\begin{equation}
    L_{x,g}=\left[
    \frac{\int|\phi_{j_g,y}(x,y_p)|^2\,\mathrm dx}
    {\int|\partial_x\phi_{j_g,y}(x,y_p)|^2\,\mathrm dx}
    \right]^{1/2}.
    \label{eq:cavity_streamwise_variation_scale}
\end{equation}
The integrals cover the entire measured streamwise interval; derivatives are calculated by grid differences, and integration uses the trapezoidal rule. This measure is invariant to the overall phase and amplitude scaling of a mode and, in data-coordinate units, quantifies a spatial variation scale rather than a wavelength. For the first two TCC groups, the corresponding $L_{x,g}$ values are 0.5017 and 0.3672, respectively, representing a 26.80\% difference.

\begin{table}[!htbp]
    \centering
    \small
    \caption{Velocity signals at the observation point and streamwise variation scales of representative cavity mode groups.}
    \label{tab:cavity_local_group_signal}
    \begin{tabular*}{\textwidth}{@{\extracolsep{\fill}}rrrrrr@{}}
        \toprule
        \shortstack{DMD frequency\\(Hz)} & \shortstack{Local peak\\(Hz)} & \shortstack{Local amplitude\\ratio} & \shortstack{940\,Hz band\\power (\%)} & \shortstack{1540\,Hz band\\power (\%)} & $L_{x,g}$\\
        \midrule
        941.87 & 937.50 & 0.509 & 79.25 & 0.63 & 0.5017\\
        1519.53 & 1535.16 & 0.392 & 0.52 & 82.02 & 0.3672\\
        478.19 & 593.75 & 1.455 & 5.49 & 2.87 & 0.3689\\
        \bottomrule
    \end{tabular*}
    \par\vspace{3pt}
    \parbox{\textwidth}{\footnotesize Note: The first two rows are the first two TCC groups; the third is the leading RMS group, also third in TCC. Local peaks and powers are computed from $y_g$, unlike the complex-coordinate $a_j$ spectra in table~\ref{tab:cavity_rank_spectrum}. Neighbourhoods are centred on target spectral peaks, not DMD frequencies.}
\end{table}
\FloatBarrier

\paragraph{Spatial organisation and cross-spectral checks of the two bands.}
\label{app:cavity_spatial_organization}
For the positive-frequency representative modes of the two groups, let $\phi_{g,y}(x,y)$ denote the complex transverse velocity component. We define the normalised amplitude, phase relative to the original probe $p=(x_p,y_p)$, and squared-amplitude-weighted streamwise centre as
\begin{equation}
    \begin{gathered}
    A_g(x,y)=\frac{|\phi_{g,y}(x,y)|}{\max_{x,y}|\phi_{g,y}(x,y)|},\qquad
    \theta_g(x,y)=\arg\!\left[\phi_{g,y}(x,y)\phi_{g,y}^{*}(x_p,y_p)\right],\\
    x_{c,g}=\frac{\sum_{m,n}x_m|\phi_{g,y}(x_m,y_n)|^2}
    {\sum_{m,n}|\phi_{g,y}(x_m,y_n)|^2}.
    \end{gathered}
    \label{eq:cavity_spatial_geometry}
\end{equation}
These measures of spatial shape use equal grid weights, as in the main text. The phase profiles are evaluated at the original observation height $y_p$ and cover all 154 streamwise grid points.

Let $\eta_x(k)=u_y'(x,y_p,t_k)$ be the pre-processed local signal at the observation height. Cross-spectra and coherence measure its frequency-domain association with the original probe $y(k)$~\citep{tinney2006slse}. Welch spectra use all 3148 samples, 512-point Hann windows and 256-point overlap, giving 11 overlapping windows and 7.8125\,Hz spacing. With convention $S_{y\eta_x}=\langle\widehat y^{*}\widehat\eta_x\rangle$, its argument is the local velocity phase relative to the probe. This sign agrees with the positive-frequency mode's relative phase. For fixed bands $B_g$ centred at 940 and 1540\,Hz with half-width 50\,Hz, define
\begin{equation}
    \gamma^2(x,\nu)=\frac{|S_{y\eta_x}(\nu)|^2}{S_{yy}(\nu)S_{\eta_x\eta_x}(\nu)},\qquad
    \overline{\gamma^2}_{B_g}(x)=
    \frac{\sum_{\nu\in B_g}S_{yy}(\nu)\gamma^2(x,\nu)}
    {\sum_{\nu\in B_g}S_{yy}(\nu)}.
    \label{eq:cavity_spatial_coherence}
\end{equation}
The left expression in~\eqref{eq:cavity_spatial_coherence} is standard magnitude-squared coherence~\citep{carter1987}; the right is its target-spectrum-weighted band average used here.
Every location uses the same target-spectrum weights. The main text reports the arithmetic mean over 153 non-reference points, excluding the reference point, whose self-coherence is identically unity. Cross-spectral phases are evaluated at maximum target power within each band, at 937.50 and 1539.06\,Hz, using the same frequencies at every location. Modal relative phases and cross-spectral phases are separately unwrapped along increasing $x$ and shifted to zero at the reference point.

Using a 1024-point Hann window with a 512-point overlap gives target spectral peaks at 937.50 and 1542.97\,Hz and mean magnitude-squared coherences of 0.836 and 0.831 over the non-reference points in the two bands. The correspondence between the two phase measures remains consistent.

\FloatBarrier

\endgroup

\subsection{Truncation-rank, candidate-threshold, window, and segment results}
\label{app:cavity_sensitivity}

Table~\ref{tab:cavity_rank_comparison} compares the results at different truncation ranks. The first two TCC groups consistently correspond to the two main target bands near 940 and 1540\,Hz, whereas the next four selected groups vary with rank.

\begin{table}[!htbp]
    \centering
    \small
    \caption{Cavity mode rankings and target-signal reconstruction at different truncation ranks.}
    \label{tab:cavity_rank_comparison}
    \setlength{\tabcolsep}{4pt}
    \begin{tabular*}{\textwidth}{@{\extracolsep{\fill}}rrrrrr@{}}
        \toprule
        \shortstack{Rank\\$r$} & \shortstack{Oscillatory\\groups} & \shortstack{TCC first\\$f$ (Hz)} & \shortstack{TCC second\\$f$ (Hz)} & \shortstack{TCC\\$E_y$ (\%)} & \shortstack{CC$_0$\\$E_y$ (\%)}\\
        \midrule
        150 & 70 & 938.21 & 1516.06 & 76.71 & 73.84\\
        200 & 94 & 941.87 & 1519.53 & 74.16 & 84.09\\
        250 & 119 & 943.23 & 1521.60 & 74.35 & 73.81\\
        \bottomrule
    \end{tabular*}
    \par\vspace{3pt}
    \parbox{\textwidth}{\footnotesize Note: Frequencies are DMD eigenvalue frequencies. Both methods select the first six groups; $E_y$ is target-signal reconstruction error.}
\end{table}

All valid complex-conjugate groups with $0<f<f_N=2000\,\mathrm{Hz}$ enter the oscillatory candidate set. The rank-200 mode set contains 94 conjugate pairs and 12 real modes, with the latter retained only in the projection onto the full set. The highest oscillatory-group frequency is 1519.53\,Hz, leaving no candidates between 1600 and 2000\,Hz. Holding the mode set and projection coordinates fixed, additional lower bounds on eigenvalue modulus, $|\lambda|_{\min}=0$, 0.3 and 0.5, retain 94, 66, and 25 candidate groups, respectively. All three give the same TCC Top-6 set and order, with a target-signal reconstruction error of 74.16\%. The main setting is 0, which imposes no additional filter. Raising the lower bound to 0.7 leaves 8 groups. Both major target bands remain, but TCC and CC$_0$ now select the same Top-6, differing only in the order of the fourth and fifth groups. Both give a reconstruction error of 72.35\%. Thus, candidate filtering can preserve the main frequency bands while altering the composition of the selected set and the relative reconstruction errors obtained with TCC and CC$_0$.

As in the cylinder case, the main comparison selects six conjugate groups, yielding 12-dimensional real-valued projection coordinates. Figure~\ref{fig:cavity_errors} reports target-signal reconstruction for $K=1$--12 groups. At $K=8$, CC$_0$ and TCC errors are 72.12\% and 74.05\%, respectively; at $K=12$, they are 69.94\% and 72.41\%. Their relative order therefore differs from that at Top-6.\par

The two-sided sweep uses $\mathcal L=\{-(Q-1)/2,\ldots,(Q-1)/2\}$. Values $Q=1$, 17, 33, 65, 129, and 257 correspond to zero lag and $\pm2$, $\pm4$, $\pm8$, $\pm16$, and $\pm32$\,ms. The largest window fixes 2892 common valid samples. Top-6 target-signal errors are 84.09\%, 73.35\%, 74.16\%, 74.16\%, 74.16\%, and 75.79\%, respectively. Direction comparisons fix 129 lags and 32\,ms total width, using $[-32,0]$, $[0,32]$, and $[-16,16]$\,ms. Their union, $[-32,32]$\,ms, again defines 2892 common valid samples. Both comparisons evaluate on all 3148 samples; direction results are given in table~\ref{tab:cavity_lag_direction}.

\begin{table}[!htbp]
    \centering
    \caption{Cavity lag-direction comparisons with equal widths and common valid samples: Top-6 target-signal reconstruction error (\%).}
    \label{tab:cavity_lag_direction}
    \begin{tabular}{lrrr}
        \toprule
        Window (ms) & CC$_0$ & TCC & MaxCC\\
        \midrule
        $[-32,0]$ & 84.09 & 75.79 & 84.09\\
        $[0,32]$ & 84.09 & 74.97 & 84.15\\
        $[-16,16]$ & 84.09 & 74.16 & 84.15\\
        \bottomrule
    \end{tabular}
\end{table}
\FloatBarrier

The original 16\,kHz snapshot sequence is divided chronologically into four equal segments of 3200 snapshots. Let $\langle\cdot\rangle_{\Omega}$ denote equal-weight spatial averaging over the measurement domain and $\mathbf{u}_k=(u_x,u_y)^T$ the two-component velocity in snapshot $k$. Define
\begin{equation}
    \begin{aligned}
        U_{\Omega}(k)&=\langle u_x\rangle_{\Omega},\\
        V_{\Omega}(k)&=\left[\left\langle\|\mathbf u_k-\langle\mathbf u_k\rangle_{\Omega}\|_2^2\right\rangle_{\Omega}\right]^{1/2},\\
        e_{\Omega}(k)&=\frac{1}{2}\langle\|\mathbf u_k\|_2^2\rangle_{\Omega}.
    \end{aligned}
    \label{eq:cavity_segment_statistics}
\end{equation}
Here, $V_{\Omega}$ is velocity RMS relative to the instantaneous spatial mean, and $e_{\Omega}$ is kinetic energy per unit mass in the two measured components. Each quantity is time-averaged within each segment. For segment means $q_s$ ($s=1,\ldots,4$), the relative range is $100(\max_s q_s-\min_s q_s)/(\frac{1}{4}\sum_{s=1}^{4}q_s)$. The three relative ranges are 2.30\%, 0.59\%, and 1.94\%, respectively. The two principal probe peaks lie in 930--960 and 1530--1550\,Hz.

Segment rankings and cross-segment reconstruction of pre-processed data follow \S\ref{sec:cavity_results}; figure~\ref{fig:cavity_revision} summarises the sensitivity comparisons. The segment Top-6 sets share 4, 4, 5, and 4 groups with the full-record Top-6, respectively. The first two omit 1519.53\,Hz, the third omits 941.87\,Hz, and the fourth retains both. Centring, scaling, regression intercept, and coefficients are fitted within each segment and applied to the remaining three. Cross-segment $E_y$ uses the combined evaluation-target mean as $\overline y_{\mathrm{ref}}$. Errors are 87.84\%, 88.75\%, 90.11\%, and 77.74\%, respectively.

\begin{figure}[!htbp]
    \centering
    \includegraphics[width=0.98\textwidth]{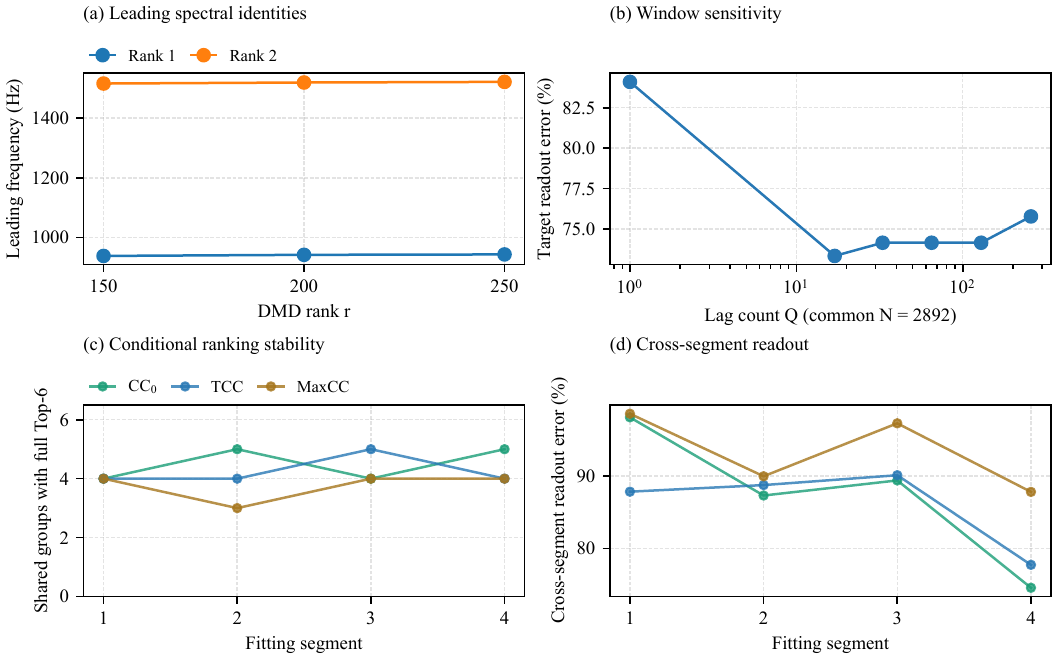}
    \caption{Cavity sensitivity to frequency bands, lag windows, and time segments. (a) First two TCC frequencies across DMD truncation ranks. (b) Two-sided window sensitivity with 2892 fixed common valid samples. (c) Number of groups shared with the full-record Top-6. (d) Cross-segment target-signal error after fitting on one segment and applying to the others. Segments share the full-flow mode set; ranking and regression are refitted within each segment.}
    \label{fig:cavity_revision}
\end{figure}

\subsection{Reconstruction comparisons and cross-terms between non-orthogonal mode groups}
\label{app:cavity_expression}

Table~\ref{tab:cavity_full_record} summarises Top-6 reconstruction over all 3148 samples. Every method retains six conjugate groups and uses the full-mode-set projection coordinates.

The full rank-200 set gives a flow-field projection error of 22.21\%. Removing only the 12 real modes while retaining 94 oscillatory groups raises fixed-coefficient truncated error to 113.35\%, indicating cancellation between real and oscillatory contributions. The corresponding correlation-operator residuals are 3.86\% and 22.99\%. With six groups, CC$_0$ and TCC give residuals of 64.39\% and 45.55\%, respectively. These results numerically illustrate the cross-group terms in Appendix~\ref{app:operator_theory}.

\begin{table}[!htbp]
    \centering
    \caption{Top-6 target-signal and flow-field reconstruction errors over all cavity samples.}
    \label{tab:cavity_full_record}
    \begin{tabular}{lcc}
        \toprule
        Method & Target-signal error (\%) & Flow-field error (\%)\\
        \midrule
        RMS & 98.79 & 205.14\\
        Kou--Zhang & 90.35 & 248.03\\
        CC$_0$ & 84.09 & 172.43\\
        TCC & 74.16 & 185.44\\
        MaxCC & 84.15 & 263.53\\
        \bottomrule
    \end{tabular}
\end{table}

\FloatBarrier

\section{Settings and results for POD mode ranking}
\label{app:pod_validation}

\subsection{Mode sets, common valid samples, and structure evaluation}
\label{app:pod_protocol}

All three cases apply the weighted SVD in~\eqref{eq:pod_weighted_svd} to the mean-removed snapshots. The synthetic and cavity fields include both velocity components with $\mathbf{W}=\mathbf{I}$; the cylinder uses transverse velocity and original cell-area weights. All snapshots construct the POD set, while the target enters only subsequent ranking. Table~\ref{tab:pod_settings} lists mode-set and sample sizes. Six cylinder or cavity DMD conjugate groups span a 12-dimensional real subspace, whereas six real POD modes span six dimensions. Lag windows are $[0,3.95]$ for the synthetic field, $[-5,5]D/U_\infty$ for the cylinder and $[-16,16]$\,ms for the cavity. Each retains its sampling increment and common-valid-sample rules. Weighted orthogonality errors $\|\boldsymbol{\Phi}^H\mathbf{W}\boldsymbol{\Phi}-\mathbf{I}\|_F$ are $3.40\times10^{-15}$, $3.63\times10^{-14}$, and $3.36\times10^{-14}$, respectively. Full-POD-set flow-field projection errors are 7.93\%, 3.19\%, and 18.89\%.

\begin{table}[!htbp]
    \centering
    \caption{Spatial dimensions, snapshot counts, truncation ranks, and common valid sample counts for POD analysis. $\eta_r$ follows~\eqref{eq:pod_energy}; cylinder squared contributions include transverse velocity only.}
    \label{tab:pod_settings}
    \begin{tabular}{lrrrrr}
        \toprule
        Case & $M$ & $N_{\mathrm{raw}}$ & $r$ & $N$ & $\eta_r$ (\%)\\
        \midrule
        Synthetic & 2688 & 1200 & 10 & 1121 & 99.37\\
        Cylinder & 14600 & 310 & 252 & 270 & 99.90\\
        Cavity & 15708 & 3148 & 200 & 3020 & 96.43\\
        \bottomrule
    \end{tabular}
\end{table}

The synthetic field uses analytical target subspaces to evaluate structural coverage. Let $\mathbf{Q}_{\mathrm{rel}}$ and $\mathbf{Q}_{\mathcal S}$ be real orthonormal bases in weighted space for the analytical target and selected-mode subspaces. For target dimension $d_{\mathrm{rel}}$, define target-subspace coverage as
\begin{equation}
    \Gamma_{\mathrm{rel}}(\mathcal S)
    =\frac{100}{d_{\mathrm{rel}}}\|\mathbf Q_{\mathrm{rel}}^T\mathbf Q_{\mathcal S}\|_F^2,
    \label{eq:pod_target_coverage}
\end{equation}
Here, $\mathbf{Q}_{\mathcal S}$ is formed from selected POD or CCD directions in weighted space. Let $\mathbf{Q}_p$ be a real orthonormal basis for analytical oscillatory component $p$. The three sustained targets have joint dimension $d_{\mathrm{rel}}=6$, while the single amplitude-modulated target has $d_{\mathrm{rel}}=2$. Coverage of an individual analytical oscillatory component is
\begin{equation}
    \Gamma_p(\mathcal S)
    =\frac{100}{d_p}\|\mathbf Q_p^T\mathbf Q_{\mathcal S}\|_F^2,
    \qquad d_p=2,
    \label{eq:pod_component_coverage}
\end{equation}
where $d_p$ is that component's real-subspace dimension. Overall target-subspace coverage uses the joint orthonormal basis of the three targets. With noise-free target subfield $\mathbf{X}'_{\mathrm{rel}}$ as reference, define
\begin{equation}
    E_{\mathrm{rel}}
    =100\frac{\|\mathbf W^{1/2}(\mathbf X'_{\mathrm{rel}}-\widehat{\mathbf X}')\|_F}
    {\|\mathbf W^{1/2}\mathbf X'_{\mathrm{rel}}\|_F},
    \label{eq:pod_target_field_error}
\end{equation}
Here, $\widehat{\mathbf{X}}'$ combines each method's selected spatial modes and weighted projection coordinates; CCD coordinates follow~\eqref{eq:ccd_comparison_basis}. The metrics $E_y$ and $E_x$ retain definitions~\eqref{eq:observable_metrics} and~\eqref{eq:field_error}, respectively.

Coverage in table~\ref{tab:pod_synthetic_ccd_metrics} is calculated directly from the analytical target space and six selected directions. Cross-response fractions use 1121 common valid samples and the $[0,3.95]$ lag window. All three reconstruction errors are evaluated on the same 1200 snapshots; target regression is fitted and evaluated on all 1200 samples. Flow-field and target-subfield reconstructions retain each method's projection coordinates. POD energy ranking and TCC-POD share the rank-10 POD set; CCD uses the first six singular directions of the full flow-field--observable correlation operator. The analytical target subfield combines three known target oscillations, whereas the full field retains background and noise.

\begin{table}[!htbp]
    \centering
    \caption{POD energy ranking, TCC-POD, and CCD with six real modes in the synthetic field. (a) Analytical target-subspace coverage; $\Gamma_1$, $\Gamma_2$, and $\Gamma_3$ correspond to angular frequencies 1.40, 2.20, and 2.90. (b) Cross-response fraction and target-signal, target-subfield, and full-flow reconstruction errors. All values are percentages.}
    \label{tab:pod_synthetic_ccd_metrics}
    \begin{tabular}{lrrrr}
        \toprule
        \multicolumn{5}{c}{(a) Analytical target-subspace coverage}\\
        Method & $\Gamma_{\mathrm{rel}}$ & $\Gamma_1$ & $\Gamma_2$ & $\Gamma_3$\\
        \midrule
        POD energy ranking & 38.67 & 54.07 & 67.87 & 1.15\\
        TCC-POD & 66.20 & 97.54 & 98.49 & 3.57\\
        CCD & 99.07 & 98.71 & 99.43 & 99.03\\
        \bottomrule
    \end{tabular}
    \par\medskip
    \begin{tabular}{lrrrr}
        \toprule
        \multicolumn{5}{c}{(b) Cross-response and reconstruction errors}\\
        Method & $\eta_C$ & $E_y$ & $E_{\mathrm{rel}}$ & $E_x$\\
        \midrule
        POD energy ranking & 72.16 & 63.20 & 217.96 & 32.43\\
        TCC-POD & 90.45 & 39.09 & 153.18 & 67.16\\
        CCD & 100.00 & 43.93 & 61.43 & 85.87\\
        \bottomrule
    \end{tabular}
\end{table}

\subsection{POD comparisons with \texorpdfstring{CC$_0$}{CC0} and MaxCC, and fixed-basis amplitude modulation}
\label{app:pod_synthetic}

Table~\ref{tab:pod_synthetic_metrics} compares selections with six real modes. POD CC$_0$, MaxCC, and TCC use 1121 common valid samples, whereas POD energy ranking uses $\mu_j$ over all 1200 snapshots. MaxCC and TCC select the same six-dimensional set and thus have identical reconstruction errors. CC$_0$ selects P5, P8, P9, P4, P1, and P10; TCC selects P5, P6, P8, P7, P4, and P1. Their target-signal errors are 8.34\% and 39.09\%, respectively, while TCC gives a slightly lower target-subfield error. These metrics distinguish target regression from subfield recovery using original coefficients. Among all six-mode subsets of the first ten POD modes, the maximum target-subspace coverage is 92.59\%, attained by P5--P10. This exhaustive ground-truth calculation bounds target-subspace coverage when retaining six modes from this fixed set.

\begin{table}[!htbp]
    \centering
    \caption{Target-subspace coverage and reconstruction errors with six real POD modes in the synthetic field. Errors use all 1200 samples; coverage is calculated directly from analytical target and selected subspaces. Criteria select individual modes, and all values are percentages.}
    \label{tab:pod_synthetic_metrics}
    \begin{tabular}{lrrrr}
        \toprule
        Ranking & $\Gamma_{\mathrm{rel}}$ & $E_y$ & $E_{\mathrm{rel}}$ & $E_x$\\
        \midrule
        POD energy ranking & 38.67 & 63.20 & 217.96 & 32.43\\
        CC$_0$ & 68.76 & 8.34 & 157.25 & 69.06\\
        MaxCC & 66.20 & 39.09 & 153.18 & 67.16\\
        TCC & 66.20 & 39.09 & 153.18 & 67.16\\
        \bottomrule
    \end{tabular}
\end{table}
\FloatBarrier

The amplitude-modulation test uses the field and observable from Appendix~\ref{app:synthetic_event_details}, projected onto the fixed POD basis constructed from sustained-oscillation snapshots. We use 800 snapshots, $\Delta t=0.05$, response delay $\delta=1.5$, and target SNR of 8\,dB. Independent real Gaussian field noise has standard deviation equal to 3\% of the noise-free field RMS. We run 200 independent noise trials, sharing the same noise across methods within each trial. Window upper limits are $0,0.25,\ldots,1.5,2.0,2.5$; the largest window fixes 750 common valid samples. Here, POD energy ranking uses mean squared modal contributions on these common samples. Table~\ref{tab:pod_event_coverage} compares two-mode target-subspace coverage for four single-mode rankings at three representative windows.

\begin{table}[!htbp]
    \centering
    \caption{Target-subspace coverage (\%) with two real modes in the fixed-POD-basis amplitude-modulation test. All criteria share the mode set and common valid samples. Values are means $\pm$ standard deviations over 200 independent noise trials, with the same noise used across methods within each trial.}
    \label{tab:pod_event_coverage}
    \begin{tabular}{crrrr}
        \toprule
        $\tau_{\max}$ & POD energy ranking & CC$_0$ & MaxCC & TCC\\
        \midrule
        1.0 & $19.36\pm0.00$ & $24.17\pm3.75$ & $14.61\pm12.29$ & $21.48\pm8.24$\\
        1.5 & $19.36\pm0.00$ & $24.17\pm3.75$ & $43.65\pm5.54$ & $40.50\pm8.70$\\
        2.5 & $19.36\pm0.00$ & $24.17\pm3.75$ & $43.65\pm5.54$ & $34.36\pm11.22$\\
        \bottomrule
    \end{tabular}
\end{table}

The maximum target-subspace coverage achievable with any two modes from this fixed POD set is 45.11\%, revealing the constraint imposed by existing spatial directions. Increasing the window limit from 1.5 to 2.5 reduces mean TCC coverage from 40.50\% to 34.36\%, while MaxCC remains at 43.65\%. Therefore, wider windows do not continuously improve coverage under these settings.

\begingroup
\raggedbottom
\subsection{Complete six-mode comparisons for the cylinder and cavity}
\label{app:pod_full_modes}

Figures~\ref{fig:pod_cylinder_modes_full} and~\ref{fig:pod_cavity_modes_full} show complete six-mode comparisons for the cylinder and cavity, respectively. These use the same rankings, spatial normalisation, and colour scales as the corresponding figures in the main text.

\begin{figure}[!htbp]
    \centering
    \includegraphics[width=0.90\textwidth]{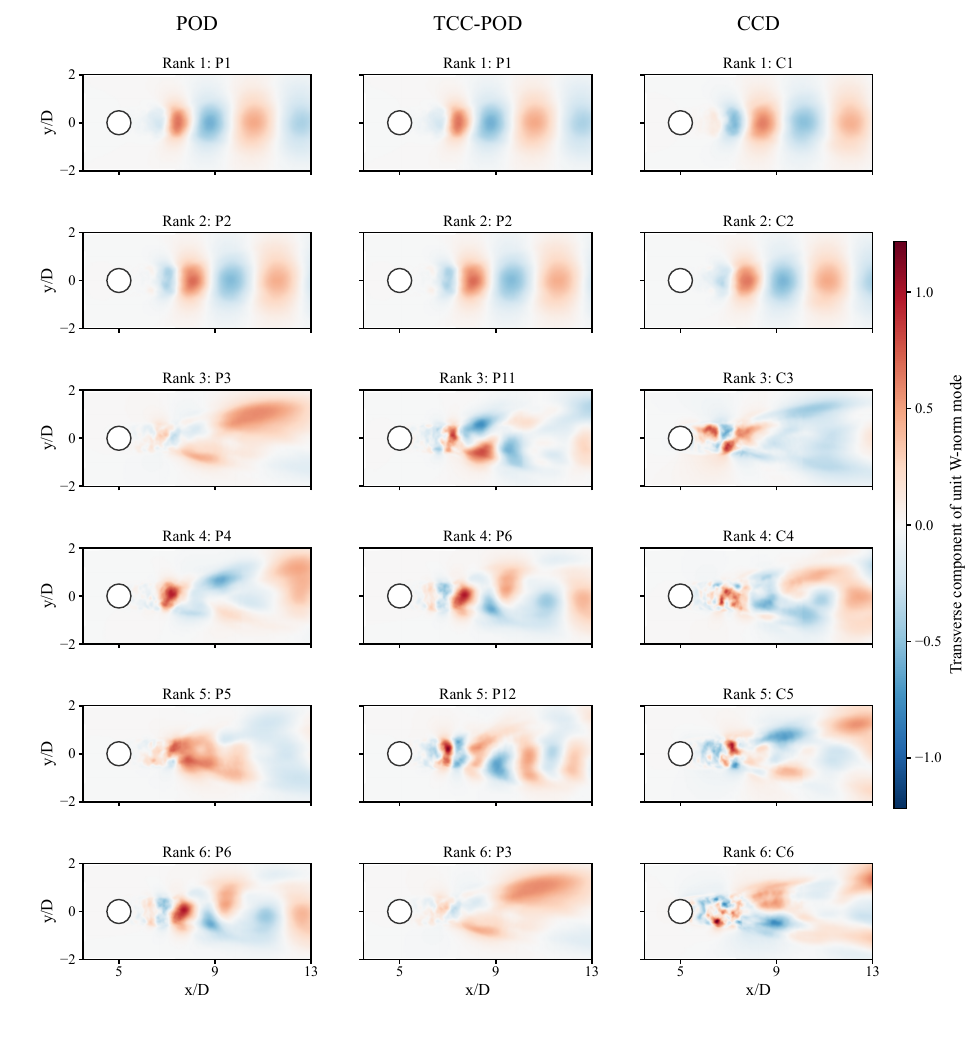}
    \caption{Complete six-mode comparison of POD energy ranking, TCC-POD, and CCD for cylinder lift fluctuations.}
    \label{fig:pod_cylinder_modes_full}
\end{figure}

\begin{figure}[!htbp]
    \centering
    \includegraphics[width=0.90\textwidth]{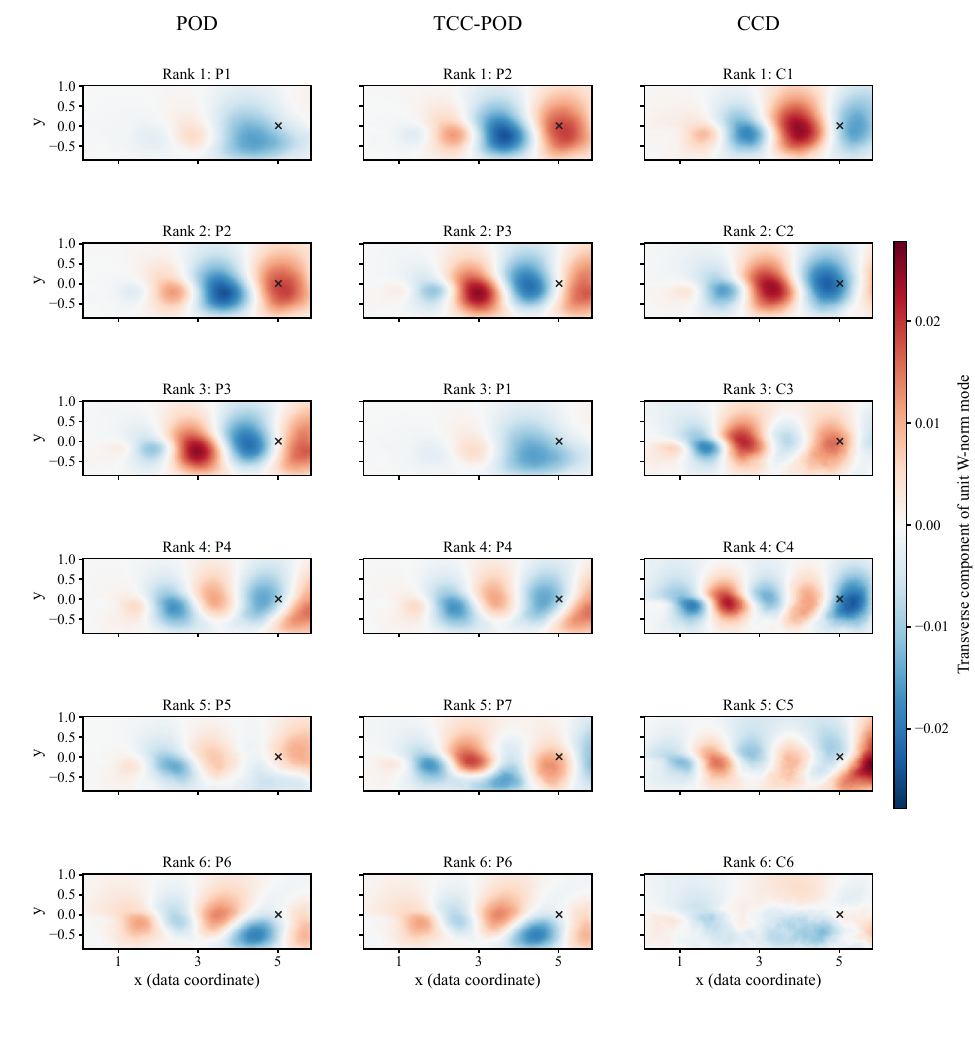}
    \caption{Complete six-mode comparison of POD energy ranking, TCC-POD, and CCD for the cavity probe target. Crosses mark the target probe position.}
    \label{fig:pod_cavity_modes_full}
\end{figure}

\subsection{Cross-responses and flow-field reconstruction}
\label{app:pod_reconstruction}

Figures~\ref{fig:pod_synthetic_metrics}--\ref{fig:pod_cavity_metrics} show complete comparison curves for $K=1$--6; six-mode results are summarised in table~\ref{tab:pod_ccd_summary}. All methods share the correlation operator and complete evaluation samples. For weighted orthogonal bases and their projection coordinates, the retained flow-energy percentage is $100[1-(E_x/100)^2]$. The absolute retained CCD cross-response is the sum of the first $K$ squared correlation singular values.

For the synthetic field, CCD gives lower flow-field errors at $K=1$--4; TCC-POD falls below CCD after including P4 and P1 at $K=5,6$. For the cylinder, all three methods capture over 99\% of the cross-response with two real modes. CCD has slightly lower flow-field errors than TCC-POD at $K=3$--5. For the cavity, the first two TCC-POD modes retain 55.14\% of the cross-response, above 47.55\% for POD energy ranking. TCC-POD flow-field errors are lower than CCD at $K=3$--6 but slightly higher at $K=1,2$. These curves show how cross-response retention and flow-field error depend on retained mode count.

\begin{figure}[!htbp]
    \centering
    \includegraphics[width=\textwidth]{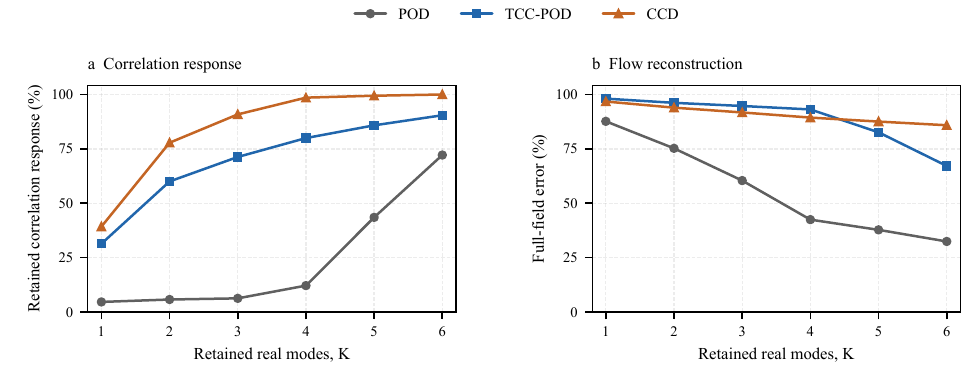}
    \caption{Cross-response and reconstruction comparisons with equal numbers of real modes for the synthetic field. (a) Retained cross-response fraction from~\eqref{eq:retained_correlation_fraction}. (b) Flow-field error over all 1200 snapshots, including both velocity components. POD energy ranking and TCC-POD share the rank-10 POD set and projection coordinates. CCD uses the correlation operator's first $K$ singular directions and their projection coordinates. Every method retains $K$ real modes.}
    \label{fig:pod_synthetic_metrics}
\end{figure}

\begin{figure}[!htbp]
    \centering
    \includegraphics[width=\textwidth]{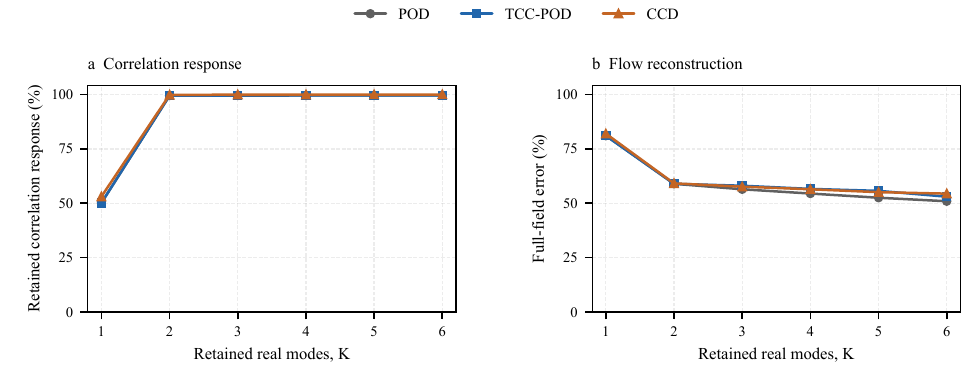}
    \caption{Cross-response and reconstruction comparisons with equal numbers of real modes for the cylinder lift-fluctuation target. (a) Retained cross-response fraction. (b) Area-weighted transverse-velocity fluctuation error over all 310 snapshots. All methods share the flow field, lift-fluctuation target, spatial weights, and 270 common valid samples; flow-field errors use all snapshots.}
    \label{fig:pod_cylinder_metrics}
\end{figure}

Six-mode lift-signal errors are reported in table~\ref{tab:pod_cylinder_signal_metrics}.

\begin{table}[!htbp]
    \centering
    \caption{Lift-fluctuation signal reconstruction error with six real POD modes in the cylinder wake. Both rankings share the rank-252 POD set; regression is fitted and evaluated over all 310 samples, with $E_y$ defined by~\eqref{eq:observable_metrics}.}
    \label{tab:pod_cylinder_signal_metrics}
    \begin{tabular}{lr}
        \toprule
        Method & $E_y$ (\%)\\
        \midrule
        POD energy ranking & 42.47\\
        TCC-POD & 41.42\\
        \bottomrule
    \end{tabular}
\end{table}

\begin{figure}[!htbp]
    \centering
    \includegraphics[width=\textwidth]{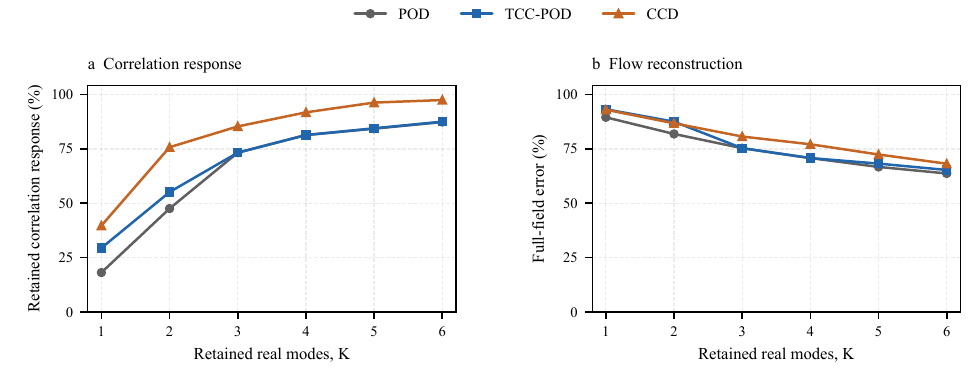}
    \caption{Cross-response and reconstruction comparisons with equal numbers of real modes for the cavity probe target. (a) Retained cross-response fraction. (b) Flow-field error for both velocity components over all 3148 snapshots. POD energy ranking and TCC-POD share the rank-200 POD set; CCD uses the same probe and 3020 common valid samples. Each reconstruction retains its original projection coordinates.}
    \label{fig:pod_cavity_metrics}
\end{figure}

Figures~\ref{fig:pod_synthetic_reconstruction}--\ref{fig:pod_cavity_reconstruction} compare original fluctuations with six-mode reconstructions, retaining each method's snapshot-wise projection coordinates and displaying transverse velocity. The synthetic snapshot is at $t=20.00$. The cylinder and cavity use middle snapshots at $38.75D/U_\infty$ and 393.5\,ms relative to their analysis start times, respectively. All four panels within each case share a symmetric colour scale and retain original velocity amplitudes.

Synthetic POD energy ranking mainly recovers broad background structures and the high-contribution upper component. TCC-POD emphasises central and below-centreline target fluctuations while retaining some broad background; CCD reconstruction is more concentrated in the three target regions. Cylinder TCC-POD and CCD both retain the main alternating wake structures, with differences in downstream contours and local near-wake fluctuations. Both correlation-based cavity reconstructions retain large-scale variations near the observation height and downstream, while small fluctuations diminish under six-mode truncation. These snapshots complement the all-snapshot errors in table~\ref{tab:pod_ccd_summary} by showing which spatial structures are retained.

\begin{figure}[H]
    \centering
    \includegraphics[width=\textwidth]{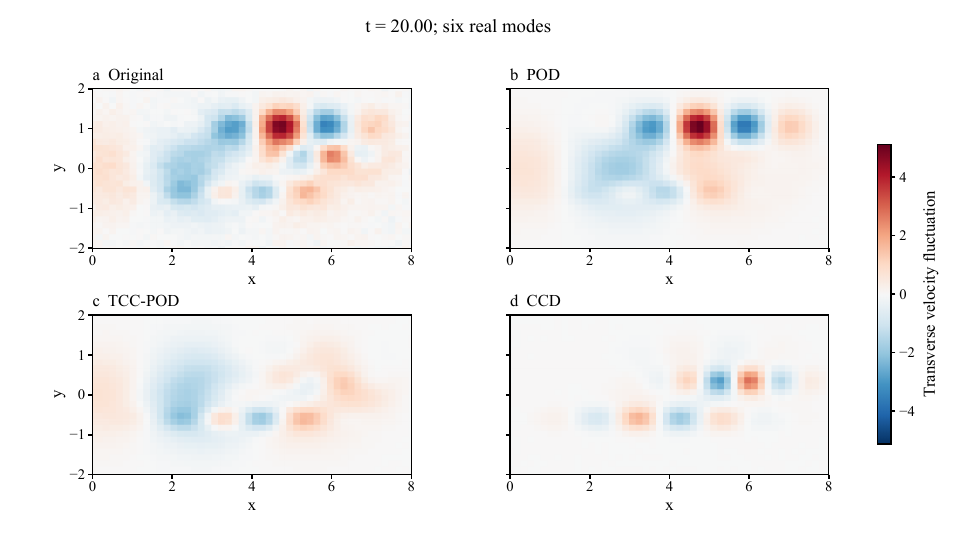}
    \caption{Synthetic fluctuation-field reconstruction at $t=20.00$ with six real modes per method. (a) Original transverse velocity fluctuations. (b) POD energy ranking. (c) TCC-POD. (d) CCD. All methods use the same snapshot and colour scale; colours show actual reconstructed transverse-velocity amplitudes.}
    \label{fig:pod_synthetic_reconstruction}
\end{figure}

\begin{figure}[H]
    \centering
    \includegraphics[width=\textwidth]{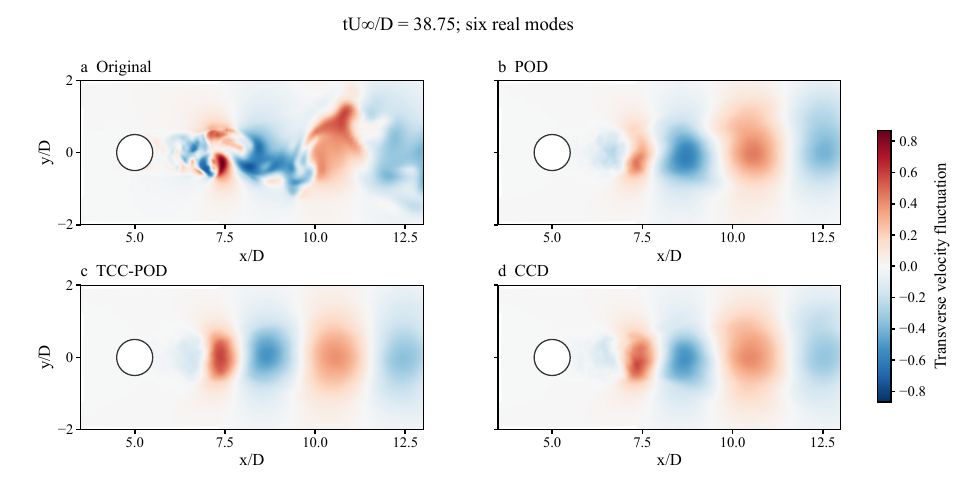}
    \caption{Cylinder fluctuation-field reconstruction at relative time $tU_\infty/D=38.75$, targeting lift fluctuations with six real modes per method. (a) Original transverse velocity fluctuations. (b) POD energy ranking. (c) TCC-POD. (d) CCD. All four panels share a colour scale and the original mesh display mask; colours represent dimensionless transverse velocity fluctuations $v'/U_\infty$.}
    \label{fig:pod_cylinder_reconstruction}
\end{figure}

\begin{figure}[H]
    \centering
    \includegraphics[width=\textwidth]{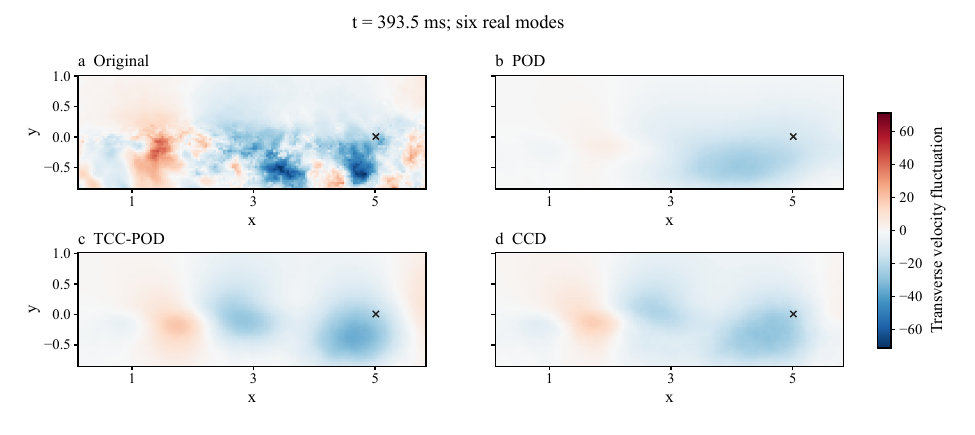}
    \caption{Cavity fluctuation-field reconstruction 393.5\,ms after the pre-processed sequence begins, with six real modes per method. (a) Original transverse velocity fluctuations. (b) POD energy ranking. (c) TCC-POD. (d) CCD. Both correlation-based methods retain the original transverse-velocity probe as target, marked by a cross. All four panels share a colour scale; colours represent transverse velocity fluctuations $u_y'$ in $\mathrm{m\,s^{-1}}$.}
    \label{fig:pod_cavity_reconstruction}
\end{figure}

\FloatBarrier
\endgroup

\end{appen}

\begin{bmhead}[Acknowledgements.]

The authors express their sincere gratitude to Professor Jing Tian from the Institute of Acoustics, Chinese Academy of Sciences, for his invaluable guidance and support throughout this study.

\end{bmhead}

\begin{bmhead}[Funding.]

This work was supported by the National Natural Science Foundation of China (No. 11904385).

\end{bmhead}

\begin{bmhead}[Declaration of interests.]

The authors declare that they have no known competing financial interests or personal relationships that could have appeared to influence the work reported in this paper.

\end{bmhead}

\begin{bmhead}[Data availability.]

The original cylinder OpenFOAM data are available through Recherche Data Gouv at \doi{10.57745/6LG9AS}~\citep{wallian2025re3900}. The cavity Challenge 2.1 training data are available through the \href{https://fluids-challenge.engin.umich.edu/}{AIAA Reduced-Complexity Modelling Community Challenge website}~\citep{schmidt2026community}. Additional data will be made available on request.

\end{bmhead}

\bibliographystyle{jfm}
\bibliography{references}

\end{document}